%% file: main.tex
\documentclass[a4paper,11pt]{article}
\pdfoutput=1 

\usepackage{jheppub} 
\usepackage[utf8]{inputenc}
\usepackage{siunitx}
\usepackage{xcolor,colortbl,listings}

\usepackage{amssymb}
\usepackage{epstopdf}
\usepackage{epsfig}
\usepackage{array}
\usepackage{lscape}
\usepackage[noabbrev,capitalise]{cleveref}
\usepackage{eurosym}
\usepackage{xspace}
\usepackage[section]{placeins}
\usepackage{longtable}
\usepackage{tabularx}
\usepackage{afterpage}
\usepackage{float}
\usepackage{lscape}
\usepackage{psfrag}
\usepackage{subfigure}
\usepackage{wrapfig}
\usepackage{lastpage}
\usepackage{fancyhdr}
\usepackage{bbding,upgreek,pifont,commath,wasysym,fourier-orns}
\usepackage{enumitem}
\usepackage[super]{nth}
\usepackage[export]{adjustbox}
\usepackage[b]{esvect}
\usepackage{environ,indentfirst}

\usepackage{tikz}
\usetikzlibrary{patterns}
\usetikzlibrary{plotmarks}
\usetikzlibrary{external}

\usepackage{amssymb,amsmath,mathtools,cases,enumitem}

\usepackage{booktabs,multirow,ragged2e}

\newcolumntype{L}[1]{>{\raggedright\let\newline\\\arraybackslash\hspace{0pt}}m{#1}}
\newcolumntype{C}[1]{>{\centering\let\newline\\\arraybackslash\hspace{0pt}}m{#1}}
\newcolumntype{R}[1]{>{\raggedleft\let\newline\\\arraybackslash\hspace{0pt}}m{#1}}
\newcolumntype{N}{@{}m{0pt}@{}}

\newcommand{\percent}{\%}
\DeclareMathOperator{\MFOM}{MFOM}

\newcommand{\lsim}{\mathrel{\mathop{\kern 0pt \rlap
  {\raise.2ex\hbox{$<$}}}
  \lower.9ex\hbox{\kern-.190em $\sim$}}}
\newcommand{\gsim}{\mathrel{\mathop{\kern 0pt \rlap
  {\raise.2ex\hbox{$>$}}}
  \lower.9ex\hbox{\kern-.190em $\sim$}}}

\newcommand{\alt}{\mathrel{\mathop{\kern 0pt \rlap
  {\raise.2ex\hbox{$<$}}}
  \lower.9ex\hbox{\kern-.190em $\sim$}}}
\newcommand{\agt}{\mathrel{\mathop{\kern 0pt \rlap
  {\raise.2ex\hbox{$>$}}}
  \lower.9ex\hbox{\kern-.190em $\sim$}}}

\newcommand{\gagamma}{g_{a\gamma}}
\newcommand{\ckcs}{counts~keV$^{-1}$~cm$^{-2}$~s$^{-1}$ }
\newcommand{\pqsym}{U(1)$_{\rm PQ}$}
\newcommand{\gae}{g_{ae}}

\newcommand{\exclude}[1]{}

\title{Conceptual Design of BabyIAXO, the intermediate stage towards the 
International Axion Observatory}

\collaboration{
    \includegraphics[height=17mm]{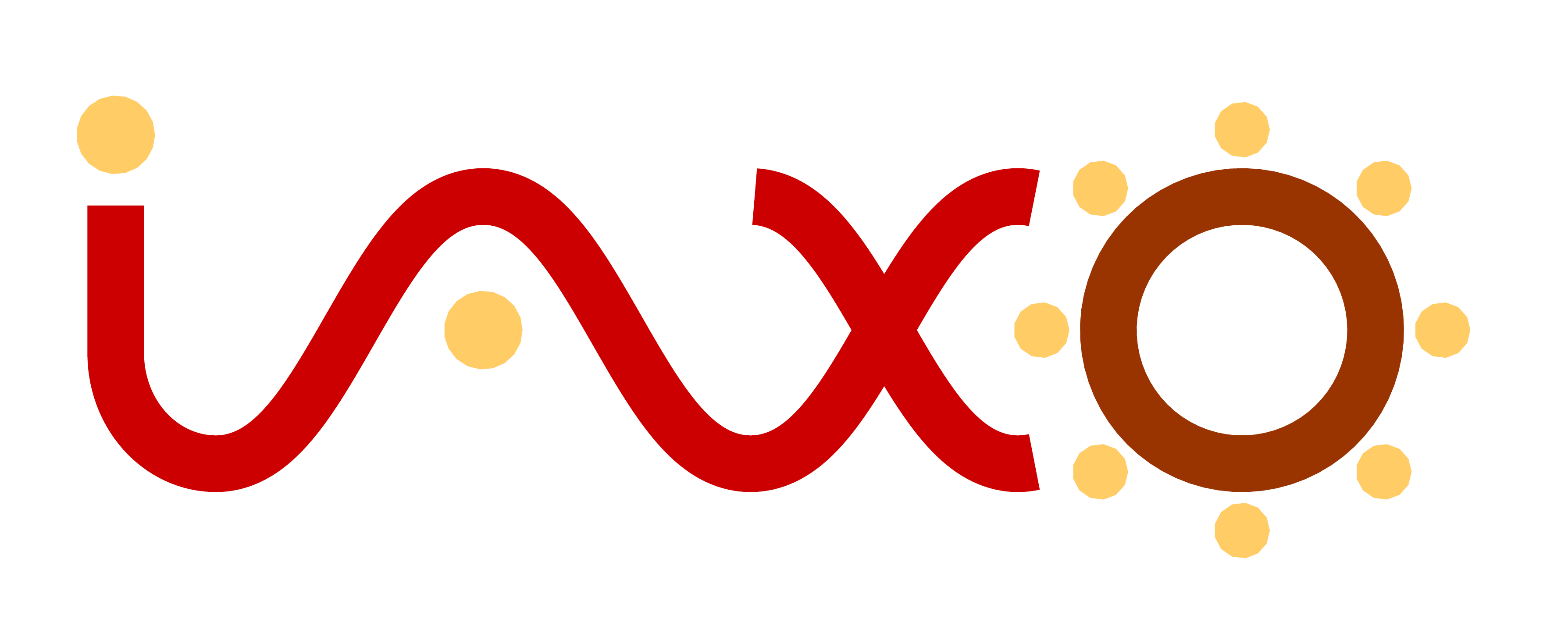}\\ [6pt] IAXO collaboration}

\input{IAXO_Authorlist.tex}

\author{\IAXOAuthorList
\\
\IAXOAffiliationList}

\emailAdd{Igor.Irastorza@unizar.es}

\abstract{This article describes BabyIAXO, an intermediate experimental stage of the International Axion Observatory (IAXO), proposed to be sited at DESY. IAXO is a large-scale axion helioscope that will look for axions and axion-like particles (ALPs), produced in the Sun, with unprecedented sensitivity. BabyIAXO is conceived to test all IAXO subsystems (magnet, optics and detectors) at a relevant scale for the final system and thus serve as prototype for IAXO, but at the same time as a fully-fledged helioscope with relevant physics reach itself, and with potential for discovery. 
The BabyIAXO magnet will feature two \SI{10}{m} long, \SI{70}{cm} diameter bores, and will host two detection lines (optics and detector) of dimensions similar to the final ones foreseen for IAXO. 
BabyIAXO will detect or reject solar axions or ALPs with axion-photon couplings down to $\gagamma \sim 1.5 \times 10^{-11}$ GeV$^{-1}$, and masses up to $m_a\sim 0.25$ eV. BabyIAXO will offer additional opportunities for axion research in view of IAXO, like the development of precision x-ray detectors to identify particular spectral features in the solar axion spectrum, and the implementation of radiofrequency-cavity-based axion dark matter setups.
}

\begin{document}

\maketitle

\section{Introduction}
\label{sec:intro}
\input{sections/BabyIAXO_intro.tex}

\section{IAXO and BabyIAXO}
\label{sec:iaxo}
\input{sections/BabyIAXO_IAXO.tex}

\section{BabyIAXO as an intermediate goal towards IAXO }
\label{sec:BabyIAXO_setup}
\input{sections/BabyIAXO_setup.tex}

\section{BabyIAXO magnet}
\label{sec:BabyIAXO_magnet}
\input{sections/BabyIAXO_magnet.tex}

\section{BabyIAXO optics}
\label{sec:BabyIAXO_optics}
\input{sections/BabyIAXO_optics.tex}

\section{BabyIAXO detectors}
\label{sec:BabyIAXO_detectors}
\input{sections/BabyIAXO_detectors.tex}

\input{sections/BabyIAXO_DM_detectors.tex}

\section{BabyIAXO structure and drive system}
\label{sec:BabyIAXO_drive}
\input{sections/BabyIAXO_SupportDriveSystem.tex}

\section{BabyIAXO site and infrastructure}
\label{sec:BabyIAXO_infra}
\input{sections/BabyIAXO_infra.tex}


\section{Conclusions}
\label{sec:BabyIAXO_conclusions}
\input{sections/BabyIAXO_conclusions.tex}
\section{Acknowledgements}

We want to devote this paper to the memory of our dearest friend, colleague and IAXO enthusiast Biljana Lakic, who passed away recently. We acknowledge support from the the European Research Council (ERC) under the European
Union’s Horizon 2020 research and innovation programme, grant agreement ERC-2017-AdG788781 (IAXO+) as well as ERC-2018-StG-802836 (AxScale), and from the Spanish Agencia Estatal de Investigación  
under grant FPA2016-76978-C3-1-P, the coordinated grant PID2019-108122GB, and the Maria de Maeztu grant CEX2019-000918-M. 
The groups at Bonn University, Heidelberg University, Mainz University and Siegen University are supported by BMBF contract 05H2018-R\&D DETEKTOREN.
The groups IRFU/CEA, LIST/CEA and IJCLAB acknowledge support from  the Agence Nationale de la Recherche (France) ANR-19-CE31-0024. The groups at INR and MIPT are supported by Ministry of Science and Higher Education of the  Russian Federation under the contract  075-15-2020-778. We also thank the Physics Beyond Colliders (PBC) initiative at CERN for its support to this work.


\bibliographystyle{JHEP}
\bibliography{bib/IAXObib,bib/detectors}


\end{document}

%% file: IAXO_Authorlist.tex
\newcommand{\UHEI}{$^{1}$}
\newcommand{\UNIZAR}{$^{2}$}
\newcommand{\CERN}{$^{3}$}
\newcommand{\CEAIrfu}{$^{4}$}
\newcommand{\INAF}{$^{5}$}
\newcommand{\IJCLAB}{$^{6}$}
\newcommand{\MIPT}{$^{7}$}
\newcommand{\INR}{$^{8}$}
\newcommand{\DTU}{$^{9}$}
\newcommand{\ICCUB}{$^{10}$}
\newcommand{\UBFQA}{$^{11}$}
\newcommand{\PNPI}{$^{12}$}
\newcommand{\UBonn}{$^{13}$}
\newcommand{\USiegen}{$^{14}$}
\newcommand{\BARRY}{$^{15}$}
\newcommand{\JENA}{$^{16}$}
\newcommand{\Columbia}{$^{17}$}
\newcommand{\CEFCA}{$^{18}$}
\newcommand{\SOLEIL}{$^{19}$}
\newcommand{\RBI}{$^{20}$}
\newcommand{\DESY}{$^{21}$}
\newcommand{\MPP}{$^{22}$}
\newcommand{\TUM}{$^{23}$}
\newcommand{\CEAList}{$^{24}$}
\newcommand{\ICREA}{$^{25}$}
\newcommand{\MIT}{$^{26}$}
\newcommand{\LLNL}{$^{27}$}
\newcommand{\SLAC}{$^{28}$}
\newcommand{\Mainz}{$^{29}$}
\newcommand{\UCT}{$^{30}$}

\newcommand{\UNIZARname}{Center for Astroparticles and High Energy Physics (CAPA), Universidad de Zaragoza, 50009 Zaragoza, Spain}
\newcommand{\BARRYname}{Physical Sciences, Barry University, 11300 NE 2nd Ave., Miami Shores, FL 33161, USA}
\newcommand{\MITname}{Massachusetts Institute of Technology, 77 Massachusetts Ave., Cambridge, MA 02139, USA}
\newcommand{\CEFCAname}{Centro de Estudios de F\'{\i}sica del Cosmos de Arag\'on, Plaza San Juan, Teruel, Spain}
\newcommand{\ICCUBname}{Institut de Ci\`encies del Cosmos, Universitat de Barcelona, Barcelona, Spain}
\newcommand{\UBFQAname}{Departament de F\'isica Qu\`antica i Astrof\'isica, Universitat de Barcelona, Barcelona, Spain}
\newcommand{\ICREAname}{ICREA, Barcelona, Spain}
\newcommand{\Mainzname}{Johannes Gutenberg University, Mainz, Germany}
\newcommand{\PNPIname}{Petersburg Nuclear Physics Institute - NRC Kurchatov Institute, Gatchina, 188300, Russia}
\newcommand{\RBIname}{Rudjer Bo\v{s}kovi\'{c} Institute, Bijeni\v{c}ka cesta 54, 10000 Zagreb, Croatia}
\newcommand{\UBonnname}{Physikalisches Institut der Universit\"at Bonn, Nussallee 12, 53115 Bonn, Germany}
\newcommand{\UCTname}{High Energy Physics, Cosmology \& Astrophysics Theory (HEPCAT) group, University of Cape Town, Private Bag, 7700 Rondebosch, South Africa}
\newcommand{\DESYname}{Deutsches Elektronen-Synchrotron DESY, Hamburg, Germany}
\newcommand{\CERNname}{CERN - European Organization for Nuclear Research, Geneva, Switzerland}
\newcommand{\UHEIname}{Heidelberg University, Kirchhoff Institute for Physics}
\newcommand{\LLNLname}{Lawrence Livermore National Laboratory, Livermore, CA, U.S.A.}
\newcommand{\Columbianame}{Columbia Astrophysics Laboratory, New York, U.S.A.}
\newcommand{\DTUname}{Technical University of Denmark, DTU Space Kgs. Lyngby, Denmark}
\newcommand{\SLACname}{SLAC National Accelerator Laboratory, Menlo Park, CA, U.S.A.}
\newcommand{\JENAname}{Institute  for  Optics  and  Quantum  Electronics,  Friedrich  Schiller  University  Jena,  Jena,  Germany}
\newcommand{\INAFname}{INAF, Italian National Institute for Astrophysics, Osservatorio Astronomico di Brera, Milano/Merate, Italy}
\newcommand{\INRname}{Institute  for Nuclear Research of the Russian Academy of Sciences, 60th October Anniversary Prospect 7A, 117312, Moscow, Russia}
\newcommand{\USiegenname}{Center for Particle Physics Siegen, Siegen University, Siegen, Germany}
\newcommand{\CEAIrfuname}{IRFU, CEA, Universit\'e Paris-Saclay, F-91191 Gif-sur-Yvette, France}
\newcommand{\CEAListname}{CEA, LIST, Laboratoire National Henri Becquerel (LNE-LNHB), F-91191 Gif-sur-Yvette, France}
\newcommand{\MIPTname}{Moscow Institute of Physics and Technology, Moscow, Russia}
\newcommand{\IJCLABname}{Universit\'{e} Paris-Saclay, CNRS/IN2P3, IJCLab, 91405 Orsay, France}
\newcommand{\SOLEILname}{Synchrotron SOLEIL, 91192 Gif-sur-Yvette, France }
\newcommand{\MPPname}{Max-Planck-Institut f\"{u}r Physik, F\"{o}hringer Ring 6, 80805 M\"{u}nchen, Germany}
\newcommand{\TUMname}{Technische Universit\"{a}t M\"{u}nchen, James-Franck-Str. 1, 85748 Garching, Germany}

\newcommand{\IAXOAffiliationList}{
\UHEI\UHEIname\\
\UNIZAR\UNIZARname\\
\CERN\CERNname\\
\CEAIrfu\CEAIrfuname\\
\INAF\INAFname\\
\IJCLAB\IJCLABname\\
\MIPT\MIPTname\\
\INR\INRname\\
\DTU\DTUname \\
\ICCUB\ICCUBname\\
\UBFQA\UBFQAname\\
\PNPI\PNPIname\\
\UBonn\UBonnname\\
\USiegen\USiegenname\\
\BARRY\BARRYname\\
\JENA\JENAname\\
\Columbia\Columbianame \\
\CEFCA\CEFCAname\\
\SOLEIL\SOLEILname\\
\RBI\RBIname\\
\DESY\DESYname\\
\MPP\MPPname\\
\TUM\TUMname\\
\CEAList\CEAListname\\
\ICREA\ICREAname\\
\MIT\MITname\\
\LLNL\LLNLname\\
\SLAC\SLACname\\
\Mainz\Mainzname\\
\UCT\UCTname\\

}

\newcommand{\IAXOAuthorList}{
A.~Abeln\UHEI,
K.~Altenm\"uller\UNIZAR,
S.~Arguedas Cuendis\CERN,
E.~Armengaud\CEAIrfu,
D.~Atti\'e\CEAIrfu,
S.~Aune\CEAIrfu,
S.~Basso\INAF,
L.~Berg\'e\IJCLAB,
B.~Biasuzzi\CEAIrfu,
P.~T.~C.~Borges~De~Sousa\CERN
% ...
}

%% file: sections/BabyIAXO_intro.tex

Axions are necessary ingredients of the Peccei-Quinn (PQ) mechanism~\cite{Peccei:1977hh,Peccei:1977ur}, the most compelling solution to the \textit{strong CP problem} of the Standard Model (SM) of particle physics. This problem arises from the fact that Quantum Chromodynamics (QCD) seems to be absent ---at the current experimental precision--- of CP-violating effects, while one would expect them. The PQ mechanism involves a new global U(1) chiral symmetry, now known as PQ (Peccei-Quinn) symmetry \pqsym
, which is spontaneously broken at a large energy scale $f_a$. The axion is the Nambu-Goldstone boson of the broken \pqsym\ symmetry~\cite{Weinberg:1977ma,Wilczek:1977pj}.

The phenomenological properties of the axion are mainly determined by the scale $f_a$ and are closely related to those of the neutral pion. Due to the \pqsym\ symmetry not being exact at the quantum level, as a result of a chiral anomaly, the axion is not massless (although it is typically very light). The axion mass $m_a$ and all axion couplings to ordinary particles (photons, nucleons, electrons) are inversely proportional to $f_a$.

In order to avoid accelerator-experiment-based constraints, $f_a$ must be much greater than the electroweak scale, thus making the axion 
very light, very weakly coupled, and very long-lived. Two classes of benchmark axion models are often referred to: KSVZ or hadronic axions \cite{Kim:1979if,Shifman:1979if} and DFSZ or GUT axions~\cite{Dine:1981rt,Zhitnitsky:1980tq}. The main difference between the KSVZ and DFSZ-type axions is that the former do not couple to ordinary quarks and leptons at the tree level. Of course, many other possible axion models have been discussed in the literature~\cite{Irastorza:2018dyq}, although the mentioned ones represent well the phenomenology at helioscopes.

Although the axion is the most studied prototype, a whole category of axion-like particles (ALPs) or, more generically, weakly interacting slim particles (WISPs) appear in extensions of the SM. ALPs can appear as pseudo Nambu-Goldstone bosons of new symmetries broken at high energy. Moreover, string theory also predicts a rich spectrum of ALPs. The phenomenology of axions and ALPs is close enough to permit their search for them with similar experiments. Of particular importance for experiments is that both axions and ALPs couple weakly to two photons, with the difference that for ALPs the coupling constant to photons $\gagamma$ is unrelated to the mass $m_a$.

Axions are just as attractive a solution to the dark matter (DM) problem as weakly interacting massive particles (WIMPs). Relic axions can be produced thermally by collisions of particles in the primordial plasma, just like WIMPs. Being quite light particles, this axion population contributes to the hot DM component. Most interesting from the cosmological point of view is the non-thermal production of axions: the vacuum-realignment mechanism and the decay of topological defects (axion strings and domain walls), both producing non-relativistic axions and therefore contributing to the cold DM. The non-thermal production mechanisms are generic to other bosonic WISPs such as ALPs or hidden photons. In both cases a wide range of parameter space (in the case of ALPs $\gagamma$-$m_a$ space) can contain models with adequate DM density, part of it at reach of current or future experiments.

Axions and ALPs have been searched for since their proposal 40 years ago, and remains elusive so far. Only recently the interest in axion searches is increasing substantially, as detection technologies have reached sensitivity to motivated regions of the axion parameters space. 
The experimental landscape is growing in diversity and size of projects~\cite{Irastorza:2018dyq}. The most prominent experimental strategies  rely on the generic axion-photon coupling $\gagamma$ and the use of strong magnetic fields where axions can convert into photons that can be subsequently detected. There are three main categories of searches, that can be distinguished depending on the source of axions: 
\begin{itemize}
\item laboratory experiments that search for 
axion-related phenomena produced entirely in the laboratory; 
\item helioscopes searching for axions that could be produced in various processes in the solar interior;
\item  haloscopes or other detectors directly sensitive to relic axions from the dark matter halo of our galaxy. 
\end{itemize}
Conceptually elegant, as they do not rely on cosmological or astrophysical assumptions, laboratory experiments cannot however reach sensitivity to the $\gagamma$ values typically expected for QCD axions. Haloscopes can reach sensitivity to QCD axions for some ranges of $m_a$ (while intense R\&D is ongoing to extend it), although relying on the assumption that dark matter is (mostly) made of axions. For subdominant axion dark matter, the sensitivity of these experiments to $\gagamma$ must be scaled accordingly.

Helioscopes~\cite{Sikivie:1983ip} look for axions and ALPs that could be emitted from the Sun and therefore do not rely on the assumption of axions being dark matter candidates. The emission of axions by the Sun is a generic prediction of most axion models and therefore helioscopes represent the only approach that combines a relative immunity to model assumptions and a competitive sensitivity to parameters largely complementary to those accessible with other detection techniques. They use strong laboratory magnetic fields to convert solar axions into detectable x-ray photons with $\sim$keV energies. Contrary to haloscopes, the helioscope sensitivity is independent on the axion mass up to relatively large values (up to $\sim$\SI{0.02}{eV}). The most advanced helioscope so far has been the CERN Axion Solar Telescope (CAST), active at CERN for more than 15 years~\cite{Zioutas:1998cc,Zioutas:2004hi,Andriamonje:2007ew,Arik:2008mq,Arik:2013nya,Anastassopoulos:2017ftl}. During its last solar axion campaign, CAST obtained the most stringent limit on the axion-photon coupling $\gagamma$~\cite{Anastassopoulos:2017ftl}, and hosted activities already in anticipation of the next generation axion helioscope. Most relevantly, this result was obtained in part thanks to the implementation of the ``IAXO pathfinder'' detection line~\cite{Aznar:2015iia}, including an x-ray telescope and a low-background Micromegas detector, both based on the same technologies proposed for International Axion Observatory (IAXO)~\cite{Irastorza:2011gs}.

Together with haloscopes, helioscopes are the only technique with proven sensitivity to explore realistic QCD axion models. 
An advantage of helioscopes is that there is a clear scaling strategy~\cite{Irastorza:2011gs} to substantially push the present sensitivity frontline to lower values of $\gagamma$ and $m_a$, a strategy that is implemented by the IAXO as detailed below.

%% file: sections/BabyIAXO_IAXO.tex

IAXO is a new generation axion helioscope designed to search for solar axions and ALPs with the axion-photon coupling $\gagamma$ and the axion-electron coupling $\gae$. Similarly IAXO will test models of other proposed particles at the low energy frontier of particle physics, like hidden photons or chameleons. In addition, the IAXO magnet has been designed to easily accommodate new equipment (e.g., microwave cavities or antennas) to search for relic axions. The discovery of the QCD axion or an ALP in the parameter space accessible by IAXO could have strong implications in other contexts of theoretical particle physics, cosmology and astrophysics.

\begin{figure}[t]
	\begin{center}
		\includegraphics[width=0.95\textwidth]{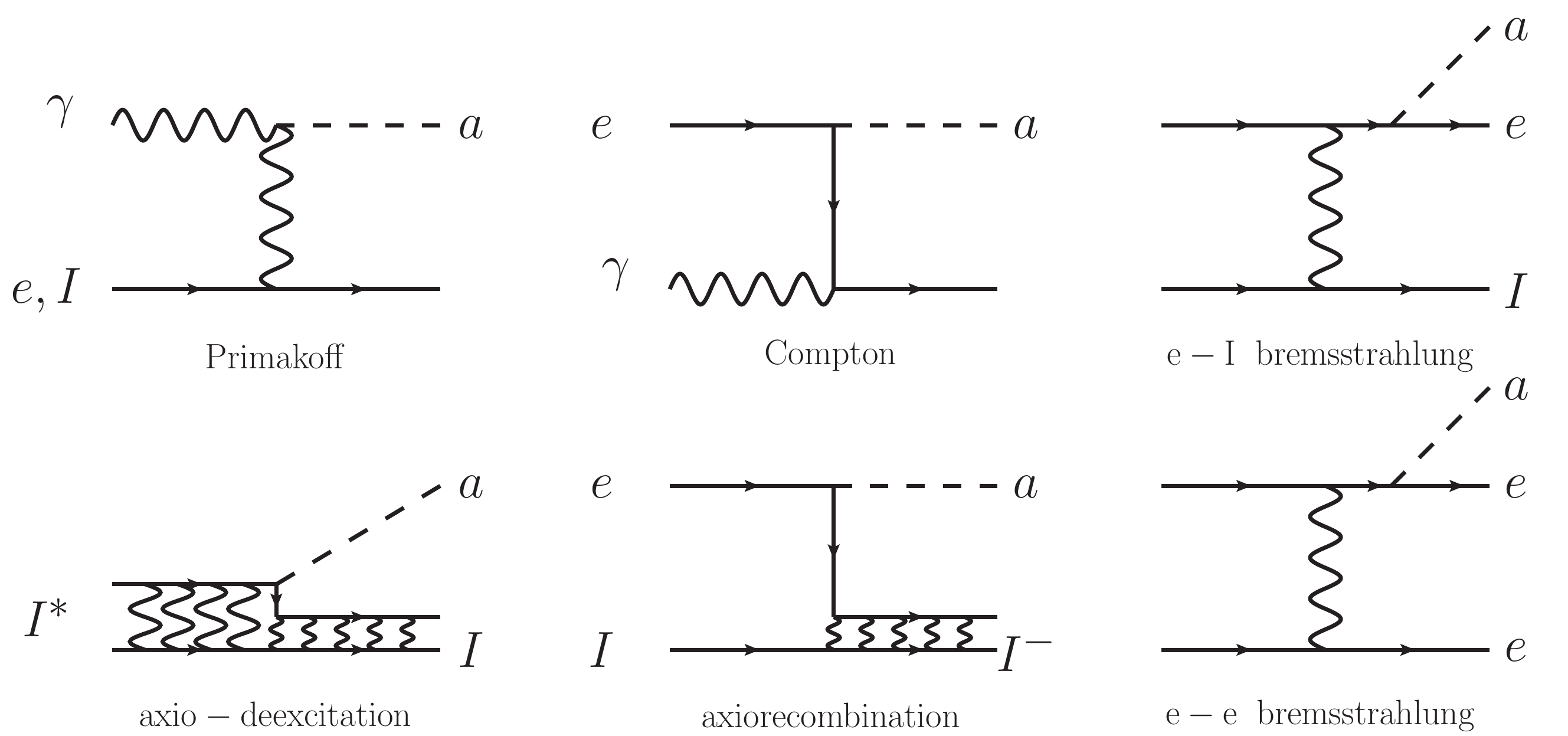} 
	\end{center}
	\caption{Feynman diagrams of the different processes responsible for axion production in the Sun. The Primakoff conversion of photons in the electromagnetic fields of the solar plasma depends on the axion-photon coupling $\gagamma$, and is present is practically every axion model. In non-hadronic models, in which axions couple with electrons at tree level, additional mechanisms of solar axion production are relevant, namely: atomic axio-recombination and axion-deexcitation, axio-Bremsstrahlung in electron-ion  or electron-electron collisions and Compton scattering with emission of an axion. Collectively, the flux of solar axions from all these latter $g_{ae}$-mediated channels is sometimes called ABC solar axions, from the initials of the mentioned processes. In the diagrams, the letters $\gamma$, $a$, $e$ and $I$ represent a photon, axion, electron and ion respectively. Figure from ref~\cite{Redondo:2013wwa}.}
	\label{fig:processes}
\end{figure}

\begin{figure}[t!]
\begin{center}
\includegraphics[width=0.8\textwidth]{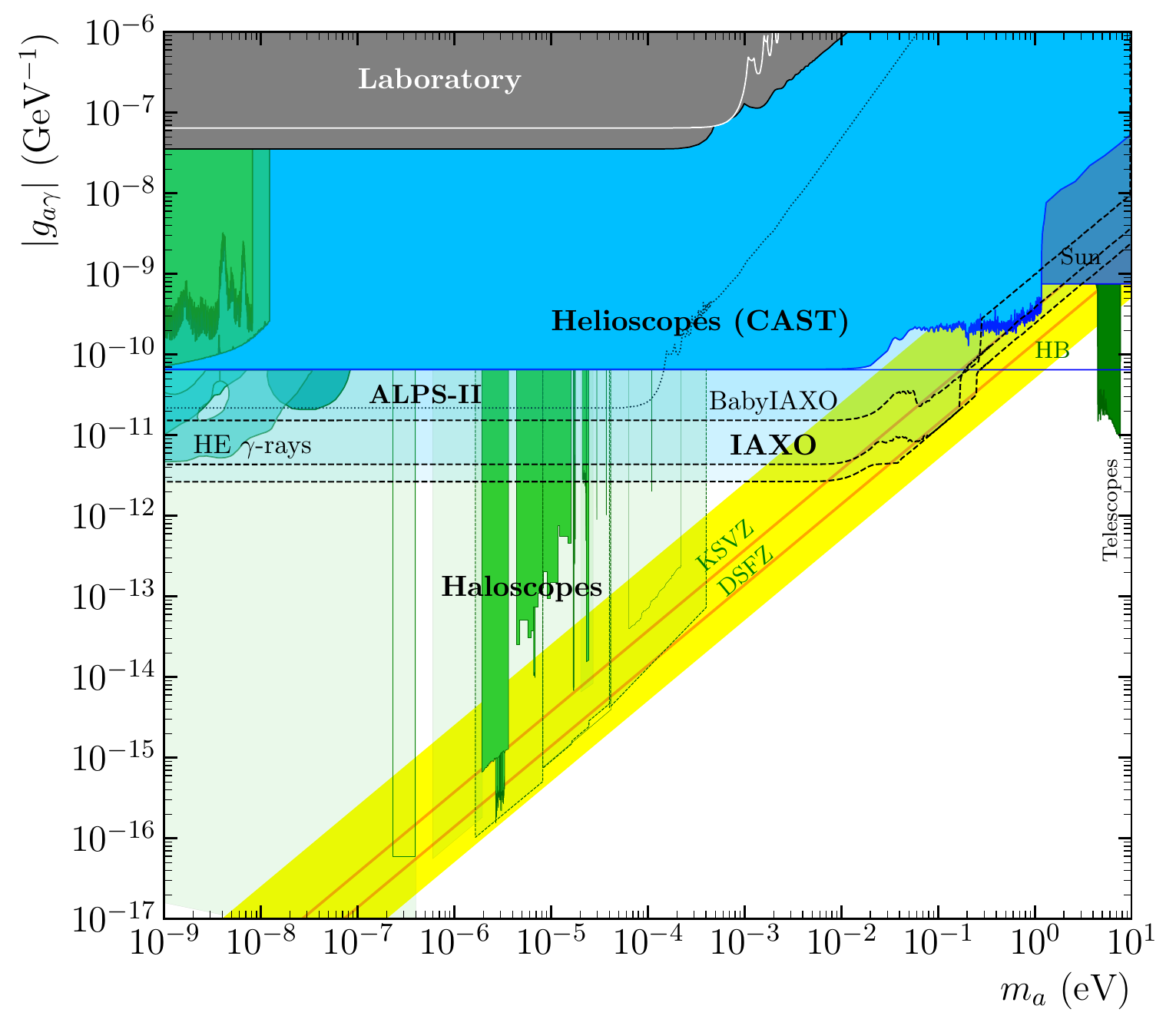}
\caption{Sensitivity plot of IAXO and BabyIAXO in the primary $\gagamma$-$m_a$ parameter space, compared with the QCD axion (yellow) band~\cite{DiLuzio:2016sbl} and other current (solid) and future (dashed) experimental and observational limits (we refer to~\cite{Irastorza:2018dyq} for details on those limits). The yellow band represent the standard QCD axion models and the orange line the benchmark KSVZ model.}
    \label{fig:sensi}
\end{center}
\end{figure}

As its primary physics goal, IAXO will look for axions and ALPs originated in the Sun (see Figure~\ref{fig:processes}) and converted to photons, by inverse Primakoff conversion, in a laboratory magnetic field. These photons are then focused by x-ray telescopes into small focal spots that are imaged with low-background x-ray detectors~\cite{Irastorza:2011gs}. The energies of these axions and ALPs fall in the range of about \SIrange{1}{10}{keV}. IAXO aims to improve the current CAST state-of-the-art by more than a factor of \num{e4} in signal-to-noise ratio (SNR). In terms of sensitivity to $\gagamma$, this represents an improvement of more than one order of magnitude with respect to current astrophysical and experimental upper bounds. This will allow to explore $\gagamma$ values down to \SI{e-12}{\per\GeV} for a wide range of axion masses. The current sensitivity curves in the $\gagamma - m_a$ plane are shown in figure~\ref{fig:sensi}, for both BabyIAXO and IAXO. For the latter, two curves are shown: the nominal projection (corresponding to the set of experimental parameters of the current IAXO conceptual design \cite{Armengaud:2014gea}) and a possible improved scenario, referred to as IAXO+, with an additional factor 10 better SNR, that could be potentially implemented after the experience of the BabyIAXO stage. The values for the main experimental parameters for both BabyIAXO and IAXO (and the latter for both nominal and improved scenarios) on which the aforementioned prospects are based, are listed in Table~\ref{tab:scenarios}. The explanation and justification of these values can be found in~\cite{physics_paper}.

\begin{table}[tb]
\footnotesize
\centering
\begin{tabular}{cccccc}
\hline  \textbf{Parameter} & \textbf{Units} & \textbf{CAST} & \textbf{ BabyIAXO} & \textbf{IAXO  }  & \textbf{IAXO+ }\\
\hline \\
 $B$           & T  &    9    & $\sim$2       & $\sim$2.5  & $\sim$3.5        \\
 $L$           & m   & 9.26       & 10       & 20    & 22    \\
 $A$           & m$^2$  & 0.003 $(^*)$   & 0.77  & 2.3  & 3.9  \\
                                                                    \\
\hline
 $f_M$         &   T$^2$m$^4$    & 21   & $\sim$230              & $\sim$6000     & $\sim$24000 \\
                                                    \\

 $b$     &          ${\rm keV^{-1}\, cm^{-2}\, s^{-1}}$ & $1\times10^{-6}$ ($^{**}$) & $1\times10^{-7}$ & $10^{-8}$& $10^{-9}$\\
 $\epsilon_d$  &      & $\sim0.6$         & 0.7      & 0.8 & 0.8   \\
 $\epsilon_o$  &    &       0.3    & 0.35      & 0.7 & 0.7   \\
 $a$             &  cm$^{2}$    & 0.15 & 2 $\times$ 0.3   & 8 $\times$ 0.15  & 8 $\times$ 0.15  \\
                                                                        \\
 $\epsilon_t$  &        &       0.12      & 0.5   & 0.5  & 0.5  \\
 $t$             & year      &      $\sim1$        & 1.5     & 3 & 5    \\
 \hline \hline
\end{tabular}
\caption{Indicative values of the relevant experimental
parameters representative of BabyIAXO as well as IAXO, compared to those of CAST. The parameters listed are the magnet cross-sectional area $A$, length $L$ and magnetic field strength $B$, the magnet figure of merit $f_M = B^2 L^2 A$, the detector normalized background $b$ and efficiency $\epsilon_d$ in the energy range of interest, the optics focusing efficiency or throughput $\epsilon_o$ and focal spot area $a$, as well as the tracking efficiency $\epsilon_t$ (i.e. the fraction of the time pointing to the sun) and the effetive exposure time. We refer to~\cite{physics_paper} for a detailed explanation and justification of these values. The values listed for CAST are indicative, as many different systems have been used throughout the experiment's lifetime, usually overlapping in time, as the four bores of the experiment were equipped with different detection lines (some of them without optics).($^*$) This is the area of the two magnet bores.($^{**}$) The detector parameters of the CAST column are those of the IAXO pathfinder system. }\label{tab:scenarios}
\end{table}

IAXO sensitivity curves are the envelope of two data taking phases, Run-I and Run-II, each of them assuming 3 years exposure time. IAXO Run-I will be performed with vacuum in the magnet bores, and will determine the sensitivity of IAXO for axion masses below 0.01 eV. IAXO Run-II will use a $^4$He buffer gas inside the magnet bores, with density continuously changed from 0 to 1 bar at room temperature. Run-II allows for improved sensitivity in the high mass range up to \SI{0.25}{eV}. The particular distribution of exposure time in the different density steps is adjusted to achieve a sensitivity down to the DFSZ $\gagamma$ for each $m_a$ values, although other distributions are possible.

The BabyIAXO sensitivity line in figure~\ref{fig:sensi} is also the envelope of two data taking campaigns, one in vacuum and one with buffer gas. A 1.5 year effective exposure is assumed for each of them. As in the case of IAXO, the gas phase exposure is distributed unequally in the various density steps to adjust the sensitivity to the KSVZ axion model $\gagamma$ for a relatively large mass range. We refer to~\cite{physics_paper} for more details of these sensitivity calculations. As can be seen from figure~\ref{fig:sensi}, BabyIAXO should reach values of $\gagamma \sim $\SI{1.5e-11}{\per\GeV} up to masses $m_a\sim$~\SI{0.02}{eV}. In the gas phase, BabyIAXO will extend to higher masses and in particular will probe KSVZ axions in the approximate range of about 0.06 to 0.25 eV.

\begin{figure}[t!]
\begin{center}
\includegraphics[width=0.8\textwidth]{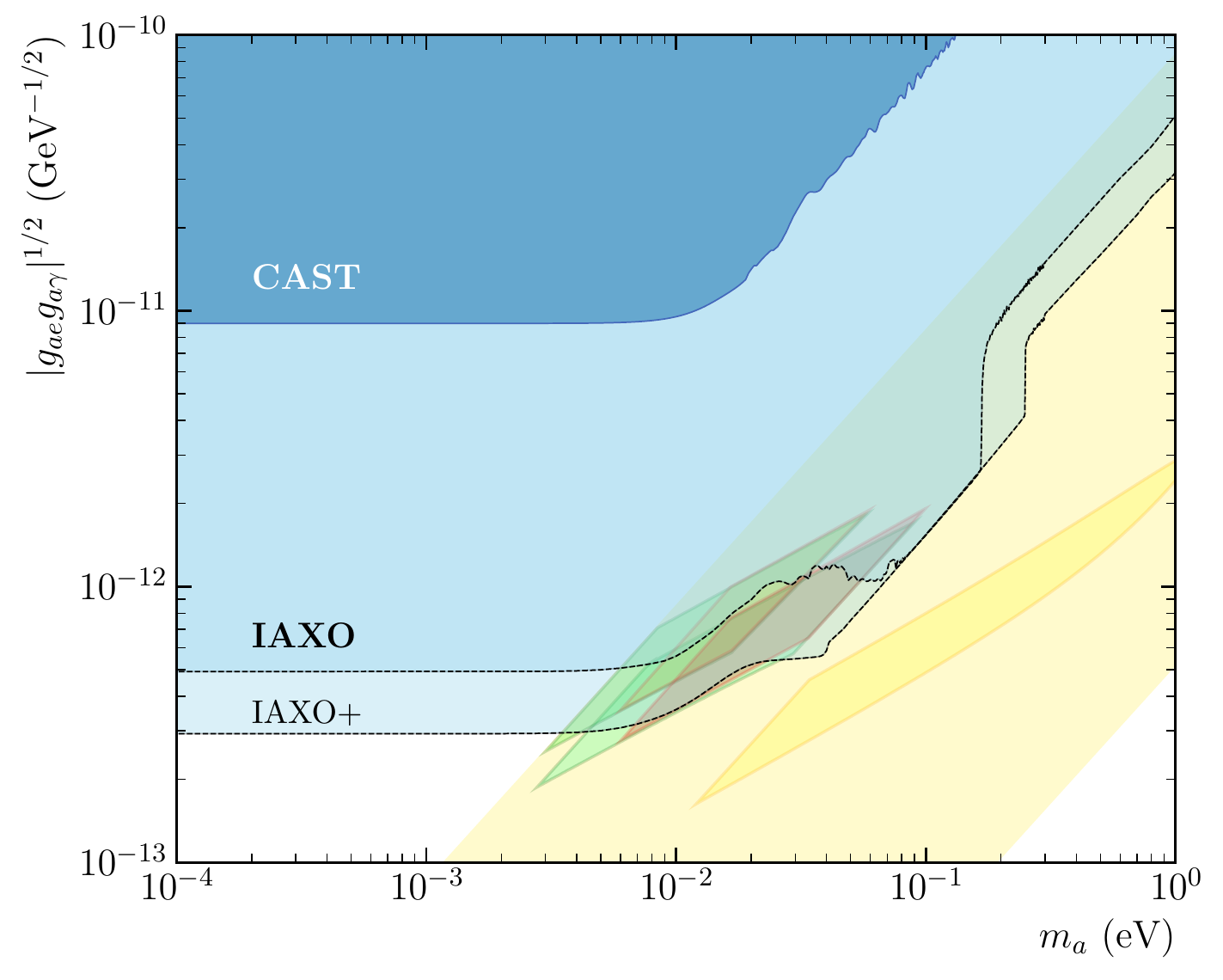}
\caption{Sensitivity plot of IAXO to BCA ($\gae$-mediated) solar axions, in the $(|\gae\gagamma|^{1/2}$-$m_a)$ parameter space. The yellow band corresponds to the QCD axion models and the diamond-shaped color regions correspond to particular QCD axion models that are able to fit all the anomalous stellar cooling observations, following~\cite{Giannotti:2017hny}.}
    \label{fig:gae}
\end{center}
\end{figure}

IAXO will also be sensitive to non-hadronic axions with a coupling to electrons $\gae \sim$~\num{e-13} because it could detect the flux of solar axions originating from (electron-ion and electron-electron) axion-Bremsstrahlung, Compton, and  axio-deexcitation of ions~\cite{Redondo:2013wwa} (together referred to as BCA reactions). The energies of these axions fall in the range of about 0.5$-$2 keV. In this case the expected signal depends on $\gae \gagamma$, the product of the electron coupling (responsible for the production in the Sun) and the photon coupling (responsible for the detection in IAXO). Not only could IAXO reach a sensitvity that will surpass the very stringent astrophysical bound on $\gae$ for the first time for an experiment, but, as shown in figure~\ref{fig:gae}, it could reach specific QCD axion models that may be hinted by anomalous astrophysical observations that are commented below. 

\subsection{Summary of IAXO physics case}

A comprehensive review of the physics case of IAXO has recently been released by the collaboration~\cite{physics_paper}. We refer to it for a detailed explanation of the theoretical, cosmological and astrophysical motivation of the search for axions in the range of parameters at reach of IAXO, as well as a detailed discussion of the physics potential of the experiment. In the following we list and briefly summarize some of the major points:

\begin{itemize}

\item IAXO will enter into completely unexplored ALPs and axion parameter space. The current experimental limit to $\gagamma$ competes with the best astrophysical bounds. IAXO will push this limit more than one order of magnitude, in a wide mass range.  At the high-mass end of the parameter region IAXO will be able to explore a broad region of realistic QCD axion models. Its sensitivity would cover axion models with masses down to the few meV scale. IAXO is the only proposed technique able to  probe a large fraction of QCD axion models in the meV to eV mass band. This region is the only one in which astrophysical, cosmological (DM) and theoretical (strong CP problem) motivations overlap~\cite{DiLuzio:2020wdo}. Already BabyIAXO will start probing unexplored parameter space, and thus there is potential for discovery already in this stage.

\item Theory predicts that the Universe should be opaque to very-high-energy, VHE, photons (i.e., photons with energies above 100 GeV) coming from distant emitters like active galactic nuclei (AGN) due to their non-negligible probability to interact with the background photons (EBL -- extragalactic background light) permeating the Universe while traversing long intergalactic distances. However, several independent observations seem to indicate that the Universe is too transparent for these photons \cite{DeAngelis:2011id,Horns:2012fx,Aharonian:2005gh,Aliu:2008ay,Teshima:2007zw}. Several authors~\cite{Horns:2012fx,Meyer:2013pny,Csaki:2003ef,DeAngelis:2008sk,Roncadelli:2008zz,Simet:2007sa,SanchezConde:2009wu,Dominguez:2011xy,DeAngelis:2011id,Rubtsov:2014uga,Kohri:2017ljt} have suggested explanations of these observations by invoking photon-ALP oscillations triggered by cosmic magnetic fields. Thus, the ALP component can travel unimpeded through the intergalactic medium and as a result the effective mean free path of the VHE photons increases. The required ALP mass is $m_a \sim$\SI{10}{\nano\eV} and coupling to photons $\gagamma \sim$ \SI{e-11}{\per\GeV}. As shown in figure~\ref{fig:sensi}, IAXO is sensitive to the entire parameter region where ALPs have been proposed to solve the VHE photon transparency problem, and BabyIAXO will already probe a large fraction of such region.
 
\item A set of independent observations of diverse stellar systems have shown an excessive amount of energy loss, indicating a lack of understanding in the current modeling of stellar cooling (see Refs.~\cite{DiVecchia:2019ejf,DiLuzio:2020wdo} for recent reviews). These deviations, often referred to as cooling anomalies, have been observed in: 1) white dwarfs (WDs), in which the cooling efficiency was extracted from the rate of the period change \cite{Isern:2010wz,Corsico:2016okh,Corsico:2012ki,Corsico:2012sh,Battich:2016htm}; 2) the WD luminosity function, which describes the distribution of WDs as a function of their brightness~\cite{Isern:1992,Isern:2008nt,Bertolami:2014wua}; 3) red giant branch (RGB) stars, in particular the luminosity of the tip of the branch\footnote{Notice, however, that the recent analysis in Ref.~\cite{Capozzi:2020cbu} indicates no exotic energy loss.}~\cite{Viaux:2013hca,Viaux:2013lha,Straniero:2020iyi}; 4) horizontal branch (HB) stars or, more precisely, the $R$-parameter, i.e., the ratio of the number of HB over RGB stars~\cite{Ayala:2014pea,Straniero:2015nvc}; 5) helium burning supergiants, more specifically, the ratio (B/R) of blue and red supergiants~\cite{Skillman:2002aa,McQuinn:2011bb}; and 6) neutron stars~\cite{Ho:2009mm,Heinke:2010cr,Shternin:2010qi}. The extra cooling mechanism via emission of light and weakly interacting axions/ALPs produced in the stellar core that are able to stream freely outside carrying energy away, provides a possible explanation of the cooling anomalies. Some of the anomalies require $\gae$ at the few 10$^{-13}$ level, while $\gagamma$ around few \SI{e-11}{\per\GeV} is needed for others. It is remarkable that particular QCD axion models featuring both $\gae$ and $\gagamma$ couplings are able to explain collectively all anomalies~\cite{Giannotti:2017hny,Giannotti:2015kwo,DiLuzio:2020wdo}. These models correspond to $m_a$ in the ballpark of \SI{10}{meV} and could be probed by IAXO, as shown in figure~\ref{fig:gae}. At lower masses, the hinted $\gagamma$ values overlap with the parameters required to explain the transparency hints. In all cases, IAXO shows a high potential to explore the relevant parameter space.


\item Axions (and ALPs) are very compelling candidates to compose the cold dark matter. The relic axion density depends on the cosmological history of the axion (in particular if the PQ symmetry breaks spontaneously before or after inflation) and on other axion model parameters, especially its mass. In general, a broad range of axion masses remains compatible with the correct relic density, including values above meV for the QCD axion that could be accessible by IAXO. Standard production of cold axions via the vacuum-realignment mechanism requires $m_a$ in the ballpark of $m_a \sim$\SI{e-5}{eV}, although if we allow for finetuning of the initial axion field value, a much larger $m_a$ range is permitted. If the PQ symmetry is restored after inflation, model-dependent contribution of axion strings and domain-wall (DW) decays to axion DM must be taken into account, and are potentially dominant. According to the most recent numerical studies~\cite{Gorghetto:2018myk,Gorghetto:2020qws}, computation uncertainties still allow for a relatively large axion mass window, the upper part of which ($m_a$ $\sim$ meV) would be at reach of IAXO. 
Moreover, models with long-lived DWs~\cite{Ringwald:2015dsf} would further increase the yield and point to QCD axions of 1 to 100 meV as those accounting for the totality of the Universe's DM.

\item In addition to the above, for all non-thermal production mechanisms the resulting relic axion density is approximately inversely proportional to the axion mass, $\rho_a \sim m_a^{-1}$, and therefore if axions are only a subdominant DM fraction the corresponding $m_a$ moves to larger values. Moreover, these mechanisms are generic to other ALP models. A large section of the parameter space accessible by IAXO contains ALP models with adequate DM density~\cite{Arias:2012az}. To summarize, although IAXO does not rely on axions to be the DM to detect them, it would be sensitive to axion models that could be part or all of the DM.

\item The EDGES collaboration has recently observed~\cite{Bowman:2018yin} an anomalously strong 21-cm absorption line in the sky-averaged radio spectrum, corresponding to the era of first stars. This has been connected with possible cooling of the primordial hydrogen gas by interaction with dark matter. In particular, a small amount of axion dark matter could explain this observation~\cite{Houston:2018vrf,Houston:2018vbk} within the context of standard models of axions and ALPs. This interpretation requires a QCD axion mass in the 100--450 meV range (slightly larger for ALPs). IAXO will have the capability to test a substantial part of the hinted region.

\item Axions and ALPs are also invoked in cosmological scenarios other than cold DM. For example, they can be also produced from thermal processes or from decays of heavy particles, contributing to the relativistic degrees of freedom in the early Universe and potentially today. 
Current cosmological observations present some tension that could suggest the presence of Dark Radiation~\cite{Bernal:2016gxb} (i.e. a cosmological relic of relativistic particles, with very weak interactions with the standard model). This could be in the form of axions or ALPs~\cite{Turner:1986tb}, whose abundance can be detectable by forthcoming cosmic microwave background (CMB) experiments~\cite{Abazajian:2016yjj,SPT-3G} especially if produced via interactions with quarks~\cite{Ferreira:2018vjj} or leptons~\cite{DEramo:2018vss}. These dark radiation axions/ALPs can feature a coupling to photons which is likely to be in the IAXO range and can thus be observed independently by IAXO for most values of the relevant parameters~\cite{Ferreira:2018vjj,DEramo:2018vss}. 

\item ALPs have masses protected by large radiative corrections and therefore it is very suggestive to use them as candidates for the inflaton field, whose slow-roll in a potential is thought to have driven primordial inflation. A recent scenario (dubbed the ``ALP miracle'') showed that, with an adequate potential and coupling to photons for reheating, an ALP with the parameters in the approximate ranges $m_a \sim$ 0.01--1 eV and $\gagamma \sim 10^{-11}$-$ 10^{-10}$~GeV$^{-1}$, accessible to IAXO, could be responsible for cosmic inflation, and simulaneously be the DM candidate~\cite{Daido:2017tbr,Daido:2017wwb}.

\item ALPs in the early Universe have also been proposed as an explanation of the observed baryon asymmetry of the Universe. A particularly interesting framework is the recently proposed ``ALP cogenesis''\cite{Co:2020xlh}, in which the ALP gets an initial field velocity that delays the beginning of the oscillation of the ALP around the minimum of the potential, and enhances the ALP abundance in comparison with the conventional misalignment mechanism. The couplings required to explain both DM and the baryon asymmetry are
much larger than those of the QCD axion. Part of the predicted region of parameter space is at reach of IAXO.  

\item It must be stressed that IAXO's sensitivity goals do not depend on the hypothesis of axion being the DM, i.e. the sensitivity only relies on the axion emission by the Sun, a firm prediction of most axions models. Therefore, in case of non-detection, IAXO will provide a robust exclusion of the corresponding regions of ALP/axion parameter space.

\item In the case of a positive detection, IAXO may also determine details of the underlying model, given sufficient signal statistics. For $m_a$ higher than around 0.02 eV, axion-photon oscillations destroy the coherence of the conversion along the magnet. This coherence can be restored by the buffer gas technique, providing a way to measure $m_a$. For masses as low as $3\times10^{-3}$~eV the onset of these spectral oscillations can be observed and used to determine the axion mass~\cite{Dafni:2018tvj}. Moreover, if the axion signal is composed by significant fractions of Primakoff and ABC solar axions, the combined spectral fitting can provide independent estimations of $\gagamma$ and $\gae$~\cite{Jaeckel:2018mbn}. To exploit these capabilities high-resolution and low-threshold detectors are preferred, and this motivates the multi-detector R\&D program being undertaken within IAXO. 

\item IAXO will also constitute a generic infrastructure for axion/ALP physics with potential for additional  search strategies. Most relevant is the possibility of implementing haloscope-like setups (e.g. based on the use of  RF cavities) to search for DM axions. The BabyIAXO magnet will already surpass the haloscope figure of merit ($B^2V$) of the magnets currently in use for haloscope searches, while the IAXO magnet will enjoy a figure of merit one order of magnitude larger.

\end{itemize}


\subsection{BabyIAXO physics case}

All of the points above stated apply to BabyIAXO, to a correspondingly lower extent. BabyIAXO will enjoy sensitivity to axions and ALPs down to $\gagamma \sim$ \SI{1.5e-11}{\per\GeV} (for $m_a \lesssim$ \SI{0.02}{eV}) and therefore will already enter unexplored space, as shown in figure~\ref{fig:sensi}. At the high mass part of this region, BabyIAXO will probe QCD axion models. In particular, making use of the buffer gas technique, BabyIAXO will find or exclude KSVZ axions between 70 and \SI{250}{meV}. Part of this region is motivated by the stellar cooling anomalies  mentioned above. Also these parameters partially overlap with the ones invoked in the ``ALP miracle'' models and in the axion dark matter interpretation of the EDGES anomalous observation, both mentioned above. 
At lower masses, BabyIAXO will probe a large fraction of the region of parameters invoked to solve the anomalous observations of the transparency of the Universe to VHE photons. 
The generic arguments involving ALPs from higher energy extensions of the SM, the ALP as dark matter, etc. are also generically applicable to BabyIAXO. To summarize, BabyIAXO will  probe a relevant region of so-far-unexplored parameter space, and a discovery is not excluded already at this stage. 


%% file: sections/BabyIAXO_setup.tex
BabyIAXO is conceived as a first experimental stage towards IAXO, with the two-fold objective of:
1) being a technological prototype of IAXO, mitigating risks and better preparing the ground for IAXO,
and 2) producing intermediate but relevant physics outcomes, allowing the collaboration to promptly move into
``experiment'' mode. These two considerations have been the guidance to define the conceptual
design for BabyIAXO. In addition, it is expected that the experience of building and operating the BabyIAXO systems will offer opportunities of improvement of one or more of the baseline experimental parameters, and therefore lead to an enhanced IAXO figure of merit with respect to the current design. The possibility of exploring incremental improvements of the different BabyIAXO subsystems will be taken into account whenever possible, provided it does not jeopardize the baseline physics program.

The BabyIAXO magnet is based on a ``common coil'' layout, i.e. two parallel flat racetrack coils with opposite current directions. In between the coils two parallel 70-cm diameter bores are placed. Unlike in the IAXO toroidal design,  the two bores are located between the racetrack sides, where a relatively high (mostly dipolar) field is produced. As discussed in section~\ref{sec:BabyIAXO_magnet}, with a length of 10 meters, the chosen configuration already provides a magnet figure of merit more than 10 times higher than that of the CAST magnet, allowing for the relevant physics outcome discussed in the previous section. At the same time, and despite the differences with the IAXO toroidal design, the BabyIAXO superconducting coils enjoy very similar engineering parameters (winding, geometry, etc.) as those of the final IAXO toroidal design, and therefore they constitute relevant technological prototypes of the latter. 

The two BabyIAXO magnet bores will be equipped with two detection lines, each one composed by an x-ray optics and a low-background detector, of dimensions and parameters representative of the ones foreseen for IAXO. In fact, comparing with the dimensions set in the IAXO conceptual design report~\cite{Armengaud:2014gea}, the BabyIAXO lines target a slightly larger diameter (\SI{70}{cm} versus \SI{60}{cm}). This decision takes advantage of the availability of XMM spare optics of \SI{70}{cm} diameter, and therefore does not constitute a risk for the physics goals of BabyIAXO. At the same time, it gives the opportunity to the collaboration to face the challenges associated with building the IAXO optics larger than the dimensions foreseen in the IAXO CDR (60~cm was then considered a safe, within-state-of-the-art, design choice) and develop the ingredients that will eventually contribute to push the figure of merit of the final IAXO. The baseline configuration foreseen for the BabyIAXO optics includes one XMM (X-ray Multi-mirror Mission) spare optics and one newly-built IAXO optics. Both devices are described in detail in section~\ref{sec:BabyIAXO_optics}. This decision is a good compromise between the best physics outcome of BabyIAXO and the opportunity to build and operate a prototype IAXO optics. 

At the focal points of both optics, BabyIAXO will host two low-background pixelated x-ray detectors. The baseline technology for these detectors are small ($\sim$\SI{6}{cm} wide and $\sim$\SI{3}{cm} thick) gaseous chambers read by pixelated planes of micro-mesh gas structures (Micromegas)~\cite{Giomataris:1995fq} manufactured with the \textit{microbulk} technique~\cite{Andriamonje:2010zz}. These detectors have been successfully used and developed in CAST and other low background applications~\cite{Abbon:2007ug}, and is the outcome of many years of development towards low radioactive background. The detector's components are built of very radiopure materials and they are surrounded by passive (lead and copper) and active (muon vetoes) shielding to reduce external radiation. A pathfinder system combining an X-ray optics of the same type as proposed for IAXO and a Micromegas detector has been operated in CAST during 2014 and 2015 with the expected performance~\cite{Aznar:2015iia}. 
The BabyIAXO Micromegas detectors constitute representative prototypes of the final detectors proposed for IAXO. The target background for these detectors are at most \SI{e-7}counts~keV$^{-1}$~cm$^{-2}$~s$^{-1}$  and possibly lower, and a threshold of at least \SI{1}{keV}. The plans to build two such detectors for BabyIAXO are described in detail in section~\ref{sec:BabyIAXO_detectors}.

In addition to the baseline detectors, a number of alternative or additional detection technologies are being developed. Although less developed that microbulk Micromegas from the point of view of low background, they outperform the latter in terms of energy threshold or resolution. GridPix detectors, similar to Micromegas detectors but built on a small CMOS pixelized readout~\cite{Krieger:2014wxa}, enjoy energy thresholds down to the tens of eV, and thus are of interest for the search of specific solar axion production channels lying at lower energies, like the ones mediated by the axion-electron coupling. Silicon Drift Detectors (SDD) offer better energy resolution with flexible and cost-effective implementations~\cite{1475-7516-2015-02-020,Mertens_2019}. Finally, bolometric detectors like Magnetic Metallic Calorimeters (MMC)~\cite{Fleischmann}, or Transition Edge Sensors (TES)~\cite{Ravera,2008JLTP..151...82C}, enjoying outstandingly low energy threshold and energy resolution, are also under consideration. Plans to test and develop all these technologies in specific detector platforms, and assess them for their potential use in IAXO, are discussed in section~\ref{sec:BabyIAXO_detectors}. Depending on the results of such developments, BabyIAXO could host one or more of these prototypes in subsequent beyond-baseline data taking campaigns. Eventually, the results of these developments will feed the final IAXO technical design. Ideally IAXO could host an optimal suite of different detection techniques, combining figure of merit and the robustness of different detectors (with different systematics). Needless to say, in the event of a discovery, the focus of the experiment would shift from ``discovery'' detectors (priority to low background) to ``high-precision measurement'' detectors (high spatial and energy resolutions). In this respect, we need to stress that the capability of (Baby)IAXO to determine the axion parameters once a positive detection is produced (see previous section) rely to some extent on the availability of high energy resolution and low-threshold detectors. If a discovery happens already in BabyIAXO, IAXO could be equipped from the start with high-resolution detectors, given that IAXO will enjoy a signal-to-noise ratio $\sim$\num{e-2} times larger than BabyIAXO.

\begin{figure}[t!] \centering
\includegraphics[width=\textwidth,trim=1cm 2cm 1cm 1cm,clip]{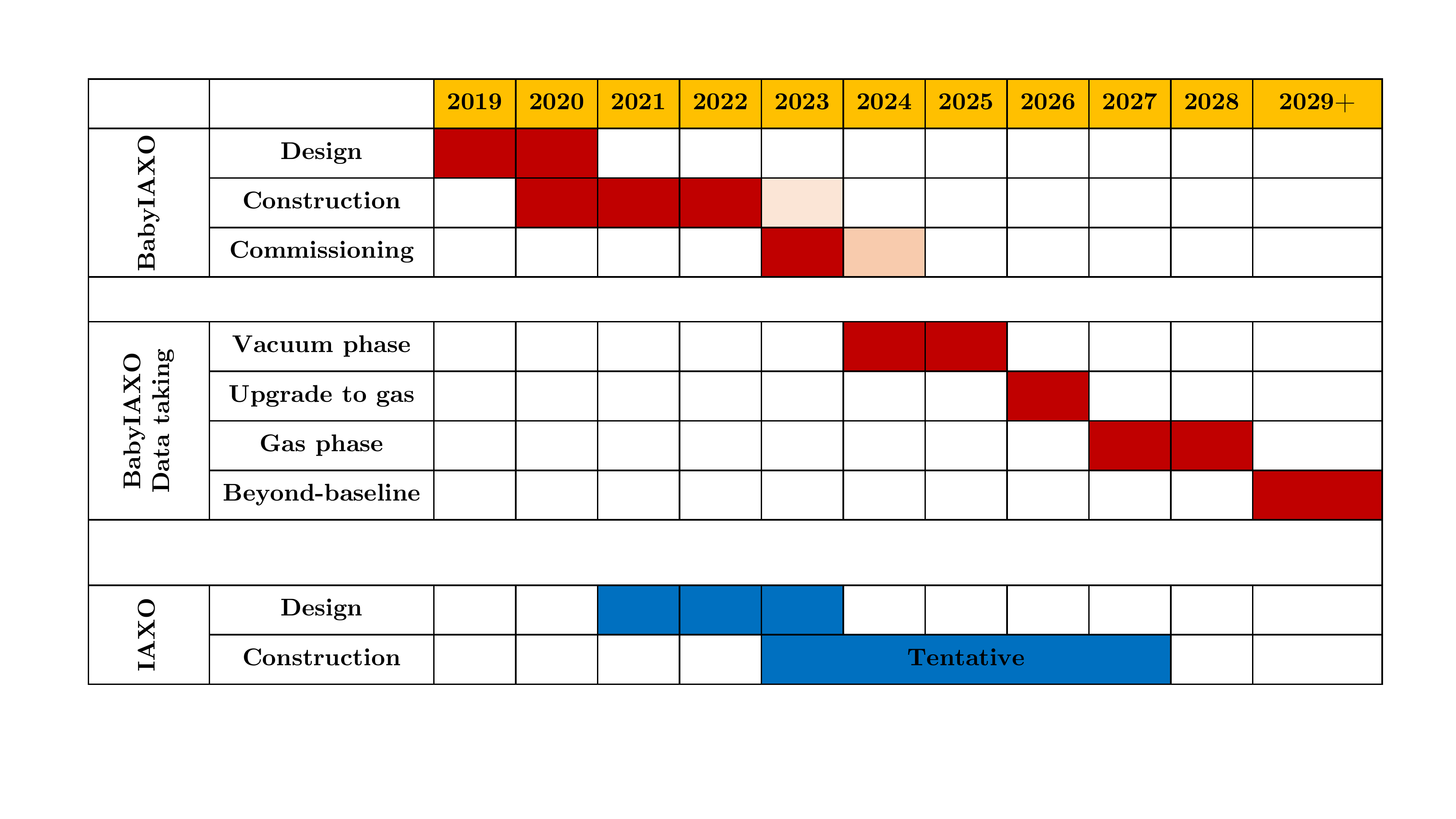}%
\caption{\label{fig:timeline} Tentative timeline for construction and operation of BabyIAXO, showing the intended start of data taking by 2024.}
\end{figure} 

Beyond serving as a prototype for everyone of the IAXO subsystems, BabyIAXO will constitute an invaluable preparatory exercise for the overall infrastructure. It will test all interfaces of relevance for IAXO, and the long-term stability of the systems. It will help create intangible resources that will be useful also for IAXO, like e.g. software and analysis tools, collaboration structure and data taking protocols. All of them will be naturally extended over to the IAXO phase.  

The BabyIAXO stage has recently been approved to be hosted at DESY, and the collaboration is already taking first steps towards its construction. The intended timeline is shown in Figure~\ref{fig:timeline}. As shown, BabyIAXO could start its first data taking in 2024. The physics prospects presented in the last section for BabyIAXO are computed assuming the baseline configuration and two data taking campaigns (one in vacuum and one with a buffer gas) of about two years of duration each. After this first physics campaign, and hopefully in parallel with the construction and commissioning of the full IAXO experiment, the BabyIAXO infrastructure will be available to assist the preparatory actions for IAXO. This actions may include tests with new detectors like the ones mentioned in the previous paragraph or new type of equipment, like RF-cavity-based setups sensitive to relic axions. The precise planning of this experimental phase will be configured once the development lines, as well as the IAXO design and construction plans, are more advanced.

%% file: sections/BabyIAXO_magnet.tex
	The main requirements for the design of the BabyIAXO superconducting detector magnet system were identified at the early stage of the project in December 2017. They comply with the guidelines exposed in the previous section combining ``prototyping'' and ``physics'' aspects, and are set in coherence with those for the full-size IAXO magnet system, namely: 1) the target performance of the magnet, in terms of the magnet figure of merit ($\MFOM$\footnote{$\MFOM \equiv f_M \sim B^2 L^2 A$ for a constant magnetic field $B$ over a lenght $L$ and a cross-sectional area $A$~\cite{Irastorza:2011gs}. More general expressions for arbitrary 2D or 3D $B$-distributions are given later.}), is set to at least 10 times higher than that of the CAST magnet $\MFOM$; 2) simple and robust design options must be chosen to allow construction within 3 to 4 years; and 3) a lowest-cost design is required to limit the magnet construction budget to a reasonable level of some $\sim$3 M\euro. 
 
\begin{figure}[!t]
\begin{center}

\includegraphics[width=1.0\textwidth]{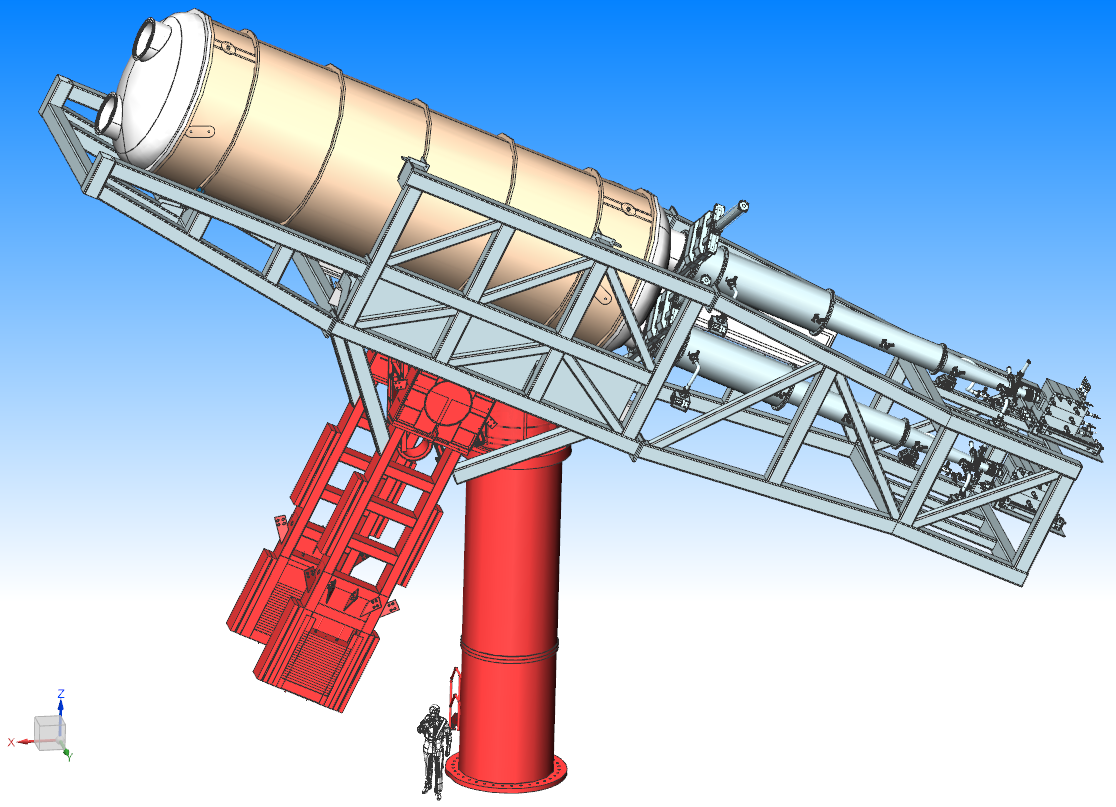}

\caption{CAD overview of the full BabyIAXO assembly, including the magnet (assuming a cylindrical cryostat, see text), telescopes, detectors, support frame and MST positioner~(in red, see section~\ref{sec:BabyIAXO_drive}). Overall system length is about 19 meters. Although both detection lines are shown of equal length (corresponding to 7.5 m of focal length of the XMM optics), in reality one of the lines will be shorter (5 m focal length, see section~\ref{sec:BabyIAXO_optics}).}
    \label{fig:drive2a}
\end{center}
\end{figure}

As a result of the assessment of various technical options for magnet and cryogenics, the principal solutions for the BabyIAXO magnet were fixed as follows, regarding:
\begin{itemize}
	\item \textbf{conductor:} NbTi Rutherford cable co-extruded with a pure aluminum matrix with \SI{2}{K} temperature margin;
	\item \textbf{coil windings:} two flat racetrack coils of \SI{10}{m} length arranged in a common-coil layout;
	\item \textbf{detection bore:} two \SI{700}{mm} free-bore tubes;
	\item \textbf{electrical operation:} persistent current mode with power supply switched off after charging;
	\item \textbf{cooling mode:} conduction cooled at \SI{4.2}{K} using cryocirculators;
	\item \textbf{cryogenics:} use of cryocoolers for cool down and stationary operation (dry cooling condition).
\end{itemize}
Details and implications of the selected approach for BabyIAXO are presented in the following sections. 
A CAD overview of the entire BabyIAXO experiment is shown in \cref{fig:drive2a}.


\subsection{Conductor and cold mass}

\begin{figure}[!b]
	\centering
	\includegraphics[width=0.5\linewidth,valign=c]{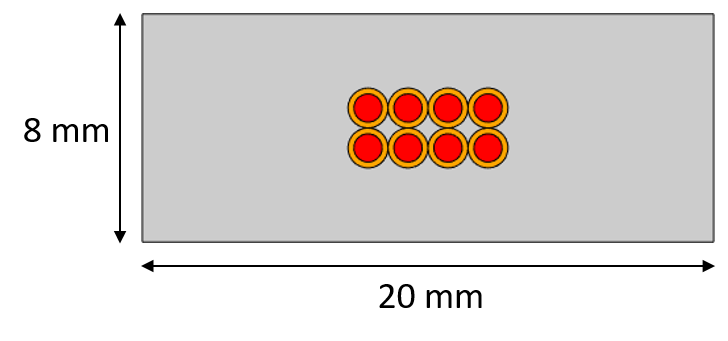}%
	\includegraphics[width=0.5\linewidth,valign=c]{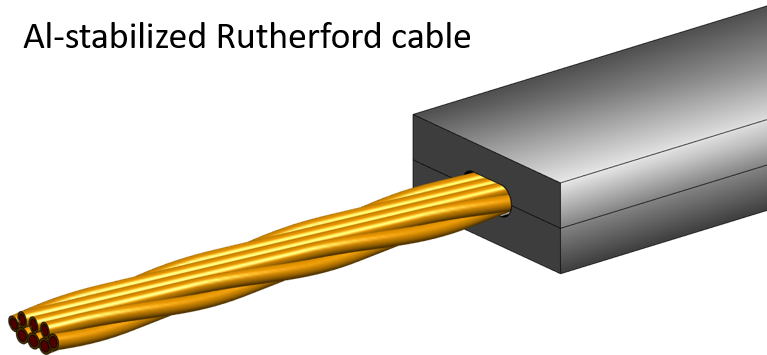}
	\caption{Cross-section (left) and artistic view (right) of the Al-stabilized 8 NbTi/Cu strands Rutherford cable.}
	\label{fig:conductor}
\end{figure}

The conductor comprises a Rutherford cable with 8 NbTi/Cu strands of the \SI{1.4}{mm} diameter and a copper/non-copper ratio of 1.0, see \cref{fig:conductor}. After the cabling stage, the flat two-layer assembly of twisted strands is co-extruded in a high purity aluminum matrix, in order to ensure sufficient thermal stabilization, a safe quench protection and mechanical properties.

Procurement of such an Al stabilized conductor is nowadays a technical and organization problem since it is only incidentally produced in a few companies and currently none of these show a qualified production line or is willing to offer one. For this reason the BabyIAXO conductor design has been adjusted to meet a specific project for which conductor production is at qualification stage. The design essentially mimics the conductor currently under development for the Panda Detector Solenoid at the Facility for Antiproton and Ion Research (FAIR). The rectangular dimensions of the Al stabilizer were adjusted from \SI{10.95}{mm}$\times$\SI{7.93}{mm} to \SI{20}{mm}$\times$\SI{8}{mm}, which corresponds to \SI{92}{\percent} relative amount of Al, \SI{4}{\percent} of Cu and \SI{4}{\percent} of NbTi in the conductor. By making use of the results of the ongoing R\&D and production start-up of the Panda conductor, a procurement solution has been created thereby saving time and budget is expected for the BabyIAXO experiment.

\begin{figure}[!htbp]
	\centering
	\includegraphics[width=0.7\linewidth]{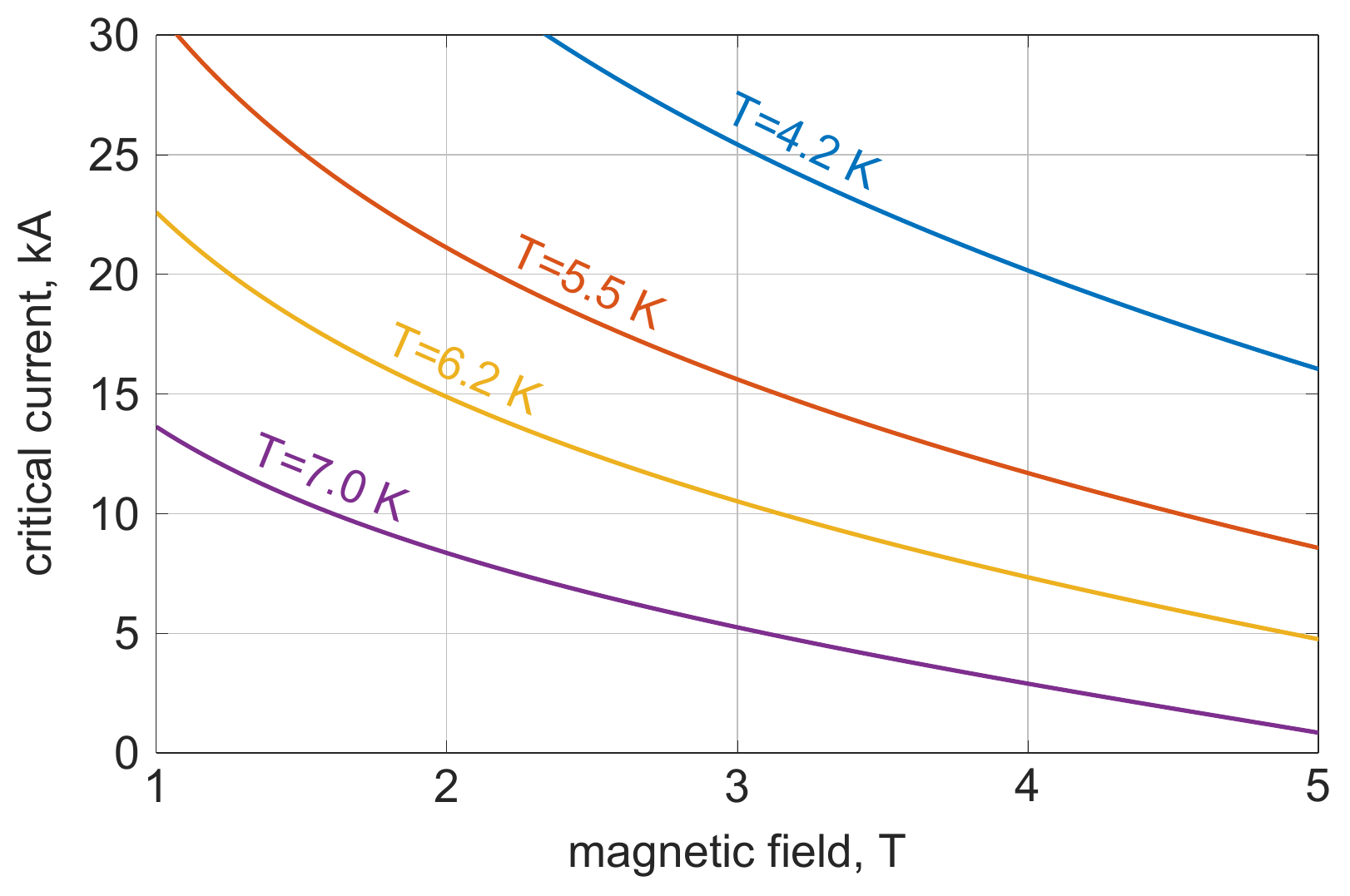}
	\caption{Dependence of the conductor critical current in kA on magnetic field in T at various temperatures between \SI{4.2}{K} and \SI{7.0}{K}.}
	\label{fig:conductorIc}
\end{figure}

The design critical current of the conductor $I_{c}$ as a function of magnetic field $B$ and temperature $T$ is shown in \cref{fig:conductorIc}. The scaling law used for the NbTi critical current density provides \SI{2.6}{kA/mm^2} at \SI{5}{T} and \SI{4.2}{K}, which is assumed some \SI{10}{\%} below the typical value of \SI{2.9}{kA/mm^2} at the operating conditions in order to take into account degradation of the superconductor properties due to cable manufacturing and conductor co-extrusion. The requested temperature margin of \SI{2}{K}, i.e. the difference between the current sharing temperature corresponding to $I_{c}$ and the operating temperature, is used for winding design in order to address a strong $I_{c}(T)$ dependence, a factor of 2 to 3 if $T$ is increased from \SI{4.2}{K} to \SI{6.2}{K} at a given $B$. This must ensure a reliable operation of the system at nominal current with a low probability of a quench in the conductor.

\begin{figure}[!t]
	\centering
	\includegraphics[width=\linewidth]{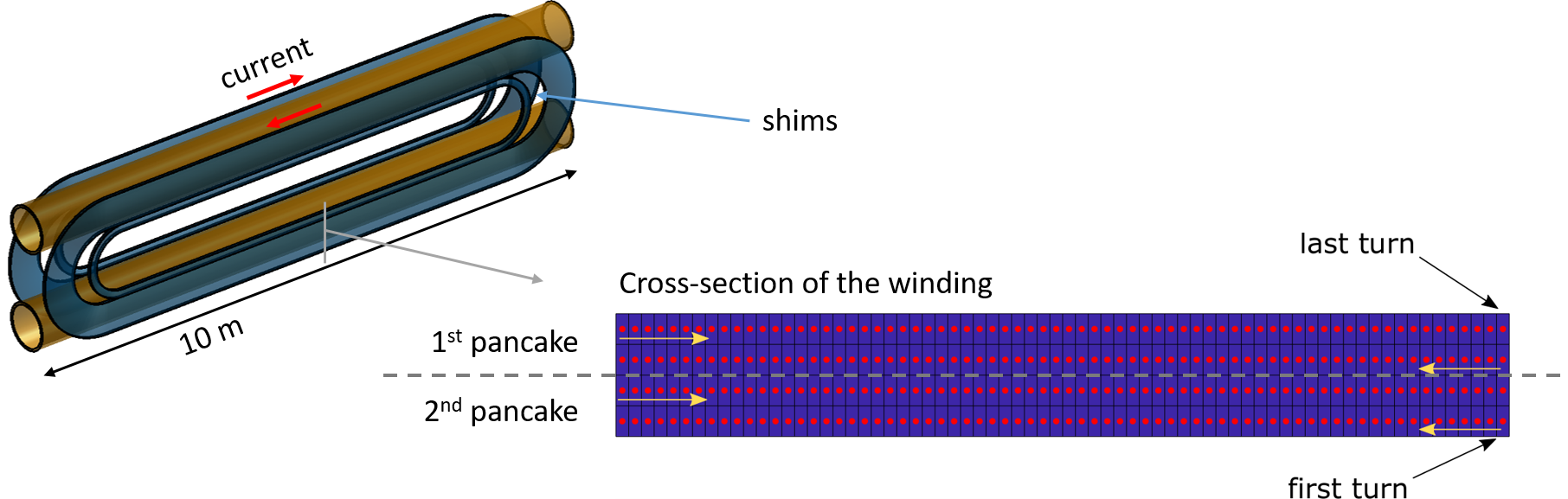}
	\caption{Schematic of the common-coil layout enclosing two user bores. The coils comprise 70 turns. In the coil heads shims or dummy turns of conductor are included in the 2 double-pancake winding for reducing the local peak magnetic field.}
	\label{fig:windingschematic}
\end{figure}

Similar to the IAXO coil windings designed as 8 flat racetrack coils arranged in a toroidal geometry, a common-coil layout was selected for BabyIAXO as shown in \cref{fig:windingschematic}. According to the comparison made of various coil layout options, which included block and saddle coils, the so-called common coil design requires about \SI{20}{\percent} more conductor for the same magnet performance, but features a very much simpler and cost efficient manufacturing process and support structure. In addition, using 2 double-pancake coil windings for BabyIAXO is in full accordance with the baseline design of the IAXO coil windings. The winding of each double-pancake starts at the inner coil radius, necessary to allow winding of both coil layers at the same time. Joints between the two pancakes and between the two coils are positioned in low magnetic field at the outer coil radius. 

The use of shims in the coil winding heads causes a reduction of the local peak magnetic field and thus effectively results in an increase of the magnet performance up to \SI{15}{\percent} at negligible extra cost. Although currently designed as a single large spacer, the layout of shims can be further optimized by using a set of smaller spacers inserted between the turns, which will further increase the performance and also is more practical from a manufacturing point of view. 

\begin{figure}[!b]
	\centering
	\footnotesize
	\includegraphics[width=0.70\linewidth,valign=c]{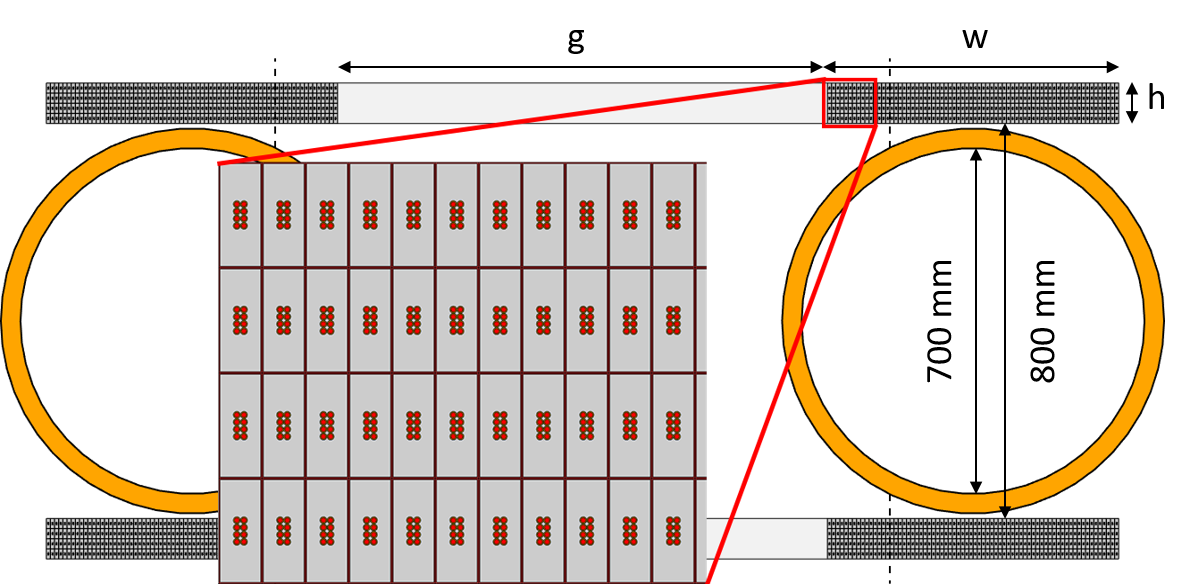}%
	\hfill%
	\begin{tabular}{@{} l c @{}}
		\toprule
		winding width, $w$ & \SI{595}{mm} \\
		winding height, $h$ & \SI{82}{mm} \\
		pole gap, $g$ & \SI{1000}{mm} \\
		\midrule		
		magnet energy & \SI{50}{MJ} \\
		inductance & \SI{1.0}{H} \\
		peak field & \SI{3.2}{T} \\
		current density & \SI{56}{A/mm^2} \\
		operating current & \SI{9.8}{kA} \\
		conductor length & \SI{11.4}{km} \\
		\midrule
		MFOM 3-D & \SI{232}{\tesla\squared\meter\tothe{4}} \\
		MFOM 2-D & \SI{326}{\tesla\squared\meter\tothe{4}} \\
		\bottomrule
	\end{tabular}
	\caption{Design of the BabyIAXO coil windings with respect to the two used bores of \SI{700}{mm} each. The insert shows a zoom of the four layer windings. The main magnet parameters are listed in the table.}
	\label{fig:windingdesign}
\end{figure}

The vertical orientation of the two coils, with two bore tubes located one on top of the other as shown in \cref{fig:windingschematic}, is preferred instead of the horizontal one with the bore tubes located side by side. This vertical orientation allows an easier layout of supports between cold mass and cryostat, thermal path symmetry for cooling and easier access for work on cryogenic parts like circulators and cryocoolers, as well as instrumentation, bus bars and current leads. It is only for the purpose of illustration that the coil windings are shown in the horizontal orientation in some of the following pictures. The main parameters of the winding pack are summarized in \cref{fig:windingdesign}.

The two \SI{10}{m}-long racetrack coils with a pole gap of \SI{1000}{mm} are placed at \SI{800}{mm} distance allowing to insert two  bore tubes of \SI{700}{mm} diameter. The total thickness of the 2 double-pancakes is \SI{82}{mm}, including the insulation around conductor of \SI{0.25}{mm} thickness. In total 70 conductor turns are used in the radial direction of the windings, resulting in a winding width of \SI{595}{mm}. Applying the temperature margin of \SI{2}{K} for the \SI{4.2}{K}  operating temperature, the nominal operating current of the winding is \SI{9.8}{kA} at a peak magnetic field of \SI{3.2}{T}, which corresponds to \SI{56}{A/mm^2} average current density in the windings. The self-inductance of the common coils layout is \SI{1.0}{H}, corresponding to a stored magnet energy of \SI{50}{MJ}. In total, \SI{11.4}{km} of conductor is needed for this magnet. 

\begin{figure}[!t]
	\centering
	\includegraphics[width=0.48\linewidth]{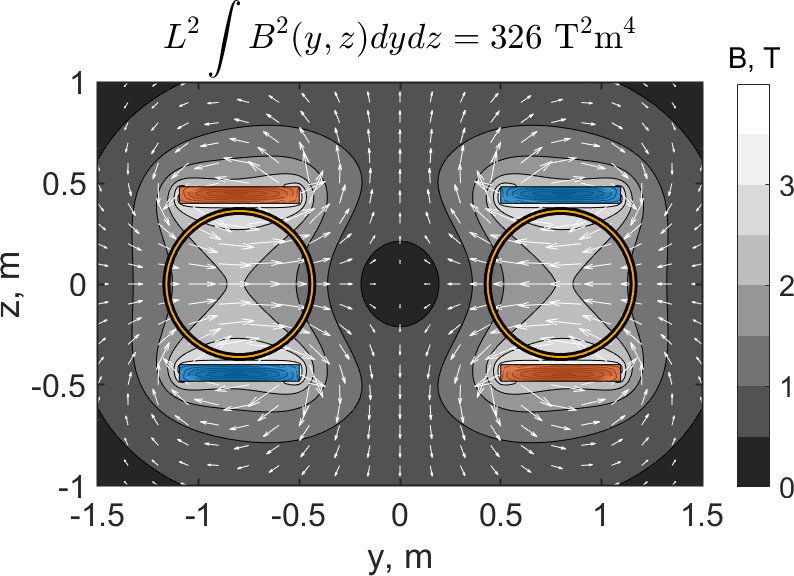}\hfill%
	\includegraphics[width=0.5\linewidth]{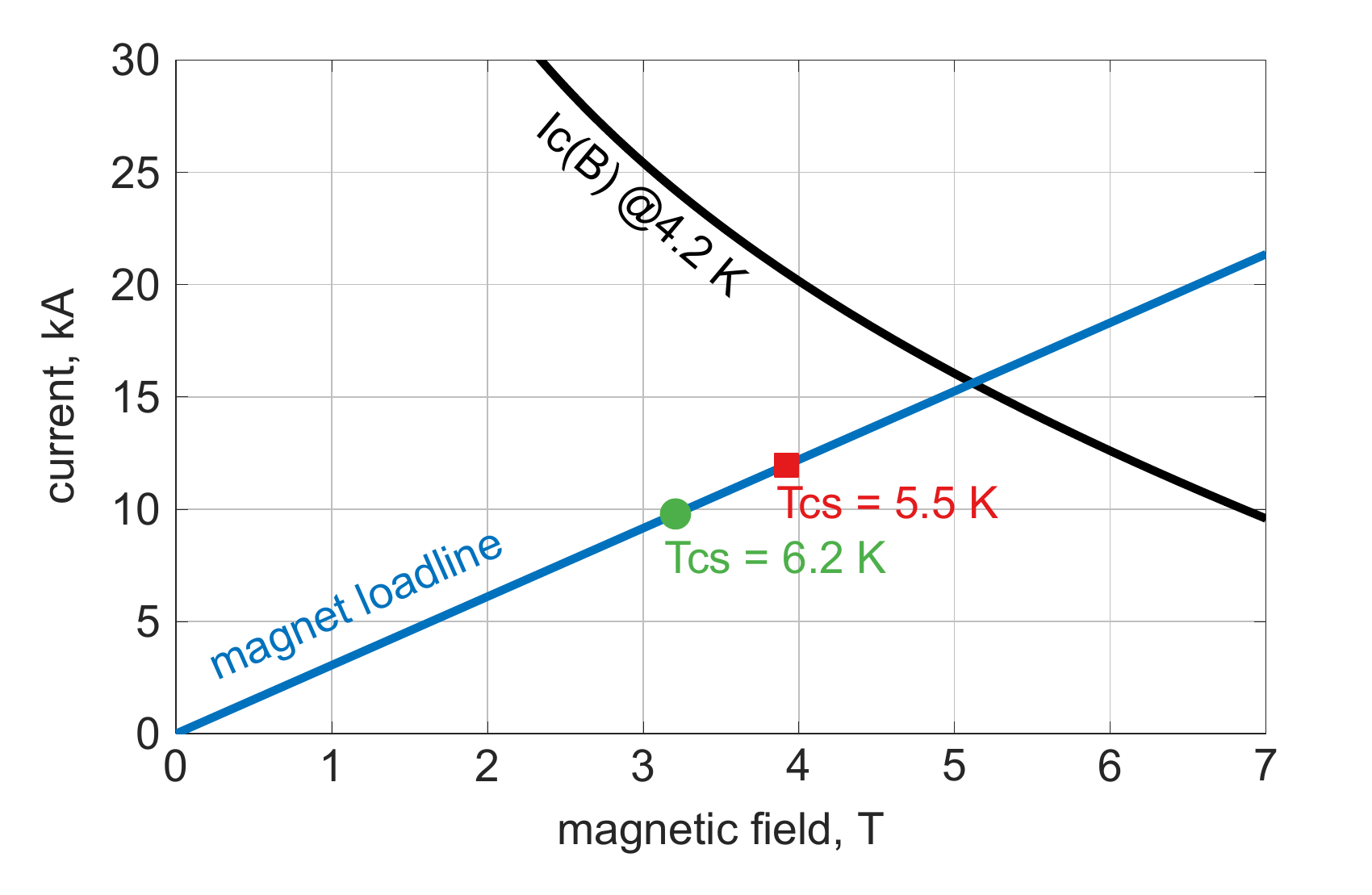}
	\caption{Left: distribution of the magnetic field in the magnet mid plane, with two bores and windings cross-section indicated (current directed outwards in red domains, inwards - in blue). Right: load line of the conductor in the coil windings with two operating points indicated for $T_{cs}=\SI{5.5}{K}$ and \SI{6.2}{K}.}
	\label{fig:winding_props}
\end{figure}

The magnet figure of merit $\MFOM$ for an axion helioscope is usually approximated to $\MFOM\equiv f_M=B^2L^2A$, for a constant $B$ over a length $L$ and an aperture area $A$. More realistic values can be obtained by integrating the distribution of the magnetic field over the space of the two bores in 2D (assuming it to be constant along $L$) or altogether in 3D, respectively. The resuls obtained for our design are: 

$$\MFOM \text{(2-D)} =L^2 \int_A B^2(y,z) \dif y \dif z={\rm \SI{326}{\tesla\squared\meter\tothe{4}}}$$

$$\MFOM \text{(3-D)} =\int_A \left(\int_L B_\perp(x,y,z) \dif x \right)^2 \dif y \dif z={\rm \SI{232}{\tesla\squared\meter\tothe{4}}}$$

\noindent where the $x$ coordinate is along the magnet length, $y$ and $z$ are the cross-section coordinates. One can see the $B(y,z)$ distribution in the magnet mid plane in the left picture of \cref{fig:winding_props}, providing \SI{2.0}{T} at the bore centers.  The lower value of $\MFOM$ obtained from the 3D calculation is due to the magnetic field decreasing from the center of the magnet towards its ends. Both values of MFOM fulfill the target requirement of \SI{200}{\tesla\squared\meter\tothe{4}}, i.e. at least 10 times the CAST performance.

As shown in the right picture of \cref{fig:winding_props}, the nominal operating current of \SI{9.8}{kA} is close to 1/3 of the conductor critical current at \SI{3.2}{T}.
Hence, there is a significant potential to eventually operate at a higher current, although the maximum current of the system can only be found during magnet testing as it strongly depends on the quality of the coil manufacturing and cooling conditions. An ultimate current for this design is set at \SI{12}{kA}, corresponding to \SI{1.3}{K} temperature margin at \SI{4.2}{K} operating temperature. Achieving ultimate current would significantly increase the magnet figure of merit performance by a factor 1.5, but also mechanical stress due to electromagnetic forces will augment by some \SI{20}{\percent}. In order to enable ultimate current in a mechanically safe way, the support structure of the coils is designed for operation at \SI{12}{kA}.

\begin{figure}[!b]
	\centering
	\includegraphics[width=0.6\linewidth,valign=c]{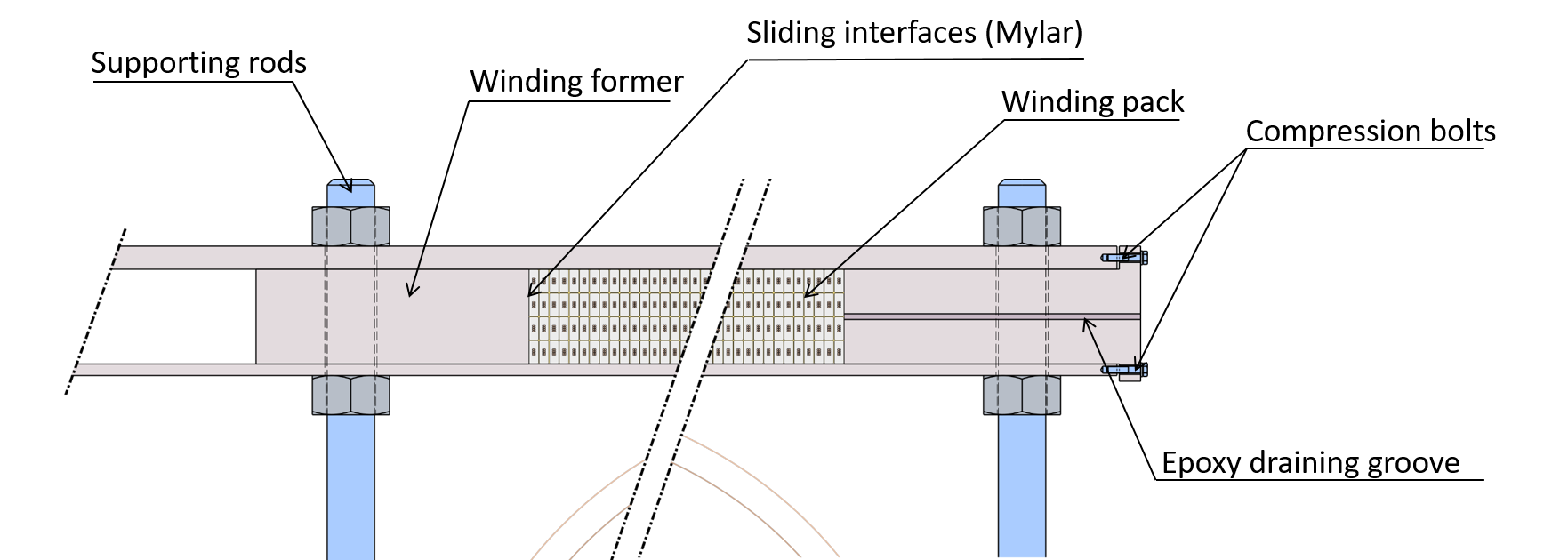}%
	\includegraphics[width=0.4\linewidth,valign=c]{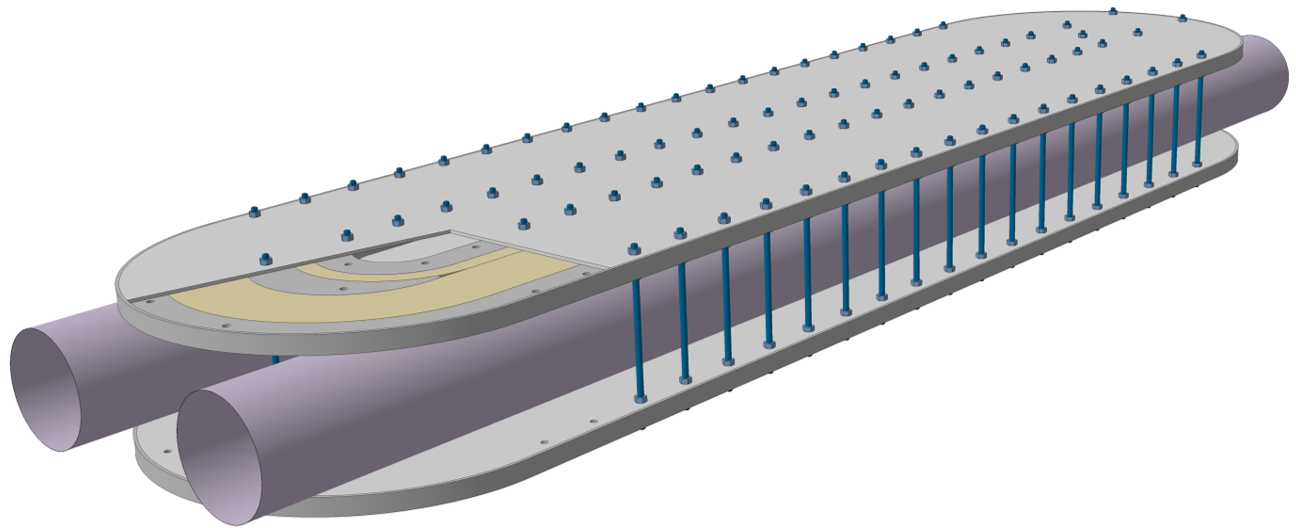}
	\caption{Left: layout of the cold mass mechanical design. Right: 3D artistic view of the cold mass (not showing all details).}
	\label{fig:cold_mass}
\end{figure}

As shown in \cref{fig:cold_mass}, the coil windings support structure is made of a sandwich of mostly flat plates of Al5083-O components to arrive at a cost efficient solution with minimum amount of machining. The conductor turns are first wrapped with pre-impregnated glass tape, then wound on a winding former and sandwiched between two flat plates, such that a sliding interface is formed between the winding pack and the supporting plates by using Mylar foil. The structure is closed with blocks equipped with an epoxy draining groove. Hence, during the curing stage of the B-stage epoxy resin, compression can be applied by tightening bolts in order to densify the windings, achieve correct dimensions and avoid voids in the windings, while excessive epoxy can be drained. This manufacturing process is fully representative for the IAXO coils.

The two coil modules are then assembled, bolted together using stainless steel rods, which handle the repelling load of about \SI{30}{MN} between the two coil modules for a magnet operation at \SI{12}{kA}. Provided each rod can withstand a \SI{0.4}{MN} load, around 80 rods are required in total. In addition, as the coils are in unstable equilibrium, a potential shear load between the coils due to their misalignment shall be resisted by bar components installed between the two coil modules near the magnet center. Preliminary results suggest that using a top plate of \SI{20}{mm} thickness, additional reinforcements with a set of beams on top brings the mechanical stress well below acceptable levels. 

\subsection{Electrical circuit}

\begin{figure}[!b]
	\centering
	\includegraphics[width=\linewidth]{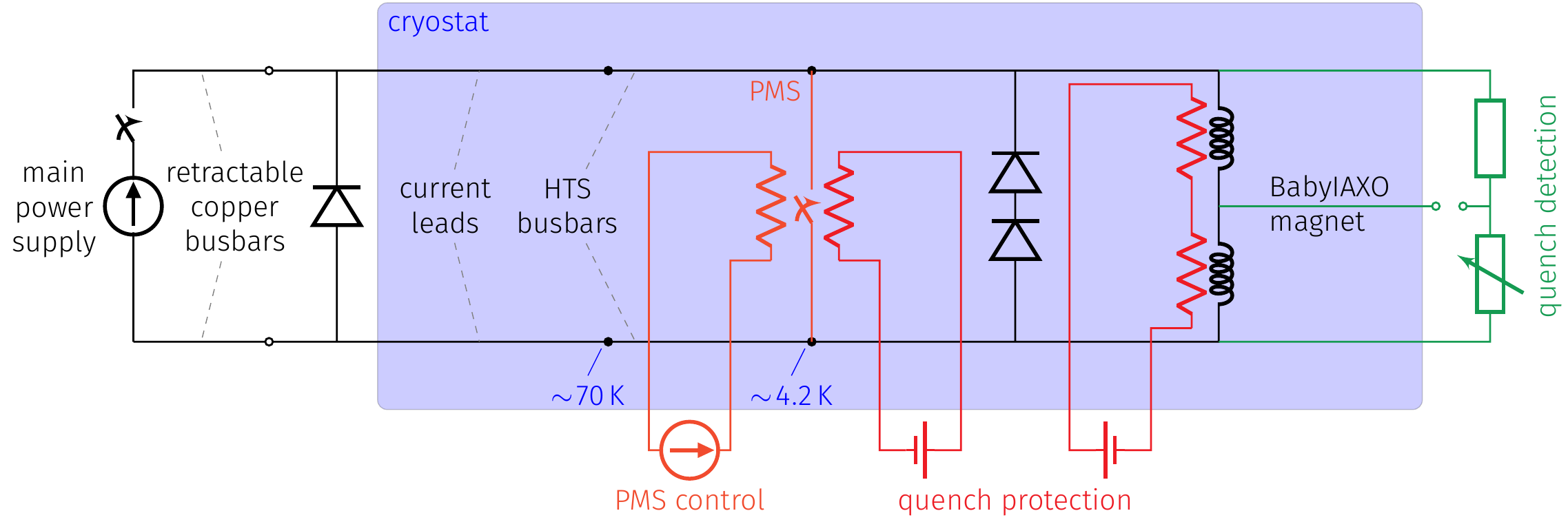}
	\caption{Electrical circuit (black) of the BabyIAXO magnet showing from left to right: the \SI{10}{kA} power converter and its current breaker, bus bars, run-down resistor-diode unit, current leads, persistent mode switch PMS (orange) with its on-off heater and quench protection heaters, cold diodes across the magnet, the magnet itself equipped with quench protection heaters, and its quench detection unit (green). The various heaters are activated with low current power supplies. The in-cryostat parts are indicated in the violet area.}
	\label{fig:electricalcircuit}
\end{figure}

The electrical circuit is shown in \cref{fig:electricalcircuit}. The main power supply feeding the magnet is connected to the cold mass through eventually retractable copper bus bars at room temperature, current leads in two sections, transferring the current from room temperature down to $\approx$\SI{70}{K} and high-temperature superconductor (HTS) bus bars from \SI{70}{K} down to \SI{4.2}{K}.

In order to minimize the cryogenic heat load at \SI{4.2}{K} and thereby the number of cryocoolers needed, persistent current mode operation is envisaged allowing to switch off the main power supply after the magnet charging. Eventually the bus bars may be disconnected as well which may simplify the sun-tracking rotation system. A thermally-controlled persistent mode switch (PMS) is introduced in the scheme for this purpose. It is made of NbTi superconductor and shortens the coils thereby creating a fully superconducting circuit with very long time constant. The resistance of the PMS is zero when it is closed, no heating applied, while it is of the order of ohms in the open state when heated up over the critical temperature of \SI{9.2}{K}. Hence, the characteristic operating cycle of the magnet is as follows:
\begin{enumerate}
	\item opening of the PMS;
	\item charging of the magnet within an hour;
	\item closing the PMS;
	\item decreasing the power supply current to zero;
	\item persistent mode operation achieved with eventually disconnected power supply.
\end{enumerate}

The main parameters of the circuit are outlined in \cref{table:ec}. The system can be charged to nominal current in 55 minutes at a ramp rate of \SI{3}{A/s} and a total charging voltage $\approx$\SI{5}{V}. Estimating the total resistance due to internal splices at \SI{1}{\nano\ohm}, the field decay rate in the stationary operation is about \SI{0.3}{\%/month}.

\begin{table}[!t]
	\caption{Main parameters of the electrical circuit.}
	\vspace{1ex}
	\centering
	\small
	\begin{tabular}{lc}
		\toprule
		Power supply voltage (depending on bus bar length) & 5 to \SI{10}{V} \\
		Maximum current & \SI{12}{kA} \\
		Nominal operating current & \SI{9.8}{kA} \\
		Ramp rate & \SI{3}{A/s} \\
		Field decay rate & \SI{<0.3}{\%/month}\\
		Regulation & $<\pm $\num{e-3}\\
		Ramp-up time & \SI{55}{min} \\
		Power supply voltage during ramp-up & $\approx$\SI{5}{V} \\
		\midrule
		Self-inductance of the magnet & \SI{1.0}{H} \\
		Inductive voltage & \SI{3}{V} \\
		\midrule
		Cross-section of brass current lead & $\approx$\SI{4000}{mm^2} \\
		Potential drop across current leads & \SI{0.2}{V} \\
		\midrule
		Cross-section of copper bus bar & \SI{2000}{mm^2} \\
		Potential drop on \SI{5}{m} length & $\approx$\SI{1}{V} \\
		\midrule
		Quench heater power supply & \SI{48}{V} battery \\
		\bottomrule
	\end{tabular}
	\label{table:ec}
\end{table}

The system has to operate at nominal current for a long time, while it can be charged rather quickly. Therefore, the current leads are ``over-current'' designed, resulting in an increased thermal load during charging when compared to classical optimized current leads, but decreased when the power supply is disconnected. However, for the full operating cycle of the magnet, this solution allows to minimize the overall thermal load during long term stationary operation. Currently, conduction-cooled current leads made of brass is the preferred option. Reduction of their cross-section by \SI{50}{\percent} from the classical optimal value for carrying stationary current is considered feasible, resulting in a \SI{50}{\percent} reduced thermal load at \SI{4.2}{K} and at zero current, while temporary increased by \SI{25}{\percent}  during the charging stage of one hour. The voltage drop at nominal current is estimated at \SI{0.2}{V}.

The HTS section of the current leads comprise a stack of HTS tapes positioned around a hollow steel tube. Using Bi2223 tapes in AgAu matrix, being optimized for the bus bar applications, would require in total some 120 tapes per lead. In order to ensure protection of the HTS tapes against quench, a mechanical switch that enables shunting of the current only in the case of overheating is currently under development.

The operation in persistent mode is envisaged, however, as alternative we seriously consider continuous current supplying mode by which the persistent mode switch is suppressed, and an extra cryocooler is installed to keep the current leads in operational conditions. This direct drive has the advantage of simplicity but requires more cooling power. A final decision will be made after having completed cost analysis and engineering stage of the persistent mode switch.

A resistor-diode unit installed at room temperature is present for serving a slow-dump of the system, while the cold diodes directly connected across the coil are needed for the protection of the persistent mode switch, for which also a separate activation heater is envisaged for redundancy reason. A self-dump is considered as the worst-case quench scenario of the windings when the magnet stored energy is completely dissipated in the cold mass. Quench protection is separately presented in the next section.

\subsection{Quench protection}

The quench process, a thermal runaway of the coil windings due to a local overheating of the conductor, is simulated using an adiabatic 3D thermal-electrical model. Heat propagation is calculated along the conductor (60 nodes per turn), in the cross-section of the windings (280 nodes) and in the coil casing (152 nodes). In total, 51840 nodes are used to represent the cold mass volume containing 2 coil modules. The magnetic field is calculated at each node and it is varied according to the operating current decreasing during the simulation due to increasing total resistance of the conductor. External cooling is not included in the model such that the entire magnet stored energy can only be released in the cold mass thereby increasing its enthalpy. In addition, since structural support components are also not included in the thermal analysis, the simulation results are considered conservative, somewhat overestimating the hot-spot temperature in the coil windings.

The thermal properties of Al alloy are taken for the coil casing, while the rule of mixtures is applied for the conductor comprising mostly Al, Cu and NbTi. The current sharing between the conductor constituents is also calculated in order to evaluate the total Joule heating at each node.

\begin{figure}[!b]
	\centering
	\includegraphics[width=0.5\linewidth,valign=c]{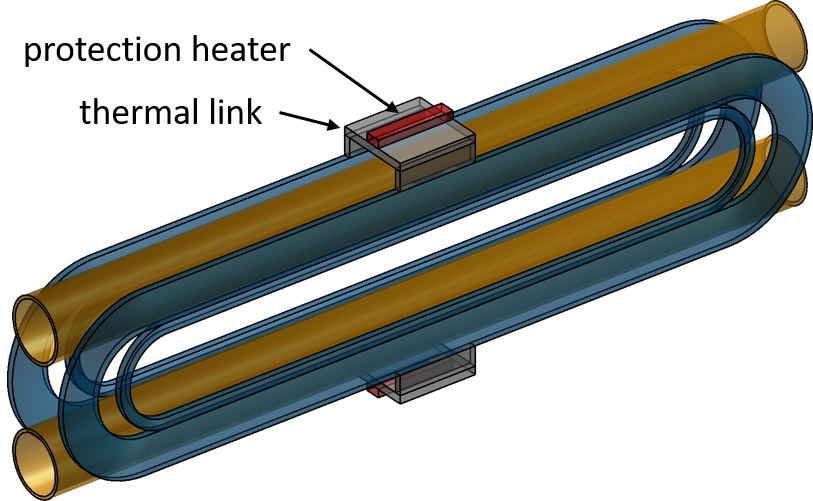}%
	\includegraphics[width=0.5\linewidth,valign=c]{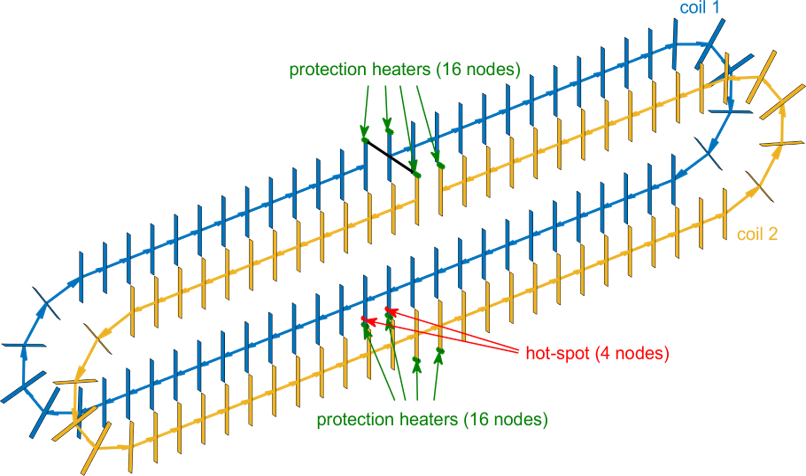}
	\caption{Left: sketch of the protection heaters installed on the cold mass, 2 units creating 4 normal zones in the two coils. Right: geometry of the quench propagation simulation model.}
	\label{fig:quench_model}
\end{figure}

The protection heaters present to introduce additional normal spots in the coil windings, thus allowing to distribute generated heat more uniformly, are installed on thermal links connected to the coil casing (see \cref{fig:quench_model}). This simplifies the construction and reduces the risk of insulation breakdown compared to installing the heaters directly on the winding pack.

The quench is initiated by applying \SI{50}{J} deposited during \SI{0.1}{s} at the selected location of the hot-spot (see \cref{fig:quench_model}). After quench is ``detected'' using \SI{0.5}{V} detection threshold, \SI{100}{W/m} (\SI{0.5}{W/cm^2} per cable face) is applied at the location of the heaters until reaching \SI{20}{K}.

\begin{figure}[!t]
	\centering
	\includegraphics[width=0.5\linewidth,valign=c]{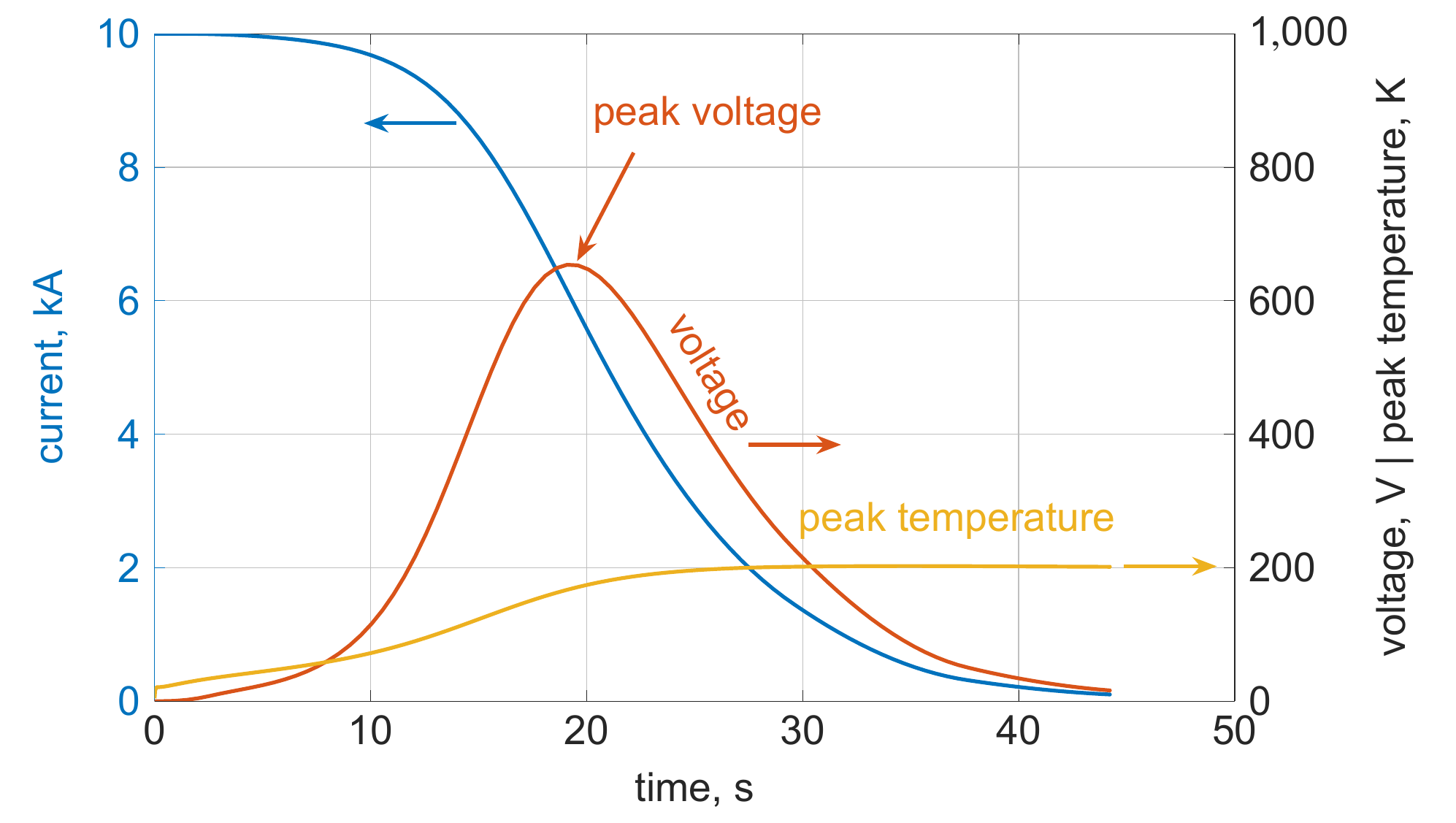}\hfill%
	\includegraphics[width=0.48\linewidth,valign=c]{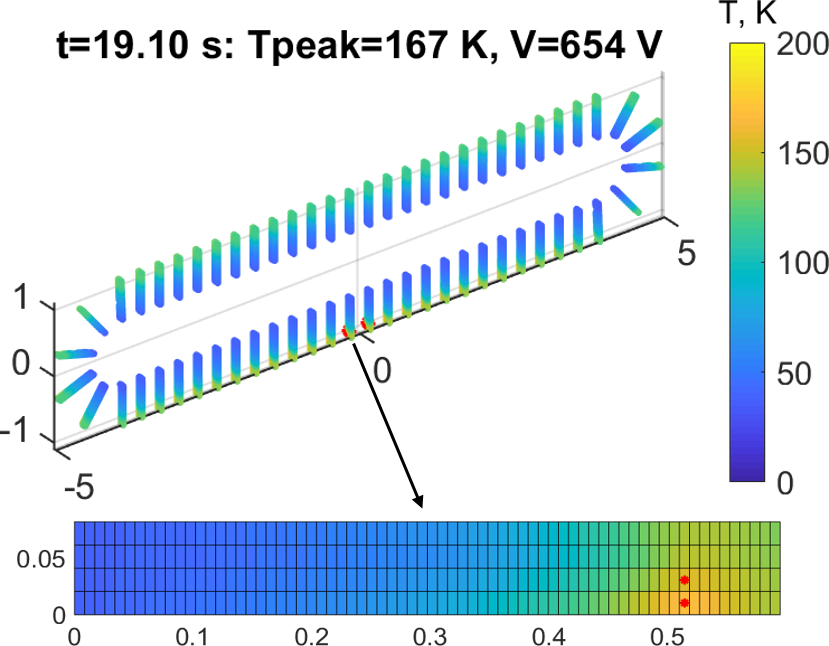}
	\caption{Left: time evolution of current, internal voltage and peak temperature of the magnet quenched at the operating current of \SI{10}{kA}. Right: temperature distribution after a quench occurrence in a single coil at the moment of reaching the peak voltage.}
	\label{fig:quench_example}
\end{figure}

An illustration of the modelling results of the magnet when operating at \SI{10}{kA} is presented in \cref{fig:quench_example}. In this particular example, one single coil without its casing and protection heaters was simulated. As the size of the normal zone increases, the current starts to gradually decrease and the total voltage reaches its maximum of \SI{654}{V} at the moment when the entire coil is in the normal state. On average, the normal zone propagates at a velocity of about \SI{7}{m}/s along the conductor and about \SI{2}{cm/s} from turn to turn across the coil windings. The peak temperature, which stays at the point of the quench initiation, saturates at \SI{200}{K}.

\begin{figure}[!b]
	\centering
	\includegraphics[width=0.5\linewidth,valign=c]{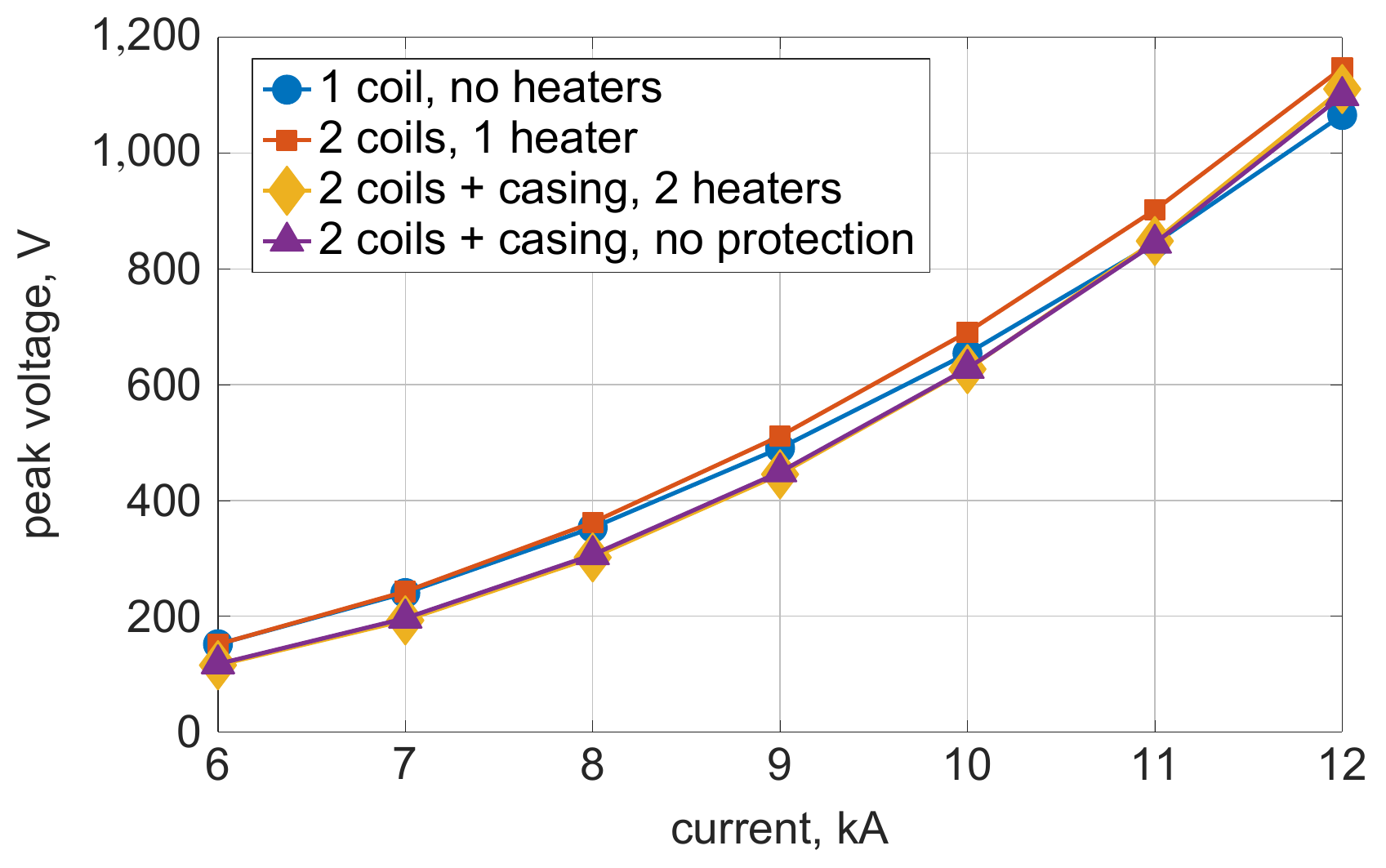}%
	\includegraphics[width=0.5\linewidth,valign=c]{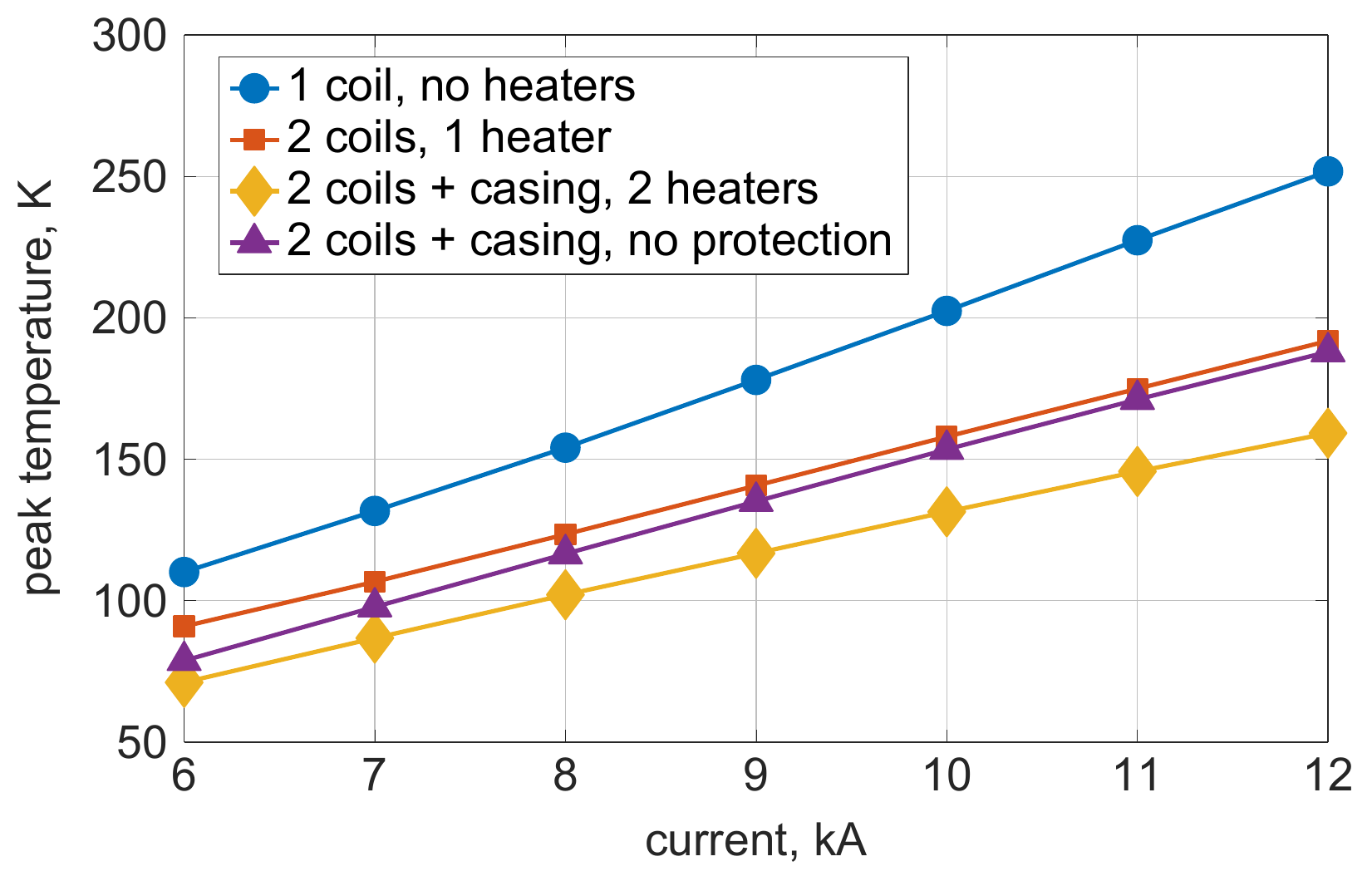}
	\caption{Peak voltage and peak temperature as a function of the operating current.}
	\label{fig:quench_comparison}
\end{figure}

The impact of the operating current on the maximum internal voltage and temperature for the various model configurations is summarized in \cref{fig:quench_comparison}. 
The peak voltage is practically independent of the considered cases, gradually increasing with current up to about \SI{1}{kV}, a value at which no insulation failures are expected. As the generated voltage increases with current, the time required to discharge the coils decreases typically from \SI{120}{s} at \SI{6}{kA} down to \SI{30}{s} at \SI{12}{kA}. The maximum temperature depends almost linearly on current, decreasing by about \SI{20}{\percent} when comparing the case of a non-protected single coil with the case of two coils protected by one heater, and by another \SI{20}{\percent} if the two coils are considered thermally coupled to the coil casing and firing of two protection heaters. In this case, the maximum temperature stays below \SI{150}{K} for the entire range of current, which is considered safe. If protection heaters would fail the peak temperature increases to some \SI{170}{K} maximum at nominal current, a value still considered safe.

In general, a moderate impact of the quench detection threshold varied in the range from 0.1 to \SI{10}{V} is observed, resulting in a maximum temperature variation of some \SI{5}{\percent}. The influence of the hot-spot location has also been investigated, showing that the maximum temperature changes by $\pm\SI{15}{\%}$ for a quench initiated at different spots along the conductor and across the winding. In addition, the ``quench-back'' effect, an additional heating in the conductor and casing due to currents induced by magnetic field changing with current, has a weak effect on the maximum temperature, decreasing it by less than \SI{5}{\percent} since the decrease of current in the windings is rather slow.

\subsection{Stray magnetic field and forces}

\begin{figure}[!b]
	\centering
	\includegraphics[width=0.45\linewidth]{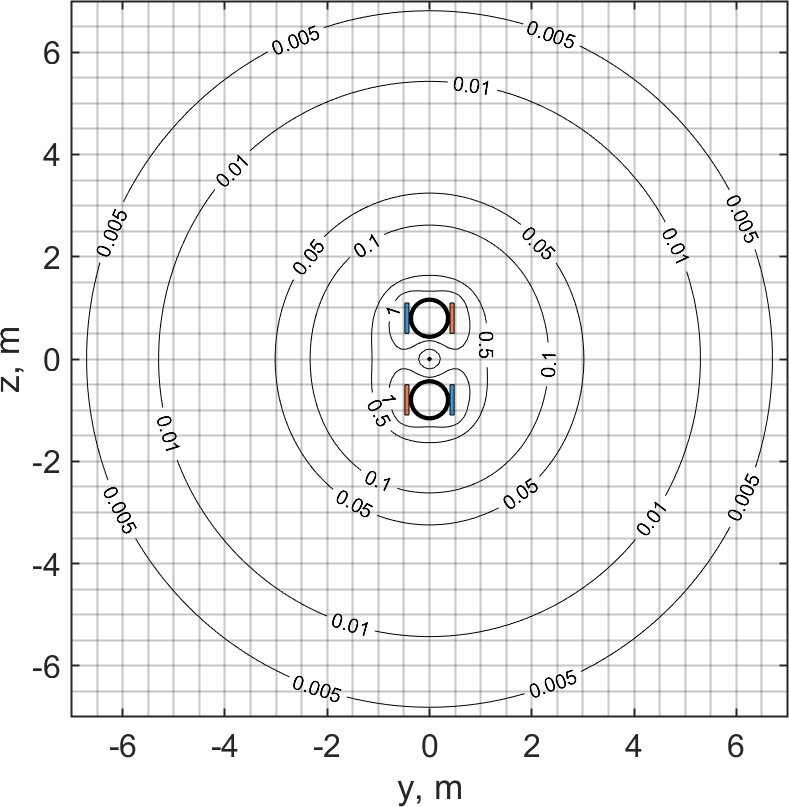}
	\caption{Map of stray magnetic field around the BabyIAXO magnet calculated in the magnet mid-plane for operation at \SI{9.8}{kA} nominal current. The contour labels are in tesla.}
	\label{fig:strayfieldmap}
\end{figure}

Magnetic field present at motors, pumps, compressors and other electrically driven rotary equipment should not exceed \SI{5}{mT} to guarantee their long term performance unaffected. For this reason, the magnetic field generated by the BabyIAXO coil windings carrying \SI{10}{kA} is analyzed at rather large distances assuming no ferromagnetic materials present. The \SI{5}{mT} requirement is fulfilled for the distance \SI{>7}{m}, see \cref{fig:strayfieldmap}.

A higher magnetic field is acceptable for the operation of cryocoolers: up to \SI{50}{mT} for their valve motors and up to \SI{0.5}{T} on the cold heads (data valid for Cryomech coolers). As presented in the following section, the cryocoolers will be installed at about \SI{1}{m} distance from the magnet center (i.e. at $x=0$, $y=\pm\SI{1}{m}$, $z=0$). One option is to use remote connection of the valve motors, which would result in some \SI{10}{\percent} cooling power performance loss and prevent their use starting from room temperature. The second option is to apply local shielding by using about \SI{18}{mm} thick iron. The second option is recommended by Cryomech, the manufacturer of the selected cryocoolers, as this customization can be provided directly by the company.

Various options of placing the five compressors required for operation of the five cryocoolers have been assessed such as attaching them to the rotation tower, installation on the rotating frame or leaving them simply on the floor. The latter option is selected for simplicity reasons after confirmation of Cryomech that the cooling performance will not be affected even if the compressor is installed at as far as \SI{100}{m} distance.

\begin{figure}[!t]
	\centering
	\includegraphics[width=0.55\linewidth,valign=c]{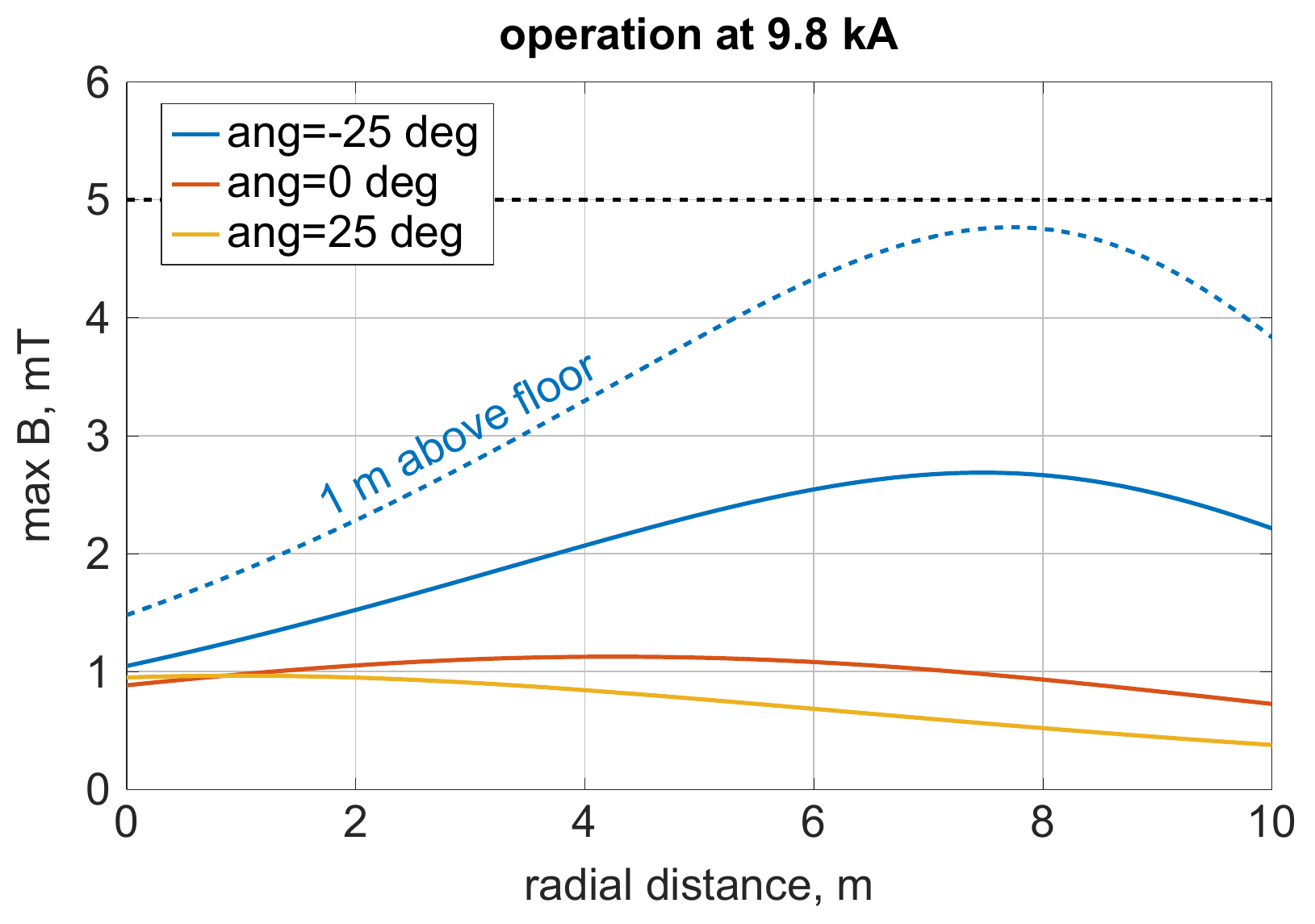}\hfill%
	\includegraphics[width=0.4\linewidth,valign=c]{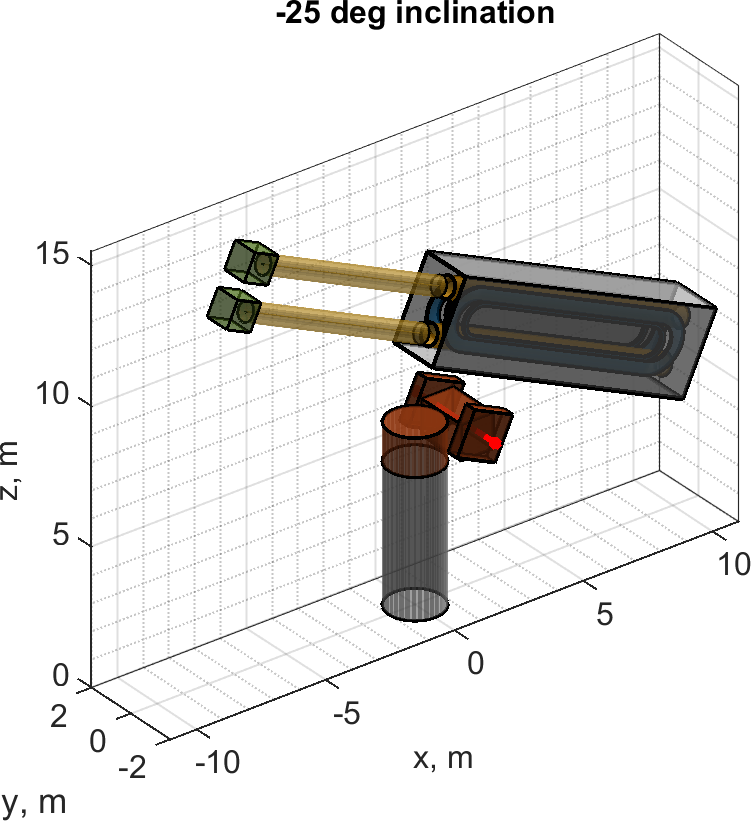}
	\caption{Stray magnetic field at $z=0$ for minimum, zero and maximum inclination of the system operating at nominal current of \SI{9.8}{kA} and at $z=\SI{1}{m}$ for the inclination at \ang{-25}.}
	\label{fig:strayfieldbmaxvsr}
\end{figure}

The stray field on the floor due to the magnet operating at \SI{9.8}{kA}, see \cref{fig:strayfieldbmaxvsr}, shows that the minimum distance is some \SI{10}{m}, at which the maximum magnetic field is below \SI{5}{mT} for all possible inclinations of the magnet. Note, that the magnetic field will be higher by a factor 1.2 in the case of operation at ultimate current of \SI{12}{kA}. In the direction to the optics and detectors, the stray field is below \SI{5}{mT} starting from half the detection line, \SI{3.7}{m} away from the cryostat end flange, and it decreases to \SI{0.5}{mT} near the position of detectors.

The magnet, optics and detectors are assembled together and installed on a non-magnetic support frame made of stainless steel. However, these components will be attached to an existing drive system (see Section~\ref{sec:BabyIAXO_drive}) essentially made of iron that will be magnetized by the magnet thereby generating a magnetic force on the magnet. Low carbon steel 1010 is used for the simulation of the drive system, whose geometry is simplified to a cylinder with wall thickness of \SI{20}{mm}, inner radius of \SI{1.0}{m} and a total height of \SI{6.5}{m} and three hollow blocks attached to it, see \cref{fig:strayfieldbmaxvsr}, right. The magnet cryostat is placed at \SI{1}{m} distance from the side blocks and it can rotate together with them as a solid body.

\begin{figure}[!t]
	\centering
	\includegraphics[width=9cm,valign=c]{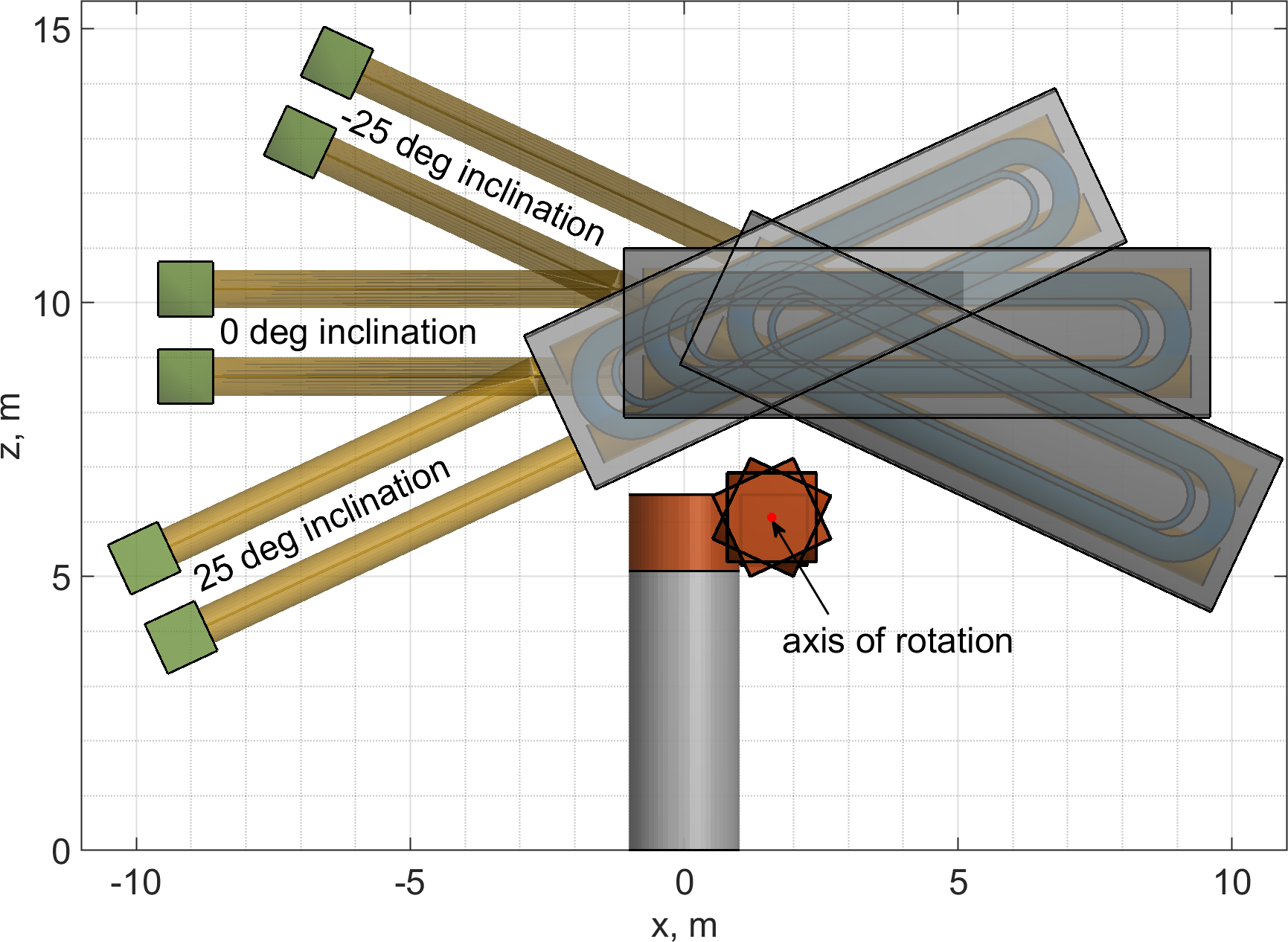}\hfill%
	\begin{tabular}{lrrr}
		\toprule
		Force & \multicolumn{3}{c}{Magnet inclination} \\
 		component & +\ang{25} & \ang{0} & \ang{-25} \\
 		\midrule
 		$F_x$, kN & +10 & -6 & -16 \\
 		$F_y$, kN & 0.0 & 0.0 & 0.0 \\
 		$F_z$, kN & -50 & -32 & -24 \\
 		\midrule
 		Total force, kN & +51 & +32 & +28 \\
 		\bottomrule
	\end{tabular}
	\caption{Left: schematic arrangement of the BabyIAXO components for the three angles of magnet inclination. Right: magnetic force on the magnet operated at \SI{9.8}{kA} due to the magnetized drive system.}
	\label{fig:strayfieldsketches}
\end{figure}

The magnetization of this iron is calculated by directing a magnetic moment of each mesh element of the tower volume along the total magnetic field generated by the magnet operated at \SI{9.8}{kA} and magnetic moments from the other mesh elements. The magnetic moments are used to calculate their magnetic field at the position of the conductor, thus the associated Lorentz force can readily be calculated. For the three inclinations analyzed, see \cref{fig:strayfieldsketches}, the largest force, equivalent to about \SI{5}{t}, is for the inclination at +\ang{25}, which corresponds to a smallest effective distance between windings and tower. It further increases to \SI{6}{t} for operation at ultimate current of \SI{12}{kA}.

In general, the obtained force load is rather high and extra cold mass supports to hold the cold mass in the cryostat in safe position without exceeding stress limits are needed. Note, that this is a consequence of using an existing rotation tower made available for the project. For a new system the rotation devices would be specified as non-magnetic. 

\subsection{Cryostat and Envelope}

\begin{figure}[!t]
	\centering
	\includegraphics[width=\linewidth]{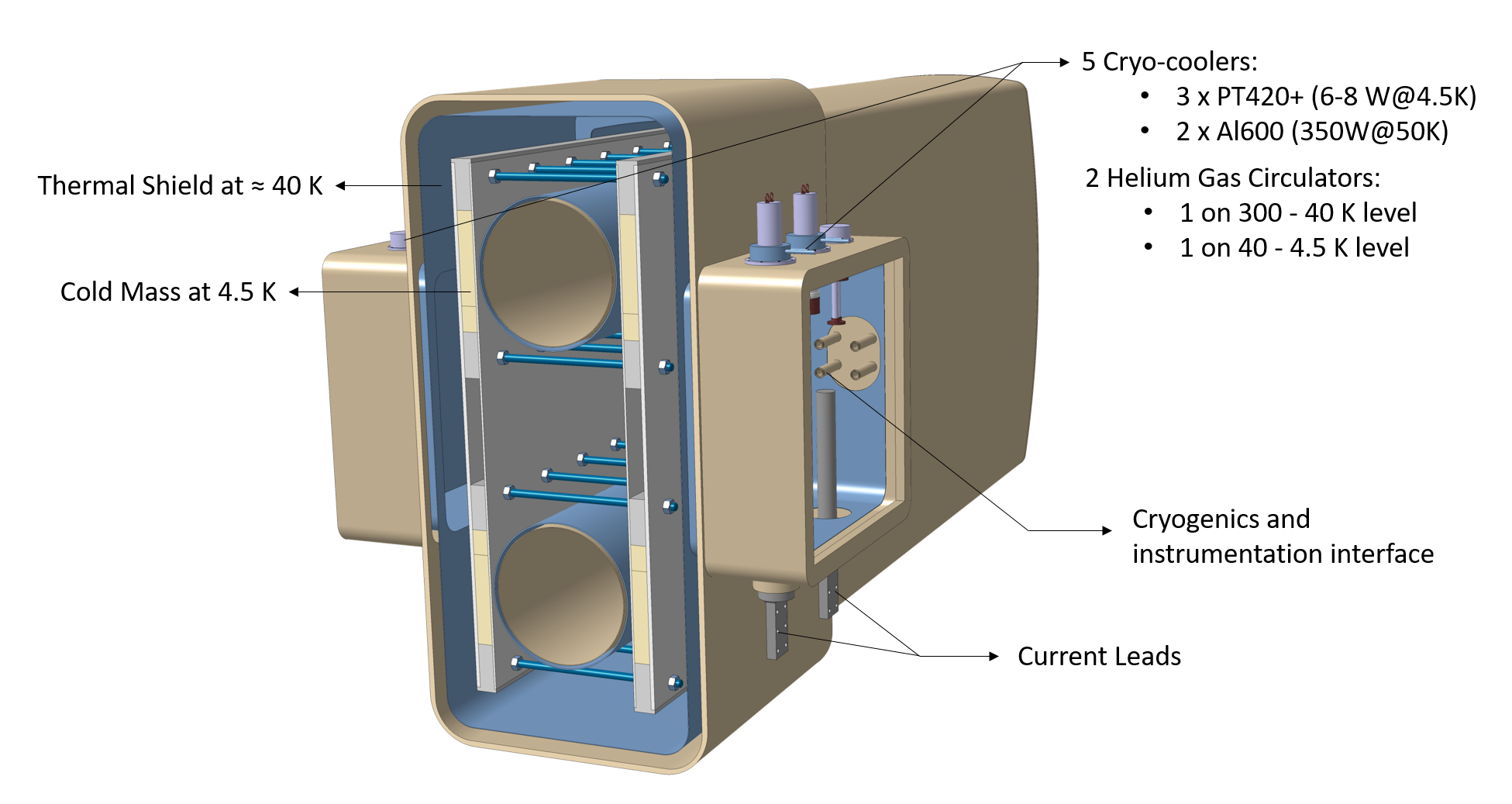}
	\caption{Cross-section of the 11 meter long BabyIAXO magnet cryostat in the center part. Clearly visible are the two 700 mm diameter free bores sandwiched between the two race track coils. All force transmitting links and services connections are positioned in a center post which is completed with two caps falling over the two cold mass extremities.}
	\label{fig:cryostat}
\end{figure}

The cryostat of BabyIAXO, shown in \cref{fig:cryostat}, comprises essentially a central post transferring the loads through the cold mass supporting and base, also including at either side service boxes with flanges, and longitudinally closed by two long caps. Essentially all components, with some exceptions like the cold mass supports, are made of Al5083-O alloy in order to reduce cost, avoid its magnetization by the magnet and minimize the overall mass, while achieving the required mechanical rigidity. 

In one service box there is access to the current leads from the bottom plate while two single-stage Al600 cryocoolers are positioned on the top plate. In the second service box there is on the top plate three two-stage PT420 cryocoolers as well as two helium gas centrifugal pumps, also called cryocirculators.  Cryogenics and instrumentation interfaces are also provided. From a maintenance and operation point of view, the service boxes have to be readily accessible. 

\begin{figure}[!b]
	\centering
	\includegraphics[width=0.5\linewidth,valign=c]{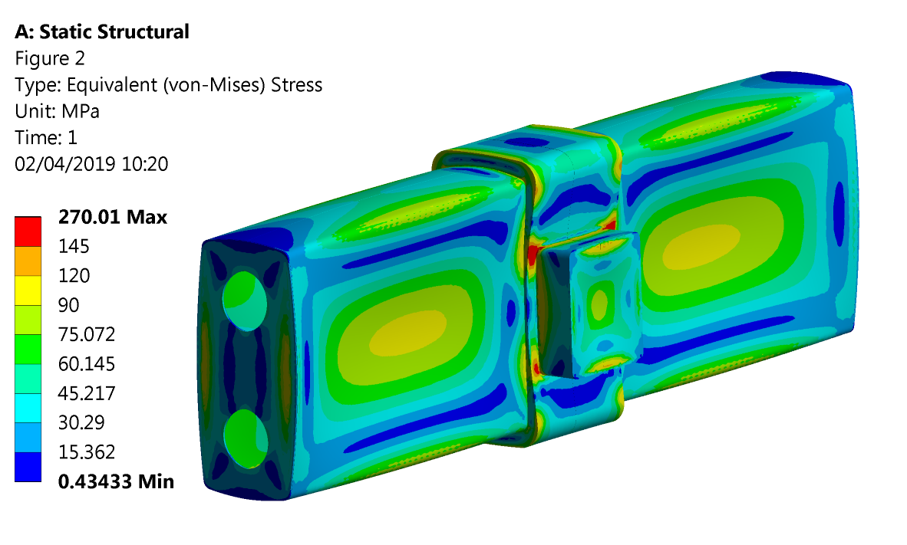}%
	\includegraphics[width=0.5\linewidth,valign=c]{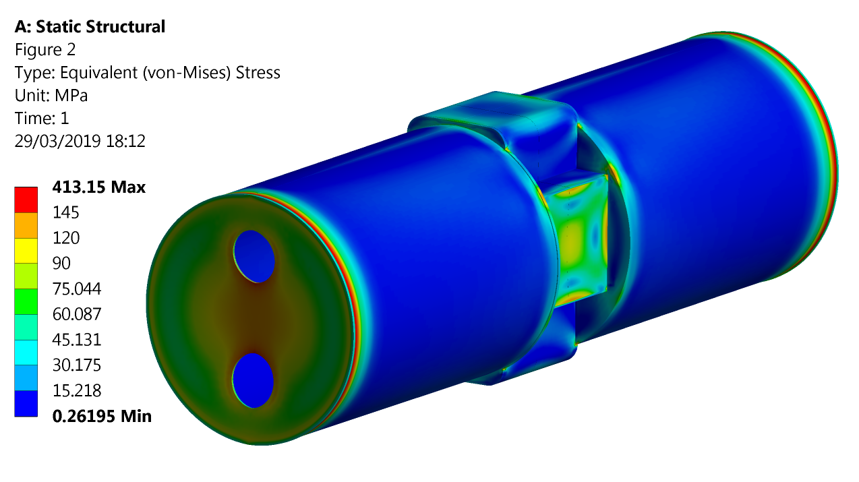}
	\caption{Von-Mises stress distribution in the rectangular (left) and circular (right) cryostats.}
	\label{fig:cryostatcomparison}
\end{figure}

Two shapes of the cryostat, rectangular box like and circular, have been investigated. A first estimate of the required amount of the material is obtained from a non-linear buckling analysis of a standalone cryostat, not accounting for the presence of the rotation system and cold mass supports. The non-linear buckling analysis takes into consideration the stress distribution from the static analysis, see \cref{fig:cryostatcomparison}, and the allowed manufacturing imperfections in the cryostat. The minimum material thickness is such that at 1.5 times the nominal load the maximum deformation is less than \SI{100}{mm} and that no buckling is observed. The material thickness is set at \SI{50}{mm} for the central box, \SI{40}{mm} for the long caps and \SI{25}{mm} for the flanges for the rectangular cryostat made of Al5083-O, while respectively \SI{40}{mm}, \SI{18}{mm} and \SI{36}{mm} for the circular one. 

Using the results of structural analysis as shown in \cref{fig:cryostatcomparison}, buckling of the structure is not observed in either option when applying about twice the ambient pressure, i.e. a factor of 2 for non-linear buckling. Note that there are regions of stress concentrations, like sharp corners, that might need reinforcement or local reshaping. In total, \SI{11.5}{t} and \SI{9.0}{t} of Al5083-O alloy are needed for the rectangular and circular cryostats, respectively. While the first option allows minimizing the footprint of the system as well as cryogenic load due to less surface, the second one has advantages of reduced mass and potentially cheaper manufacturing. Using stainless steel instead of Al5083-O allows reducing the material thickness in all three cryostat components, but it still causes an increase in the total mass by a factor of 2. The shape and material used for the cryostat are still under design and cost optimization. While a rectangular shape is assumed in some of the following illustrations, a cylindrical shape is depicted in the overview drawing of figure~\ref{fig:drive2a}.

\begin{figure}[!t]
	\centering
	\includegraphics[width=.9\linewidth]{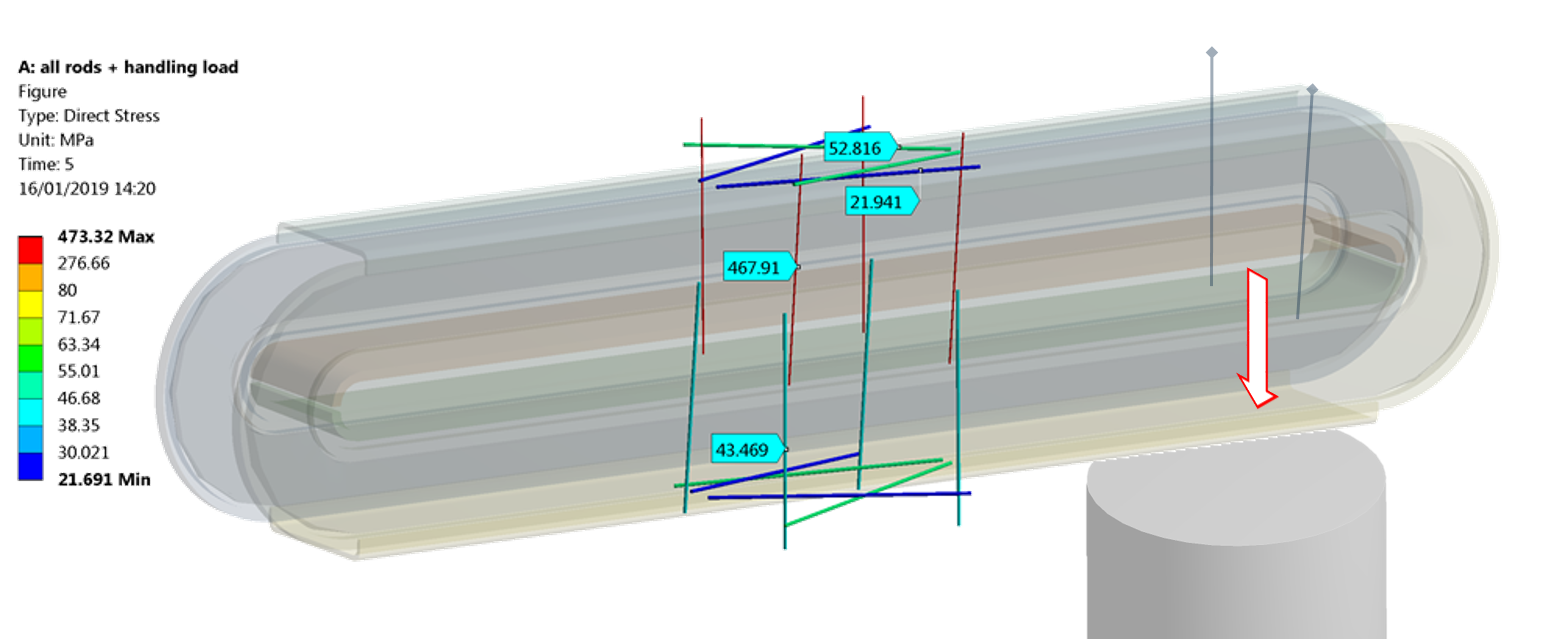}\\[1ex]
	\begin{tabular}{lccccc}
		\toprule
		& Material & Length & Diameter & \# of rods & Function \\
		\midrule
		Top vertical rods & Ti alloy & \SI{1.60}{m} & \SI{12}{mm} & 4 & Gravity support \\
		Bottom vertical rods & Permaglas & \SI{1.60}{m} & \SI{24}{mm} & 4 & Vertical centering \\
		Longitudinal rods & Permaglas & \SI{1.85}{m} & \SI{26}{mm} & 8 & Inclination support \\
		\bottomrule
	\end{tabular}
	\caption{Design parameters and stress analysis of the multi-rods suspension system of the BabyIAXO cold mass. In total 16 rods are foreseen positioned in the center post. In addition a few extra rods, not yet dimensioned, will be added for transferring the Lorentz force load between cold mass and rotation tower.}
	\label{fig:coldmasssupport}
\end{figure}

The cold mass is attached to the cryostat by means of cold-to-warm tie rods, see \cref{fig:coldmasssupport}, the best solution for achieving minimum heat loads at \SI{4.2}{K} as compared to a mixed system of rods and support posts. Four top vertical rods made of Ti alloy take about \SI{15}{t} gravity load of the cold mass, resulting in \SI{470}{MPa} tensile stress. Vertical centering of the cold mass is provided by four bottom rods made of Permaglass. Extra rods, resisting the attractive force between cold mass and iron structure present in the drive system, are also schematically illustrated. The design of these extra suspension rods is yet to be completed when details of the iron in the drive system are available. Given the estimated loads, no problems in terms of cryostat construction and thermal loads are foreseen.

Finally, lateral loads due to inclination are taken by eight longitudinal rods of Permaglass, which have to be slightly pre-stressed to ensure that they are always in tension. This cold mass suspension system fulfils the requirement of less than \SI{0.2}{mm} displacement when the cold mass is inclined at \ang{+-25}. Although the heat load can further be reduced by using glass or carbon fiber reinforced tubes, such option has been discarded due to higher manufacturing complexity and cost. 

\begin{figure}[!t]
	\centering
	\includegraphics[width=\linewidth]{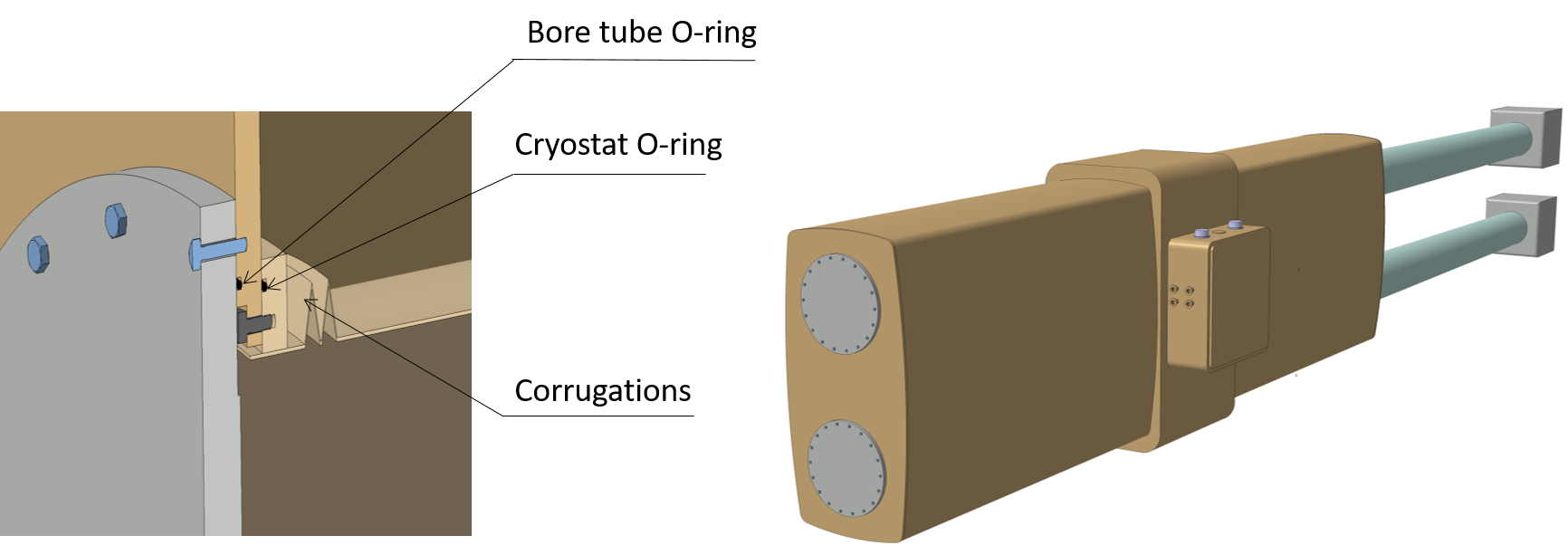}
	\caption{Layout of bore tube front closing flanges of the BabyIAXO cryostat, warm bore diameter of \SI{700}{mm}; and inserted corrugated interface bellow between bore tube and cryostat end plate.}
	\label{fig:cryostatbore}
\end{figure}

The cryostat bore tubes are made of \SI{1}{mm} thick stainless steel. The total mass of the two tubes is estimated at \SI{0.4}{t}. The user bore volume needs to be in vacuum or, in a second stage, filled with a buffer gas at low temperature. For this purpose additional flat heaters have to be installed on the surface of the tubes to prevent condensation and ice formation. As illustrated in figure~\ref{fig:cryostatbore}, the bore tubes can be vacuum sealed using an O-ring and a corrugated interface is included to accommodate the difference in thermal contraction of bore tube when at a lower temperature in the case of heater failure, and the rest of the cryostat at room temperature. 

At the front face of the magnet, two flanges can be bolted to close the bores. On the other side of the magnet, special gate valves have to be installed for the same purpose and allowing the interface separating bore tubes and optics. 

\begin{figure}[!b]
	\centering
	\includegraphics[width=\linewidth]{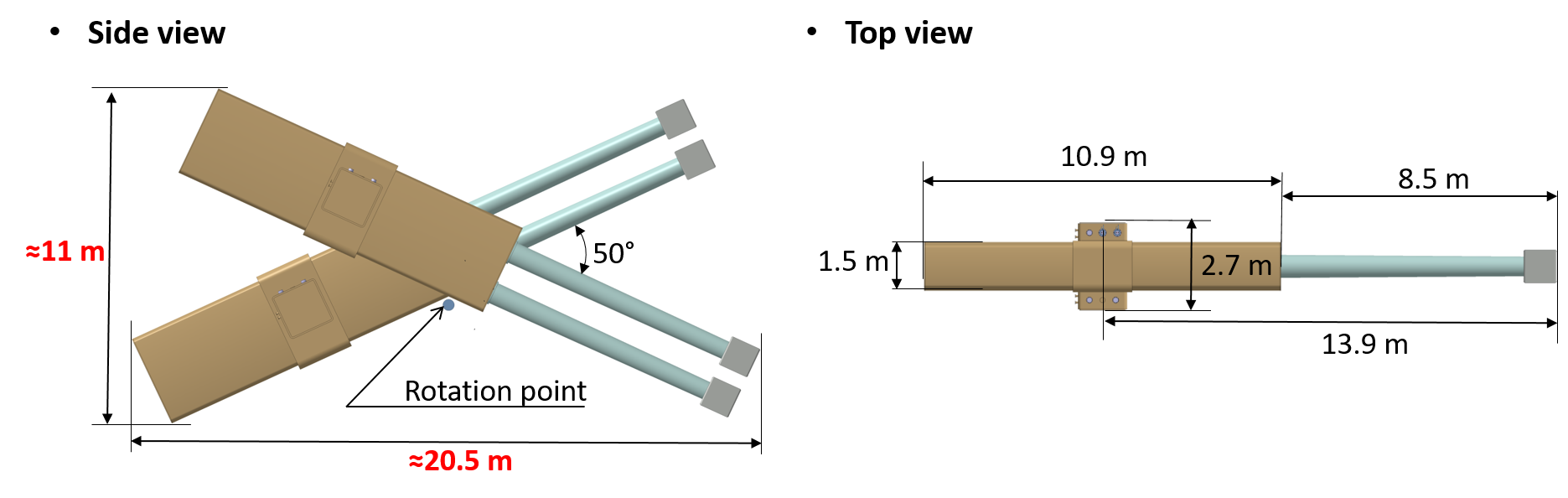}
	\caption{Main dimensions in side view and top view of the BabyIAXO cryostat with optics and detectors attached.}
	\label{fig:envelope}
\end{figure}

When accounting for the system inclination of \ang{+-25}, the overall dimensions of the moving system comprising cryostat, optics and detectors are about \SI{20.5}{m} in length and \SI{11}{m} in height, see \cref{fig:envelope}. Positioning the axis of rotation near to the geometric center of the assembly leads to a load unbalance of about \SI{700}{\kilo\newton\meter}, which has to be taken by the drive system. The overall mass of the BabyIAXO magnet is currently estimated at \SI{25}{t}.

Magnet, optics and detectors are assembled in line on a support beam that is attached to the drive system. Flexible chains comprising lines for compressors, current, vacuum and controls for the magnet services have to be routed along the beam and around the rotation tower to make 360 degree rotation possible. In addition sufficient supports for the chains and ground clearance is needed for this rotation.

\subsection{Cryogenics}

Various cryogenic options have been considered for BabyIAXO aiming at cost minimization and accounting for the absence of nearby cryogenic infrastructure at the provisioned sites at DESY. In view of the rather low heat load for stationary operation, see figure~\ref{table:heatloads},  of about  \SI{420}{W} for the thermal shield operated at \SI{45}{K} and less than \SI{4}{W} on the cold mass at \SI{4.2}{K}, a ``dry'' cooling system based on using cryocoolers has been selected.

\begin{table}[!t]
	\caption{Cryogenic heat loads of BabyIAXO.}
	\centering
	\vspace{1ex}
	\begin{tabular}{l*{2}{C{3.5cm}}}
		\toprule
		\multirow{2}{*}{Source} & \multicolumn{2}{c}{Heat load, W} \\
		\cmidrule{2-3}
		& @\SI{45}{K} thermal shield & @\SI{4.2}{K} cold mass \\
		\midrule
		Radiation & 160 & 2.2 \\
		Conduction in support structure & 2.1 & 0.1 \\
		Pair of current leads at \SI{0}{A} & 260 & 1.0 \\
		\midrule
		Total net & 420 & 3.3 \\
		Total for cryo-design (x2) & 840 & 6.6 \\
		\bottomrule
	\end{tabular}
	\label{table:heatloads}
\end{table}

It provides a simple ``plug-in'' operation at rather low cost when comparing to using cryogenic liquids, which would require a significant investment in a small helium refrigeration plant, recovery system and transfer lines given the absence of such infrastructure at the HERA South or East Halls at DESY. The disadvantage of a relatively slow cool-down of the \SI{15}{t} cold mass of some 17 days instead of a few days has to be accepted. Using cryocooler is an elegant solution forcing the designer to minimize the heat loads and increasing the efficiency of using the available cooling power. 

\begin{figure}[!t]
	\centering
	\includegraphics[width=\linewidth]{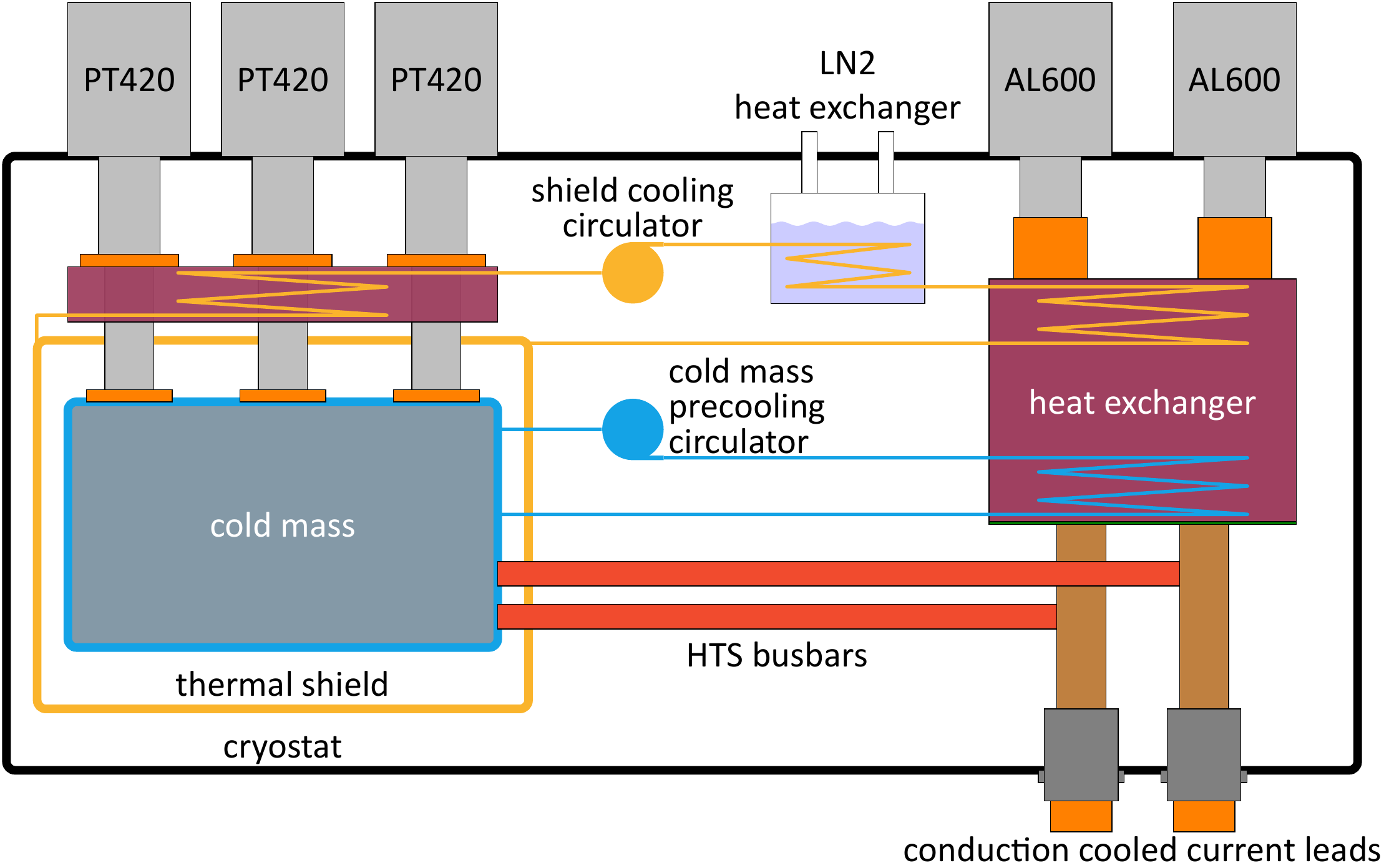}
	\caption{Layout of the innovative BabyIAXO ``dry'' cryogenic system based on using cryocoolers.}
	\label{fig:cryogenicslayout}
\end{figure}

\cref{fig:cryogenicslayout} shows the principal layout of the cryogenic system developed and optimized especially for BabyIAXO. It essentially comprises: 

\begin{itemize}
	\item two single-stage Cryomech AL600 Gifford-McMahon cryocoolers, maintaining the thermal shield and current leads at \SI{45}{K}, and delivering the main power for cooling down from room temperature to \SI{45}{K};
	\item three two-stage Cryomech PT420 pulse tube cryocoolers, maintaining the cold mass at \SI{4}{K} and helping to cool-down the cold mass from room temperature to \SI{4}{K};
	\item two helium gas circulators, key components for delivering the cooling power from the cryocooler cold heads to the cold mass and thermal shield; 
	\item a liquid nitrogen heat exchanger, which is normally off, but it can support the AL600 as a backup source of the cooling power and eventually be used to speed-up the magnet cool-down.
\end{itemize}

\begin{figure}[!b]
	\centering
	\includegraphics[width=0.9\linewidth]{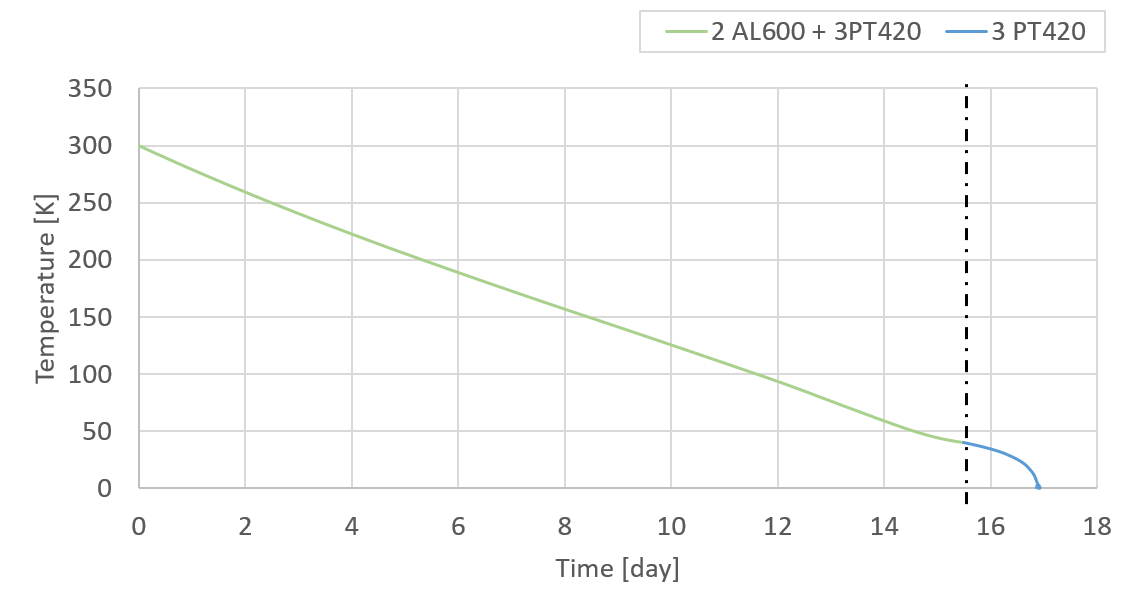}
	\caption{Cooling-down of the BabyIAXO cold mass by five cryocoolers within 17 days (not using the LN2 heat exchanger). At \SI{45}{K} the cold mass is disconnected from the single stage AL600 machines and cool down continues using the three PT420 only.}
	\label{fig:cryogenicscdtime}
\end{figure}

The cooling down of the system is as follows. All five cryocoolers will be switched on starting from room temperature. For a precooling stage of the cold mass from 300 to \SI{45}{K}, a certain amount of the AL600 cooling power is delivered by circulating helium gas circulating in pipes between the heat exchanger on the cryocooler heads and the cold mass. Together with the second stages of PT420, connected to the cold mass through thermal links, \SI{45}{K} is reached in 15 days, see \cref{fig:cryogenicscdtime}.

Next, the helium gas from the cold mass precooling circulator has to be pumped out and cooling of the cold mass is continued only by means of the three units PT420, which takes another 2 to 3 days to reach \SI{4.2}{K}. At the same time the two Al600 units keep the shield and current leads at \SI{45}{K} level. In total, the time required to cool down the cold mass is estimated at 17 days. Detailed hydraulic analysis of the cooling circuit is to be performed.

Using helium gas circulating in cooling pipes is also crucial to distribute the cooling power rather uniformly to the thermal shield around the 10-m long cold mass and bore tubes. In contrast to the cold mass itself, heat conductivity of the thermal shield is rather low over the entire temperature range and a single localized cooling source would not be effective.

Note that the cooling power of the AL600 cryocooler is about \SI{300}{W} at \SI{45}{K} and it increases strongly at higher temperatures up to roughly \SI{1.5}{kW} at room temperature. The PT420 cryocooler is capable to provide \SI{2}{W} at \SI{4.2}{K} on the second stage, which increases to some hundreds of watts at room temperature, and from \SI{55}{W} at \SI{45}{K} up to \SI{400}{W} at room temperature on the first stage. Concerning the reliability of using cryocoolers, increase of the operating temperature from \SI{4.2}{K} up to \SI{6.2}{K} due to \SI{10}{W} of heat loads would take some 20 minutes in the case of a power cut with all cryocoolers switched off. This time is considered to be sufficient to start backup diesel power sources.

After the cool down, the cryogenic system has to cope with the extra heat load when charging the magnet, which causes a drastic increase of the heat load due to the conduction cooled current leads up to about \SI{1}{kW}. This is because their cross-section is made purposely 40 to \SI{50}{\percent} smaller than for an optimal stationary current lead, for which the heat load at full current is just twice higher than that at zero current. This adjusted optimization allows to minimize the overall heat load during one full operating cycle of the magnet, i.e. short charging and very long idle mode. Hence, a temperature rise from \SI{45}{K} is expected during 1 h of charging, but it will stay below \SI{60}{K} because the \SI{1.5}{t} of thermal shield at \SI{45}{K} can absorb excessive heat in the system at a rather low increase of its temperature. As a result, the HTS bus bars have to be designed for a full current operation at temperature up to \SI{70}{K}.

The proposed cryogenic solution is unique for application in large superconducting detector magnets. Given the cost of cryocoolers of around 60~k\euro\ per unit and cost of a cryocirculator of about 30~k\euro, a complete cryogenic solution can be realized for BabyIAXO for less than 400~k\euro\ which is considered very cost-efficient.

\subsection{Mains, vacuum pumps, cooling water and Magnet Control System}

The electrical circuit of BabyIAXO requires a power converter rated at least for \SI{10}{kA} current and 5 to \SI{10}{V} terminal voltage, thus 50 to \SI{100}{kW} mains power. Each cryocooler consumes \SI{11.5}{kW}, in total \SI{58}{kW} for the five units.

Two standard oil diffusion pumps, connected to the vacuum vessel through a short bellow accommodating the \ang{+-25} inclination, are used to establish vacuum conditions for normal operation with magnetic field on. In addition, a turbo pump is foreseen as a back-up solution when magnetic field is off, allowing intervention on the diffusion pumps under vacuum. Additional parts are all-range vacuum gauges and various valves.

Water cooling is required for the operation of the diffusion pumps, for the cryocooler compressors and for the room temperature terminals of the current leads, which produce excessive heat during the 1-h magnet charging.

The magnet is controlled by a standardized Magnet Control System featuring control of the electrical circuit components, vacuum system, cryogenics, as well the diagnostics and safety systems.





%% file: sections/BabyIAXO_optics.tex
\subsection{General considerations for the IAXO and BabyIAXO optics}
In order to utilize the full potential of a strong magnet, the use of x-ray optics is crucial for any axion helioscope. Telescopes enable focusing a potential x-ray signal from axion-to-photon conversion into a small spot and therefore allow for the use of a small-area detector which is essential to achieve the ultra-low background levels required for (Baby)IAXO. While boosting the signal-to-noise ratio is one of the main advantages of x-ray telescopes (XRTs), they also open up the possibility to simultaneously measure signal (in the spot area) and background (outside the focal spot region) while operating in axion-sensitive conditions.

Generally speaking, the performance of an x-ray optic can be characterized by three basic properties: (1) the point spread function (PSF) of an XRT determines the shape and size of the observable focal spot; (2) the throughput, $\epsilon_{\rm{O}}$, refers to the amount of incident photons properly focused by the optic; and (3) the field-of-view (FOV) describes the extent to which a given optic can focus off-axis photons properly. 
For IAXO as well as BabyIAXO, the large entrance pupil and the energy band led to the consideration of grazing-incidence reflective optics over alternative approaches such as refractive lenses. The design of these reflective XRTs, which rely on the principle of total external reflection of keV-range photons, requires a careful optimization of the geometric layout of the optic as well as its mirror coatings to maximize the XRT figure of merit $f_{O}$, which improves with increasing throughput $\epsilon_{\rm{O}}$ and decreasing focal spot area $a$:
\begin{equation}
f_{O}=\frac{\epsilon_{\rm{O}}}{\sqrt{a}}
\end{equation}
Optimizations can be challenging, since, for example, to achieve the smallest spot area $a$, the optics should have a focal length, $F$, that is as short as possible due to the fact that the spot area grows quadratically with the focal length of the optic ($a\propto F^{2}$). At the same time, the throughput needs to be maximized and depends on individual mirrors having the highest possible x-ray reflectivity. This reflectivity increases with decreasing graze angle, $\theta$, and since $F\propto\frac{1}{\theta}$, the focal length of any XRT for axion research should be as long as possible to achieve the highest throughput conflicting with the focusing requirement as outlined above. The optimization of the optical design is further complicated by the fact that throughput, PSF and FOV of an XRT have a complex dependence on both the incident photon energy $E$ and the graze angle $\theta$. Additionally in the case of implementation in an axion helioscope, maximization of the telescope performance needs to simultaneously take into account the axion spectrum and the detector response to achieve the best possible sensitivity for the experiment.

Since the x-rays produced via the conversion of axions to photons in the (Baby)IAXO magnet have the same directionality as the incident axions due to conservation of energy and momentum, the FOV of the XRT needs to be just slightly larger than the inner $3$ arcminutes ($\approx$~\SI{0.9}{mrad}) of the solar disk. Most axions produced in the Sun originate from this region. Required imaging performance for BabyIAXO is more relaxed than for space x-ray aplications. Other factors on the other hand, such as high throughput, cost-effectiveness and low-risk fabrication, are important selection criteria for the fabrication technique of the two BabyIAXO telescopes and eventually the eight final IAXO optics.  

The x-ray astronomy community has significantly advanced the technology of reflective x-ray optics over the past half century. Even though requirements and optimization of such XRTs for axion helioscopes differ from those for astrophysical applications, it is still possible to greatly benefit from the developments that led to mature technologies. For IAXO and BabyIAXO, the concept of segmented, (hot or cold) slumped-glass optics was adopted as the baseline fabrication approach for several reasons. First, the technology is mature and has been developed by members of the IAXO collaboration, most recently for the NuSTAR satellite mission~\cite{nustar2013}. Second, this approach facilitates the deposition of either single-layer or multi-layer reflective coatings. The use of multi-layer coatings could potentially enhance the throughput and allow for optimization of the spectral response of the BabyIAXO XRT. Third, this fabrication technique is less expensive than most others and, fourth, a rather modest focusing of the central $3$-arcminute core of the Sun is achievable. Just recently a second approach has been added to the baseline approach: replicated aluminum foil optics~\cite{Okajima:01,Okajima:02,Okajima:03,Okajima:04}. Although other technologies may offer better resolution than these two baseline approaches, they would not produce a significantly smaller focused spot of the solar core given the extend of this axion source and their effective area would not match that of segmented optics either; factors that are key for the optics figure of merit of BabyIAXO. 

\begin{figure}[b!] \centering
\includegraphics[width=0.6\textwidth]{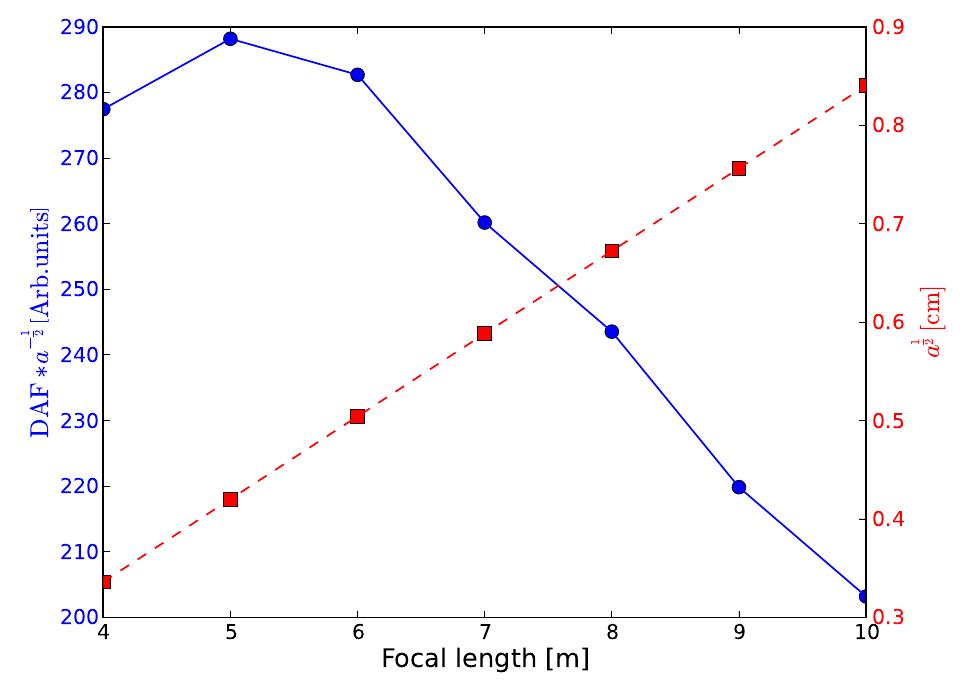}\hspace{2pc}%
\caption{\label{fig:focusoptim} Value of the focal spot size $\sqrt{a}$ (red squares and dashed line, right axis) and the figure of merit $f_{\rm{DO}}$ (blue circles and solid line, left axis) versus focal length $f$. The optimal figure of merit is found for $f = 5$~\si{m}.}
\end{figure}

Both optics as proposed for IAXO are based on approximations to Wolter-I-type telescopes~\cite{Wolter52}. Mirror shells can be nested confocal and coaxial, making it possible to achieve high efficiency for the telescope. Reflectivity can be further enhanced by making use of Bragg's law resulting in constructive interference of the incoming photons. This could be accomplished by coating the mirror substrates with a multi-layer. Multi-layers consist of periodic or non-periodic structures of alternating thin film layers of two or more materials (high-mass metals, such as for example tungsten, and low-mass spacers, like silicon) deposited on an optical substrate. In the design process of the IAXO optics, the optical prescription and reflective coatings are being identified by a systematic search of a multidimensional parameter space that accounts for detector efficiency, axion spectrum, optics properties and recipes of the reflective coatings. The combined figure of merit for XRT and detector together, $f_{\rm{DO}}$, is then computed and the optical prescription and multi-layer recipes with the highest $f_{\rm{DO}}$ are chosen as the nominal IAXO optics design. Figure~\ref{fig:focusoptim} shows $\sqrt{a}$ and $f_{\rm{DO}}$ as a function of the focal length demonstrating the optimization process for these parameters. 
From this, the optimal focal length for the IAXO optics is found to be $f=5$~\si{m}. This and further considerations, as detailed in Ref.~\cite{Armengaud:2014gea}, determined the initial design proposed for the IAXO optics. An IAXO pathfinder optic~\cite{Aznar:2015iia,Anastassopoulos:2017ftl} was fabricated, tested and installed in the CAST experiment. This prototype optic consists of a \SI{30}{\degree} segment of a full-revolution NuSTAR-like segmented-glass optic. With results from its characterization and operation further refinements were implemented in the optics design for the BabyIAXO and IAXO telescopes. The key design parameters for the baseline segmented-glass optic telescopes are listed in table~\ref{tab:IAXOoptic}. 
\begin{table}[t] \centering 
\small
\begin{tabular}{ll}
\hline
Telescopes & \num{8} \\ 
\textit{N}, Layers per telescope & \num{123} \\
Segments per telescope & \num{2172} \\
Geometric area of glass per telescope & \SI{0.38}{m^2} \\
Focal length & \SI{5.0}{m} \\
Inner radius & \SI{50}{mm} \\
Outer radius & \SI{300}{mm}  \\
Minimum graze angle & \SI{2.63}{mrad} \\
Maximum graze angle & \SI{15.0}{mrad} \\
Coatings (multilayers) & W/B$_{4}$C or Ir\\
Pass band & 1--10 keV \\
IAXO, Nominal, \SI{50}{\%} Encircled energy function (\SI{50}{\%} EFF, HPD) & \SI{0.29}{mrad} \\
IAXO, Enhanced, \SI{50}{\%} EFF (HPD) & \SI{0.23}{mrad} \\
IAXO, Nominal, \SI{80}{\%} EFF & \SI{0.58}{mrad}  \\
IAXO, Enhanced, \SI{90}{\%} EFF & \SI{0.58}{mrad}  \\
FOV & \SI{2.9}{mrad}  \\ \hline
\end{tabular}
\caption{\label{tab:IAXOoptic} Main design parameters of the IAXO x-ray telescopes.}
\end{table}
We note that the optimization process continues and any additional results from the pathfinder, which continues to acquire data in the CAST experiment at CERN, as well as ongoing R\&D work will be used to further improve the telescopes for BabyIAXO and IAXO. In particular to  fully cover the proposed BabyIAXO magnet bore diameter. 

\subsection{Optics for BabyIAXO}
The BabyIAXO magnet will have two \SI{70}{cm}-diameter ports available and thus two optics will be employed to maximize the science return of this experiment. Our baseline approach is to cover one of the two magnet bores with a custom-designed BabyIAXO optic with performance similar to a final IAXO optic and the other one with an existing flight-spare module from the X-ray Multi-mirror Mission (XMM) Newton, one of ESA's most successful flagship missions currently flying in space.
\subsubsection{Custom x-ray telescope for BabyIAXO}
\begin{figure}[b!] \centering
\includegraphics[width=0.8\textwidth]{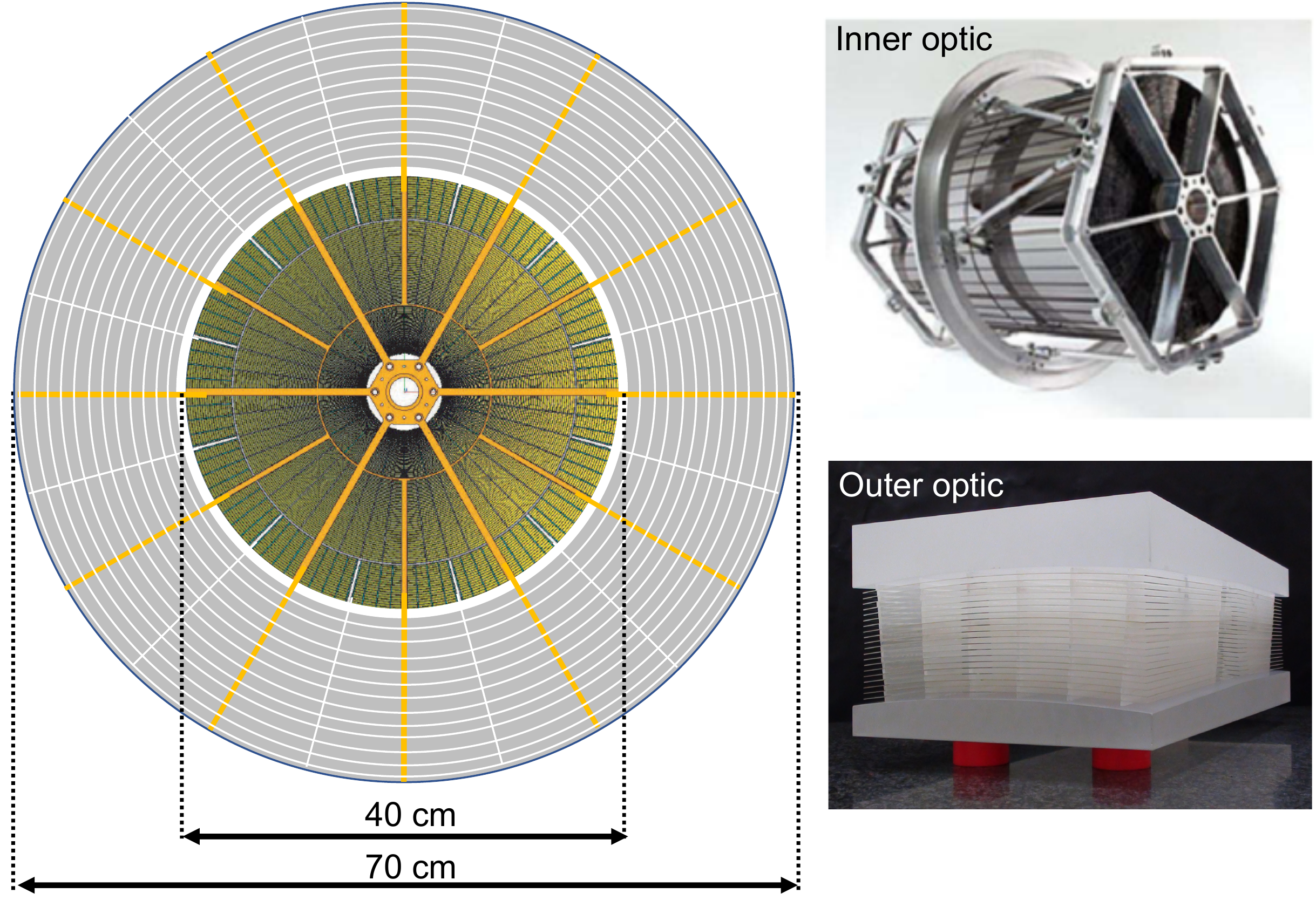}\hspace{2pc}%
\caption{\label{fig:IAXOXRT} Left: Schematic view of the BabyIAXO optic, including the hexagonal “spider” structure that will be used to mount the optic to the magnet bores. The optic consists of an inner and outer part using two different glass-slumping and optics assembly technologies. The thousands of individual mirror segments are visible. Right: the top image shows the NuSTAR optic which forms the base of the inner part of the BabyIAXO optic, while the bottom image displays the cold-slumped glass technology that is going to be used to build the outer part of the axion telescope.}
\end{figure}

The baseline x-ray optic for BabyIAXO is an x-ray telescope that is as close as possible in dimensions and performance to the final IAXO optic layout (see Fig.~\ref{fig:IAXOXRT} for a schematic of x-ray telescope) utilizing a segmented-glass approach. Simultaneously, this will allow for the optimization of two different techniques used to build segmented-glass optics and their integration into a single large-diameter optic.

\begin{figure}[t!] \centering
\includegraphics[width=0.8\textwidth]{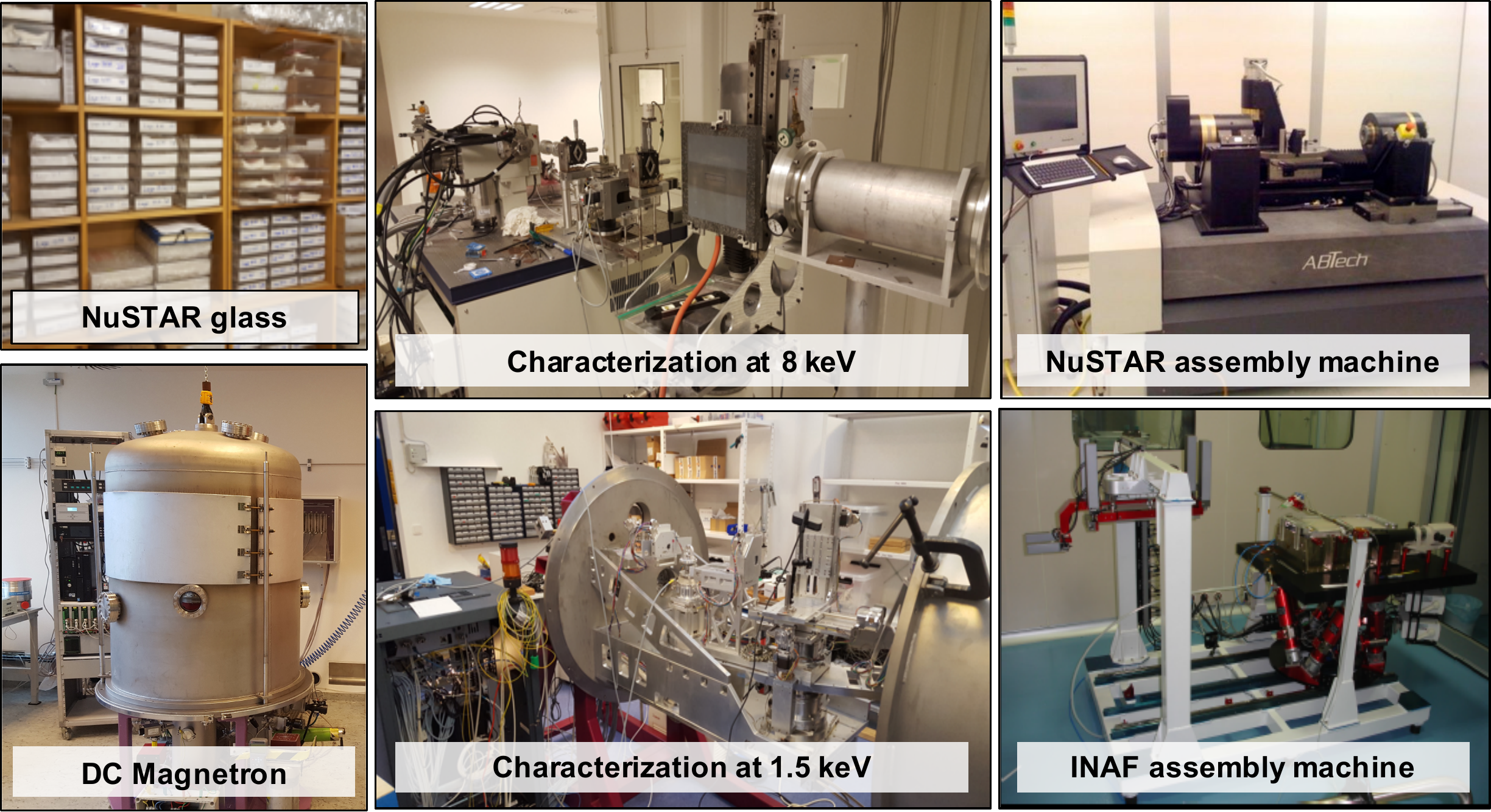}\hspace{2pc}%
\caption{\label{fig:Coatingfacilities} Available infrastructure to fabricate and characterize the BabyIAXO multi-layer-coated optic. Top left: NuSTAR leftover flight glass in storage. Bottom left: DC magneton sputter coater to deposit a variety of multi-layer coatings on curved substrates. Top middle: DTU calibration facility at \SI{8}{keV}. Bottom middle: DTU calibration facility at \SI{1.5}{keV}. Top right: NuSTAR precision assembly machine used to fabricate the IAXO pathfinder optics and suitable to fabricate the central part of the BabyIAXO and IAXO optics. Bottom right: Precision assembly machine for the outer part of the BabyIAXO and IAXO optics using the cold-slumped glass approach.}
\end{figure}

The inner part of the custom BabyIAXO optic ($r=$~5--20 cm) will rely on the same technology as the NuSTAR telescopes, i.e. hot-slumped, multi-layer-coated glass mounted in the NuSTAR precision assembly setup. This will make use of all existing NuSTAR infrastructure, which will help reduce costs significantly. The outer part of the BabyIAXO optic from a radius of \SI{20}{cm} to \SI{35}{cm} will be covered using cold-slumped glass and its assembly technology that has been recently demonstrated at INAF~\cite{2016SPIE.9905E..6UC}. In this way, final risk reduction for both technologies can be performed, while covering a maximal amount of BabyIAXO's cross-sectional area and thus maximizing the FOM for the experiment. Both technologies are mature and no major challenges are anticipated for their hybrid-assembly. The complete, segmented-glass, reflective optic with its multi-layer-coated segments will be placed inside a vacuum vessel that is attached to the magnet via a gate valve and a flexible bellow to allow for alignment of the instrument and will connect on the downstream-end to the detector placed in the focal plane. The telescope will focus the parallel beam of photons from axion conversion in the magnetic field volume into a small (few \si{mm^2}) spot on the detector. The XRT vacuum vessels will be installed on the telescope platform of the BabyIAXO magnet with the help of hexapod mounts, allowing for precise alignment of the optic with respect to the reference axis that runs through the center of the magnet bore and points directly at the center of the solar disk. An x-ray source mounted at the end of the magnet located opposite the telescope can be used to monitor the focal spot stability over time. In order to fabricate the BabyIAXO optic quickly and cost-efficiently, the production leverages investments and expertise at LLNL, DTU, MIT and Columbia University in the technology and construction used for NASA's NuSTAR mission, as well as the advanced technology developments for cold-slumped, segmented-glass optics at INAF. Figure~\ref{fig:Coatingfacilities} shows some of the existing and available facilities and infrastructure to build the BabyIAXO optics, including glass storage, facilities for coating, characterization and assembly of optics.
The path to building the custom BabyIAXO telescope is straightforward:
\begin{enumerate}
\item {Finalization/fine-tuning of BabyIAXO XRT in terms of the geometric design and the mirror coating recipe to optimize the BabyIAXO science reach}
\item Fabrication of central part of optics, including the following steps:
	\begin{itemize}
	\item Determination of optimal use of existing NuSTAR flight-spare glass;
	\item Preparation, cleaning and cutting of glass substrates using the existing NuSTAR infrastructure;
	\item Calibration of former NuSTAR coating setup (also used for current Athena mission) and test coating of multi-layer mirrors at fully available coating facility of IAXO collaborators;
	\item Coating of final BabyIAXO mirror segments;
	\item Assembly of optics on custom-made NuSTAR Lathe CNC machine fully available for BabyIAXO use, which was already employed to fabricate the IAXO pathfinder optic.
	\end{itemize}
\item Fabrication of outer part of optics by means of:
	\begin{itemize}
	\item Cold-slumping of glass needed for large-radii parts of the BabyIAXO optics;
	\item Securing the shape of the slumped-glass via gluing back-support braces and quality check;
	\item Deposition of single or multi-layer coating and characterization of shell performance (might be done before cold-slumping step);
	\item Precision assembly of outer layers to obtain co-centric, co-aligned outer optics ring.
	\end{itemize}
\item {Assemble complete telescope by integrating inner and outer optics into a single instrument with common optical axis;}
\item{Full characterization in terms of focal length, spatial and spectral performance of the completely assembled optic at an x-ray test facility, potentially end-to-end including BabyIAXO detector;}
\item{Implementation, alignment and commissioning of the x-ray optic at final experimental site of BabyIAXO.}
\end{enumerate}

\subsubsection{XMM flight-spare x-ray telescope for BabyIAXO}
\begin{figure}[b!] \centering
\includegraphics[width=0.6\textwidth]{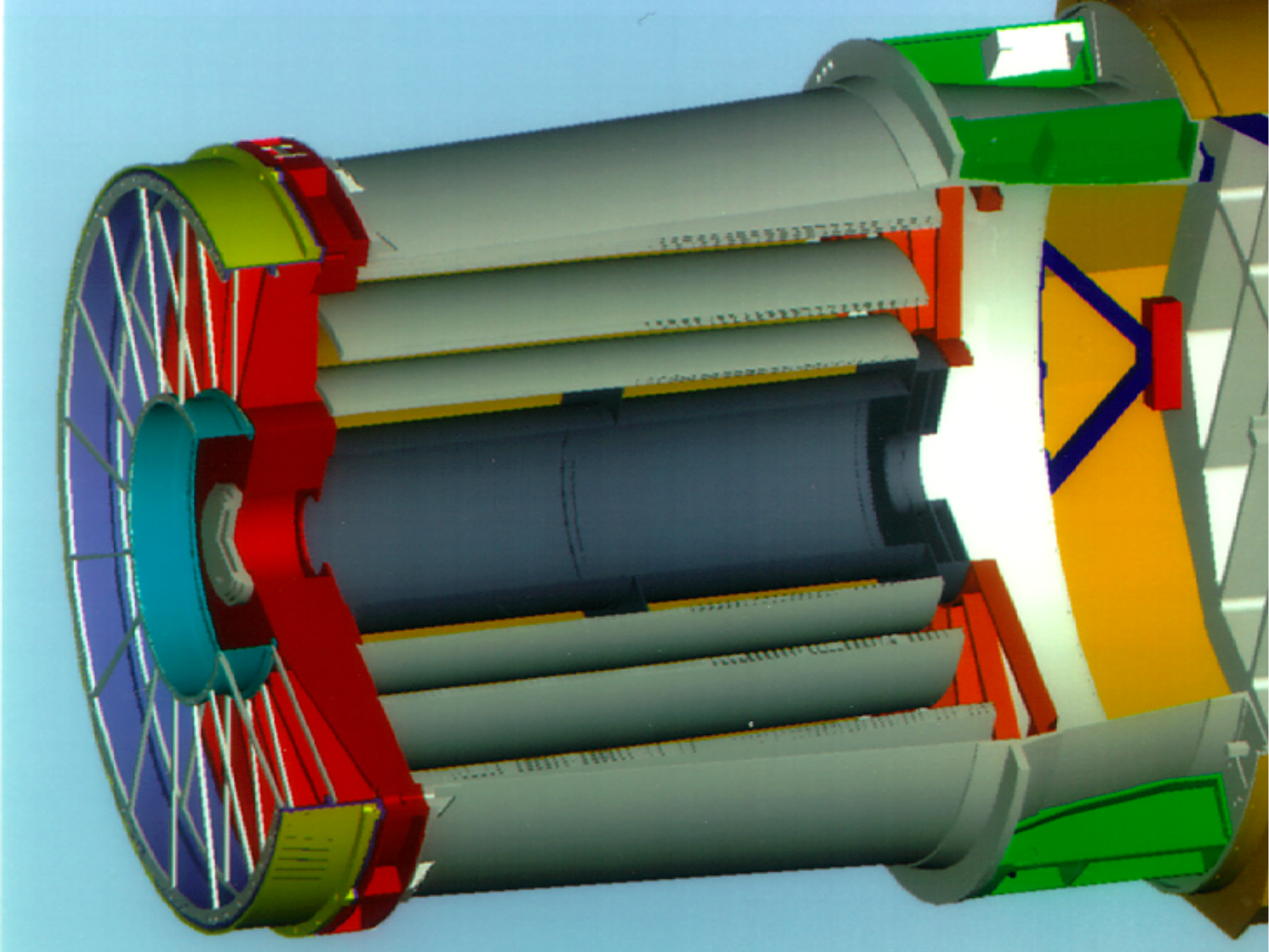}\hspace{2pc}%
\caption{\label{fig:xmmxrt} 3D model of XMM Newton telescope showing a few of its 58 full-revolution nickel shells that have been coated with gold.}
\end{figure}

\begin{figure}[b!] \centering
\includegraphics[width=1.0\textwidth]{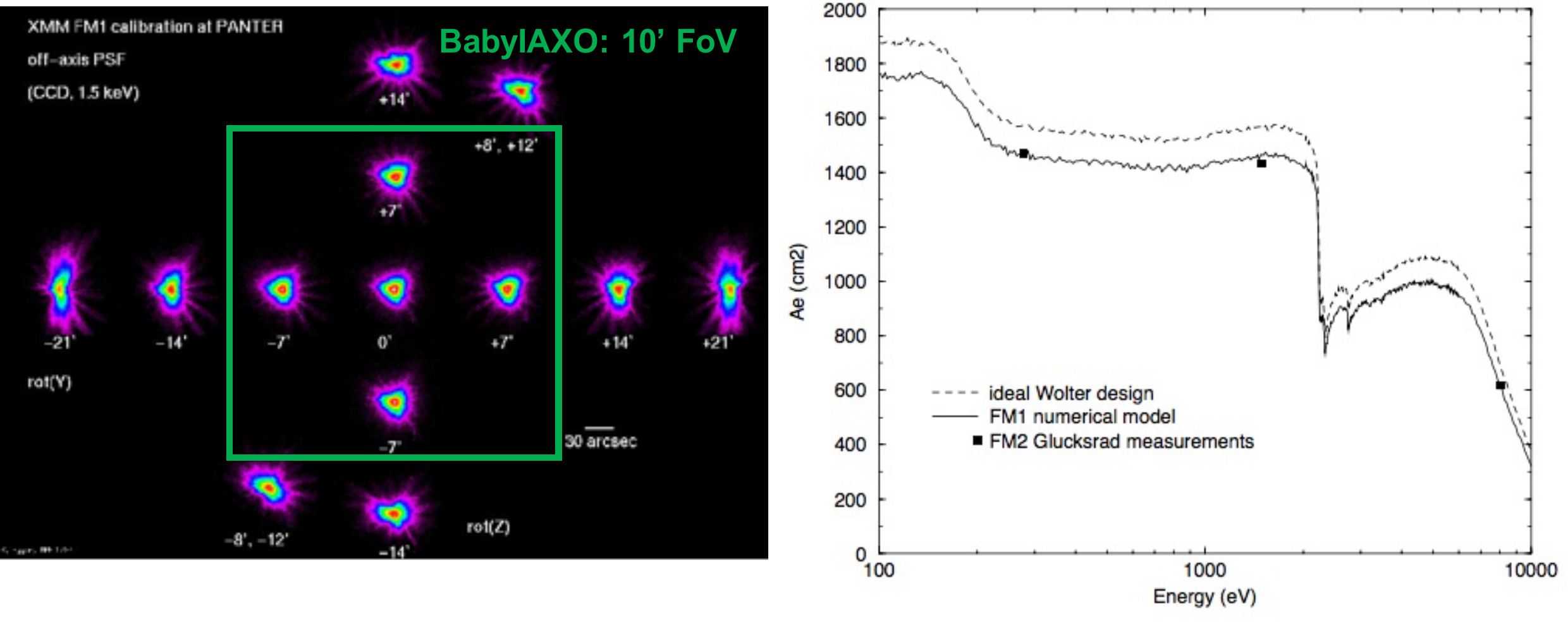}\hspace{2pc}%
\caption{\label{fig:XMMPSFEA} Left: Spatial resolution of XMM's FM1 in terms of the point spread function on- and off-axis. The (Baby)IAXO field of view is indicated in green and the HPD as well as the $90\%$ of the enclosed energy fraction are of the order (or below) $1$~arcmin, which fully satisfies the BabyIAXO experiment resolution requirements. Right: Effective area of one of the XMM optical modules (representative for all FMs) as measured at the PANTER X-ray Test Facility in a setup simulating illumination by a source at infinity (Gl{\"u}cksrad configuration). On-ground measurements and simulation match the observed on-orbit performance. Also included the effective area of the ideal Wolter design configuration for comparison (dashed line).}
\end{figure}

In order to obtain the highest sensitivity to detect solar axions with BabyIAXO, the experiment has been designed to comprise two large diameter bores (\SI{70}{cm} diameter each). For the second available bore, the use of an existing XMM Newton~\cite{Jansen:2001bi} flight spare telescope is envisioned, which matches the BabyIAXO bore dimensions well with its largest shell radius of \SI{35}{cm}. Two such optics exist and they are currently in storage at MPE's PANTER X-ray Test Facility in Neuried, Germany as property of ESA. These XMM telescopes (Flight module (FM) 1 and 5) consist of 58 Wolter I grazing-incidence mirrors each and the shells are nested in a co-axial and co-focal configuration (see Tab.~\ref{tab:XMM} for technical specifications and Fig.~\ref{fig:xmmxrt} for a 3D model of one such telescope).
\begin{table}[t] \centering 
\small
\begin{tabular}{ll}
\hline
Layers of telescope & \num{58}  \\
Focal length & \SI{7.5}{m} \\
Inner radius & \SI{153}{mm} \\
Outer radius & \SI{350}{mm} \\
Length of parabolic mirror & \SI{300}{mm}  \\
Length of hyperbolic mirror & \SI{300}{mm} \\
Mirror thickness & \SIrange{0.47}{1.07}{mm}  \\
Packing distance & \SIrange{1}{5}{mm} \\
Mirror module mass & \SI{425}{kg} \\
Reflective coating & Gold (\SI{250}{nm})\\
Pass band & \SIrange{0.15}{15}{keV} \\ \hline
\end{tabular}
\caption{\label{tab:XMM} Main design parameters of the XMM x-ray telescopes.}
\end{table}
The performance of FM1 and FM5 is comparable to the XMM optics on-board the ESA flagship mission launched in December 1999, currently in its 20th year of successful operation. The focal length of these optics is \SI{7.5}{m}, which is larger than the nominal \SI{5}{m} focal length for the IAXO baseline optics, but can be accommodated in the BabyIAXO setup. The gold-coated nickel shells in a Wolter-I configuration consist of a parabolic and a hyperbolic part of \SI{300}{mm} length each. The mirror thickness varies from \SIrange{0.47}{1.07}{mm} from innermost to outermost shell and the packaging distance ranges from \SIrange{1}{5}{mm}. This results in a total mass of each complete XMM mirror module of about \SI{425}{kg}, which is expected to be significantly heavier than the custom IAXO optics ($\lesssim$\SI{100}{kg}), but will be implemented in the BabyIAXO setup while keeping the strict alignment requirements due to the rigidity of the experiments magnet support structure and the location of the optics relatively close to the center of mass. In terms of performance, the XMM telescope matches the requirements for BabyIAXO well, even though it has not been specifically optimized for axion research. The spatial resolution of the XMM telescope provides a \SI{15}{arcsec} half power diameter (HPD, i.e. diameter of the circular area that encompasses \SI{50}{\%} of the total photon flux) and the overall effective area is \SI{1500}{cm^2} at \SI{2}{keV} and \SI{900}{cm^2} at \SI{7}{keV} for a single module. Figure~\ref{fig:XMMPSFEA} shows the XMM optics performance in terms of these two quantities.
XMM mirrors are most efficient in the energy range from \SIrange{0.1}{10}{keV}, which nicely coincides with the region of interest in energy for solar axions, with a maximum around \SI{1.5}{keV} and a pronounced edge near \SI{2}{keV} (the Au M-edge). As flight spare modules, FM1 and FM5 have been extensively characterized at the Centre Spatial de Liege (CSL) Vertical Facility and MPE's PANTER X-ray Test Facility and they perform similar to the actual flight optics. A detailed simulation model has also previously been validated using ground calibration data and was able to predict on-orbit performance well. Ground characterization of large optics can be challenging, especially in x-rays, due to the finite source distance in available calibration facilities, which, for example, can result in an underestimate of the actual on-orbit efficiency of the optics. However, by utilizing well-thought out setups and advanced ray-trace simulations these issues can be successfully addressed and actual optics performance can be predicted precisely. In the case at hand, a Gl{\"u}cksrad setup allowed for the illumination of small segments of the optic with quasi-parallel light. The results for all of the measured optical subgroups or segments are then added up to result in the total performance of the telescopes.

For the BabyIAXO case the two most relevant optical quantities are the spatial resolution (point spread function/PSF) that determines the spot area and the effective area (throughput) which appears in the optics FOM in terms of efficiency. For optimal results, the efficiency needs to be as high as possible while the focal spot area is to be minimized. For all XMM optics, the PSF varies only slightly with energy and the change as function of off-axis position over the BabyIAXO field of view is not a concern since the experiment is imaging the \SI{3}{arcmin} center of the solar core. In comparison to the custom IAXO optics the focal spot area for the XMM optics is expected to be larger (\SIrange{0.3}{0.7}{cm^2} for XMM as compared to \SI{0.2}{cm^2} for custom optic) taking into account the energy dependence of the PSF for XMM. This is due to the fact that the angular spot size for XMM and custom IAXO optics is similar, but XMM's focal length is larger at \SI{7.5}{m} than the baseline optics (\SI{5}{m}). The effective area of the XMM telescopes has been well studied both in terms of vignetting (off-axis behavior) and as a function of energy on-axis. For BabyIAXO the total effective area is about \SI{70}{\%} of a custom IAXO optic with \SI{60}{cm} diameter. 

In terms of costs the use of the XMM telescope would not incur optics construction costs for the collaboration, but installation and maintenance in the experiment are foreseen. ESA will provide a list of requirements to keep the optic contamination-free (e.g. the use of thin mylar windows) and avoid major mechanical modifications that would prevent any future use of the telescope by ESA after the completion of the BabyIAXO experiment. None of the discussed requirements poses a risk to the operation in BabyIAXO or to the science return of the experiment. An immediate re-calibration of FM1 and/or FM5 is not considered necessary prior to implementation in BabyIAXO, and end-to-end calibration including the final BabyIAXO detector might be considered. FM1 and FM5 have been frequently characterized and used over the past two decades with virtually no observable change in performance, such that a re-calibration of the telescopes alone (i.e. without the final detectors) is considered optional. 
The use of an XMM telescope in BabyIAXO allows to maximize the experiment's science return. At the same time, it also permits the collaboration to focus on the optimization of a single BabyIAXO custom optic, i.e. all efforts can be concentrated on remaining risk reduction and performance optimization instead of the production of multiple optics in parallel.

\subsubsection{Alignment and Implementation of optics in BabyIAXO}
In order to implement and align both the custom BabyIAXO and the XMM optics in the BabyIAXO setup, the telescope weight and precision requirements have been crucial input parameters for the design of the magnet support structure. The connection of magnet-to-optics will consist of large, pneumatic gate valves and a bellow between the gate valve and each telescope will enable the proper alignment procedure. The optics themselves will be located inside separate vacuum vessels and positioning will take place using hexapod structures outside these vessels. This approach provides the advantage that no mechanical vacuum feed-through controls will be needed. The optical setup will be mounted on the telescope platform that will be attached to the magnet support structure but independent of the longer-lever-arm detector platforms, making it feasible to fulfill precision positioning requirements. The optics can then be connected to the focal plane detectors by means of vacuum tubes.

The alignment procedure for the BabyIAXO custom telescope has been tested and verified with the IAXO pathfinder system at CAST. A theodolite will be used to determine the center axis of the magnet bores (which points at the center of the solar disk) by means of two cross-hair fiducials located inside the magnet bore. The custom optic will feature alignment fiducials inside the central mandrel marking the telescope's optical axis and allowing for a positioning of the telescope within \SI{0.01}{\degree} or better with respect to the pointing axis through the magnet. A laser that has been co-aligned with the theodolite can be used to verify the focal spot position and for determining the best location for the focal plane detector. Alignment of the XMM telescope will be done in a similar way making use of the existing fiducials used for XMM alignment at PANTER and inside the spacecraft along with well-established alignment protocols. It is important to keep in mind that tilting the telescopes after alignment due to e.g. mechanical bend of the support structure and telescope platform has to be minimized, strictly controlled and monitored, since a tilt between the telescope and the magnet will not only result in loss of telescope efficiency but also move the position of the focal spot in the detector plane, which might lead to partial signal loss if the signal spot becomes obscured by the detector strong-back around the central opening (\SI{8}{mm} diameter). In-situ tracking of the spot position via the structural movement might be used to partially mitigate this challenge. Modifications of the detector strong-back to allow for a larger central region are also currently under study.

%% file: sections/BabyIAXO_detectors.tex
\subsection{BabyIAXO requirements and state of the art}

\begin{figure}[b!]
\begin{center}
\includegraphics[width=15cm]{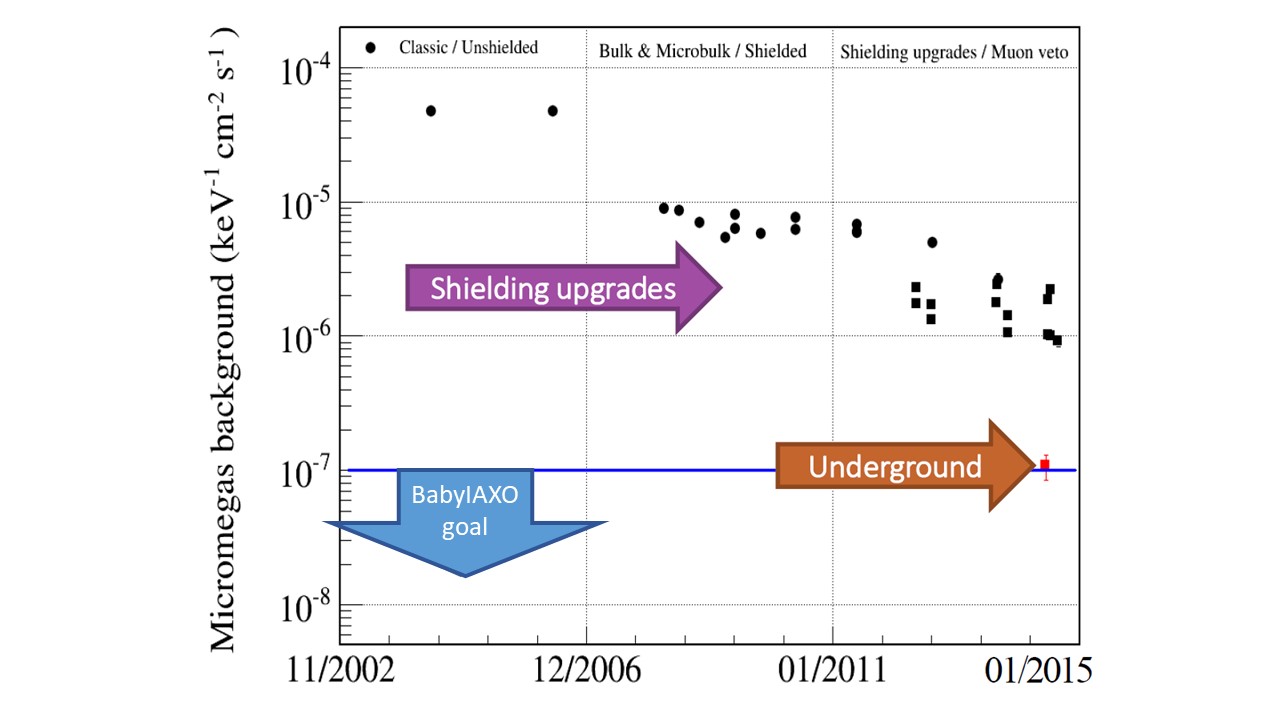}
\caption{Evolution of the Micromegas background levels in different periods of the CAST data taking. The last point on the right represents the background level obtained in LSC under special shielding conditions. The background level targeted for BabyIAXO is at most of \SI{e-7}counts~keV$^{-1}$~cm$^{-2}$~s$^{-1}$. 
}
 \label{fig:background-evolution}
 \end{center}
\end{figure}

Considering that the conversion rate of axions to photons is determined by the magnet parameters and that the fraction of photons arriving to the detector system is determined by the efficiency of the optics, the possibility to reach the sensitivity on very low coupling constants of axions resides in the availability of detectors with high detection efficiency  and very low background in the range below \SI{10}{keV}. It is useful to define a simplified detector figure of merit ($f_{D}$) as:
\begin{equation}
f_{D}=\frac{\epsilon}{\sqrt{b}}
\end{equation}
where $\epsilon$ is the detection efficiency and $b$ is the normalized background of the detector (in area, time and energy bin). This motivates the development of efficient x-ray detectors with ultra-low background levels. To reach the required background levels,  low background techniques like the use of shielding, radiopurity screening of detector components and advanced event discrimination strategies based on the detailed topological information of background events need to be combined.

The baseline detection technology in BabyIAXO are small Time Projection Chambers (TPC) with pixelated Micromegas readouts built with the microbulk technology~\cite{Andriamonje:2010zz} and surrounded by a passive (mostly copper and lead) and active (scintillating plastics) shielding. These detectors have been object of intensive low-background development during the last years, in particular within the CAST experiment, where they have shown an overall detection efficiency in the region of interest (RoI) of around 60\%-70\%. To this number contributes the detector window transmission (see below in section \ref{sec:windows}), the quantum efficiency and the efficiency of the offline analysis cuts. The quest for lower background levels, achieved after progressive understanding of background sources responsible for energy depositions in the RoI, refinement of event analysis methods and improvements of the detector shielding, is represented in Fig. \ref{fig:background-evolution}.

\begin{figure}[t!]
\begin{center}
\includegraphics[width=9cm]{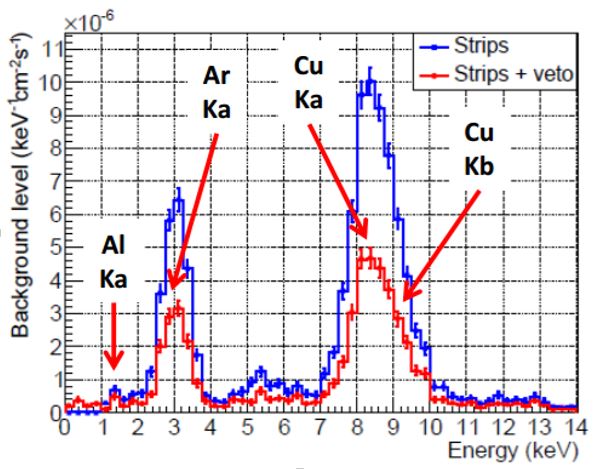}
\caption{Background spectra obtained during the 2013-2015 data taking in the CAST experiment. Copper fluorescence peaks can be clearly seen as well as the escape peak of Ar at \SI{3}{keV}.  The difference between the two spectra is the effect of the muon veto.}
 \label{fig:background}
 \end{center}
\end{figure}

As an outcome of these developments, the current state-of-the-art in terms of background level of microbulk Micromegas detectors is summarized by the following two experimental results:

\begin{enumerate}
\item  The latest microbulk Micromegas detector in operation in CAST, during the last solar axion data taking campaign in 2013--2015, achieved a background level of $10^{-6}$\,\ckcs\ in the RoI of BabyIAXO. This detector was part of the ``IAXO pathfinder system'', coupled to an x-ray telescope. It was conceived as a technological pathfinder combining the optics and detector operating in real data taking conditions~\cite{Aznar:2015iia}. The spectral distribution of the background is shown Fig.~\ref{fig:background}, and represents the lowest level achieved in CAST. The effect of muon vetoing in this detector allows to reject $\sim$50\% of the background in the RoI.  
 
\item A replica of the above mentioned detector was installed in a test platform in the Canfranc Underground Laboratory (LSC) where the muon flux is suppressed by a factor \num{e-4} with respect to surface levels. The background level achieved was $\sim$\SI{e-7}counts~keV$^{-1}$~cm$^{-2}$~s$^{-1}$~\cite{Aune:2013pna}. This value was not improved by a thicker passive shielding, suggesting that: 1) this level is representative of the intrinsic limitations of the current design and 2) the CAST result is dominated by cosmics-related events.

\end{enumerate}

In addition to these two main experimental results, and complementing them, a series of simulation-based studies have allowed to obtain some preliminary understanding of the background sources. We can summarize our current understanding in the following semi-quantitative statements:

\begin{itemize}
\item The main background component in the CAST Micromegas detector are muon-induced events, related to x-rays fluorescence produced in materials very close to the detector. This is supported by: 1) the presence of the \SI{8}{keV} Cu fluorescence peak and the \SI{3}{keV} Ar escape peak in the spectrum (see Fig.~\ref{fig:background}); 2) the factor $\sim$\num{2} reduction (especially in the mentioned peaks) by the veto anticoincidence; and 3) the additional reduction factor when going underground. The two last points are reconciled if one takes into account the relatively poor coverage (about $\sim$\SI{50}{\%}) of the muon veto in CAST, consequence of the space constraints of the experiment. 
\item
The experimental result obtained underground is compatible with  simulations indicating that the intrinsic contamination of the detector and surrounding materials should account for a background level of about $\sim$\SI{5e-8}counts~keV$^{-1}$~cm$^{-2}$~s$^{-1}$~\cite{CristinaMaster}. There is also some simulation-based evidence of external gamma-induced secondaries entering via the small non-shielded solid angle at the pipe towards the magnet, which could account for the difference between this number and the experimentally observed underground, although quantitative conclusions are uncertain. Apart from this, the passive shielding is sufficient to stop environmental gammas down to negligible levels, in agreement with the tests underground mentioned above.

\item The same simulations indicate that most of the intrinsic radioactive background comes from $^{39}$Ar in the gas of the chamber. A Xe-based detector should enjoy a reduced intrinsic background of $\sim$\SI{e-8}counts~keV$^{-1}$~cm$^{-2}$~s$^{-1}$, for the same radioactive environment. 

\item The effect of other external components, like neutrons or high-energy gammas at surface seem to be negligible at the level of the current CAST background level, but further work is needed to establish whether they are relevant at lower levels \SI{<e-7}counts~keV$^{-1}$~cm$^{-2}$~s$^{-1}$.

\item The presence of fluorescence peaks in the spectrum is suggestive of the ability of current offline~\cite{Irastorza:2015geo,Aznar:2015iia} event analysis to effectively filter out a much larger non-X-ray event population. It must be stressed that the raw trigger of these detectors in CAST is about $\sim$\SI{1}{Hz}, while the final background rate goes down to $\sim$\SI{1}{c\per\hour} rate (\SI{0.003}{c\per\hour} inside the telescope spot area). 

\end{itemize}

The above considerations lead us to consider for BabyIAXO a detector design largely based on the last CAST detectors, but with a number of improvements. A substantially improved muon veto system should allow to bring the detector background to a level of $\sim$\SI{e-7}counts~keV$^{-1}$~cm$^{-2}$~s$^{-1}$, and we consider this a realistic target for the BabyIAXO detectors. Additional improvements beyond this level are possible, following improvements in shielding and veto extensions towards the pipe to the magnet, moving to a Xe-based operation and new electronics. The final effect of these improvements in the background level remains to be quantified, but could potentially lead to the $\sim$\SI{e-8}counts~keV$^{-1}$~cm$^{-2}$~s$^{-1}$ levels. Details of the roadmap to produce detectors for BabyIAXO are given below. Whatever the performance achieved for BabyIAXO detectors, the studies  towards lower background levels will continue along the lines described below, to improve the detector figure of merit for the final IAXO.

\subsection{Micromegas detectors for BabyIAXO}
\begin{figure}[t]
\begin{center}
\includegraphics[height=5cm]{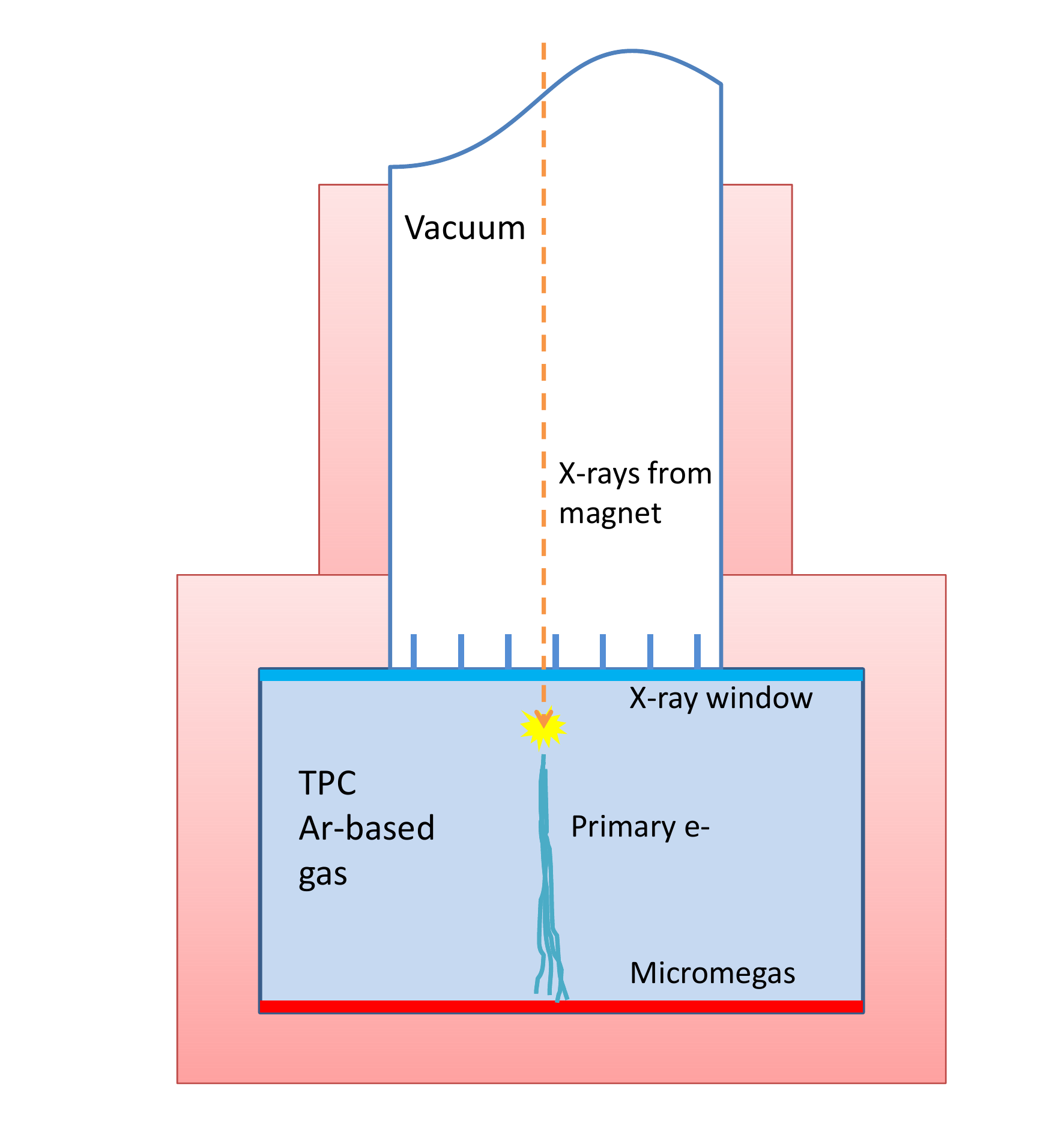}
\includegraphics[trim=0.5cm 0cm 0.8cm 0cm, clip=true,height=5cm]{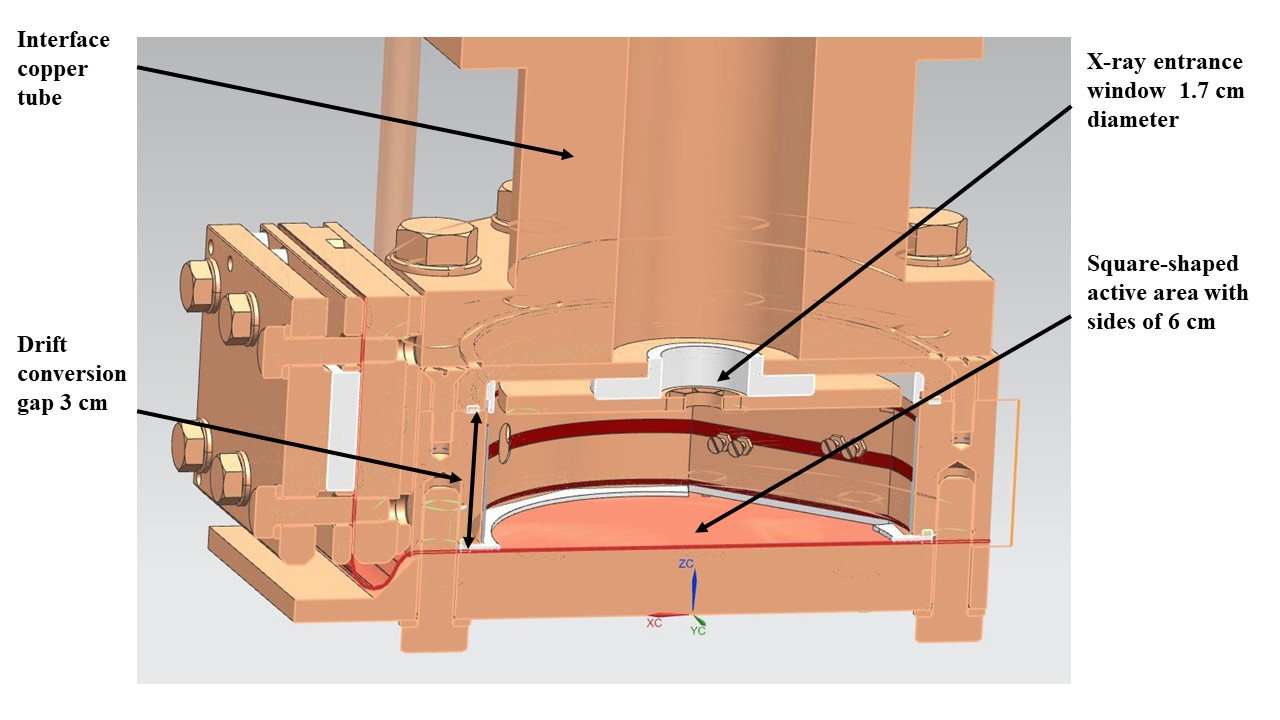}
\caption{\label{fig:det_concept} Left: Scheme of the detection principle of Micromegas detectors. Right: Design of the BabyIAXO detector prototype.}
\end{center}
\end{figure}

The inner design of the Micromegas detectors proposed for BabyIAXO remains similar to the last generation of CAST detectors.  The basic design is illustrated in Fig.~\ref{fig:det_concept}. The detector is a small TPC with a Micromegas readout at the anode, and whose cathode faces the magnet bore from where signal X-rays enter the detector. The conversion volume of the chamber is fixed to efficiently stop the signal photons, while minimizing background, and typically has 3 cm thickness and is filled with \SI{1.4}{\bar} Argon in addition to a small quantity of quencher (e.g. \SI{2}{\%} isobutane). An alternative gas filling of \SI{500}{\milli\bar} of Xe (plus isobutane) will be considered in the proposed tests. The X-rays coming from the magnet enter the conversion volume via a gas-tight window made of \SI{4}{\micro\m} aluminized mylar foil. This foil is also the cathode of the TPC, and it is supported by a metallic strong-back, in order to withstand the pressure difference with respect to the magnet’s vacuum system. 

Special care has been taken in the manufacturing of the detector to be as radiopure as possible. The body and the chamber of the detector are made of \SI{18}{mm} thick radiopure copper (Cu-ETP) walls. All the gaskets consists of radiopure PTFE. A kapton field shaper has also been installed, to increase the uniformity of the drift field and reduce border effects. The field shaper is externally covered by a \SI{3}{mm} thick PTFE coating in order to block the copper fluorescence from the body of the detector. The microbulk-type readout plane consist of a 2D strip pattern of 120 strips per axis at a pitch of \SI{500}{\micro\m} covering a surface of $6\times6$\,cm$^2$. 

\begin{figure}[b!] \centering

\includegraphics[width=10.5cm]{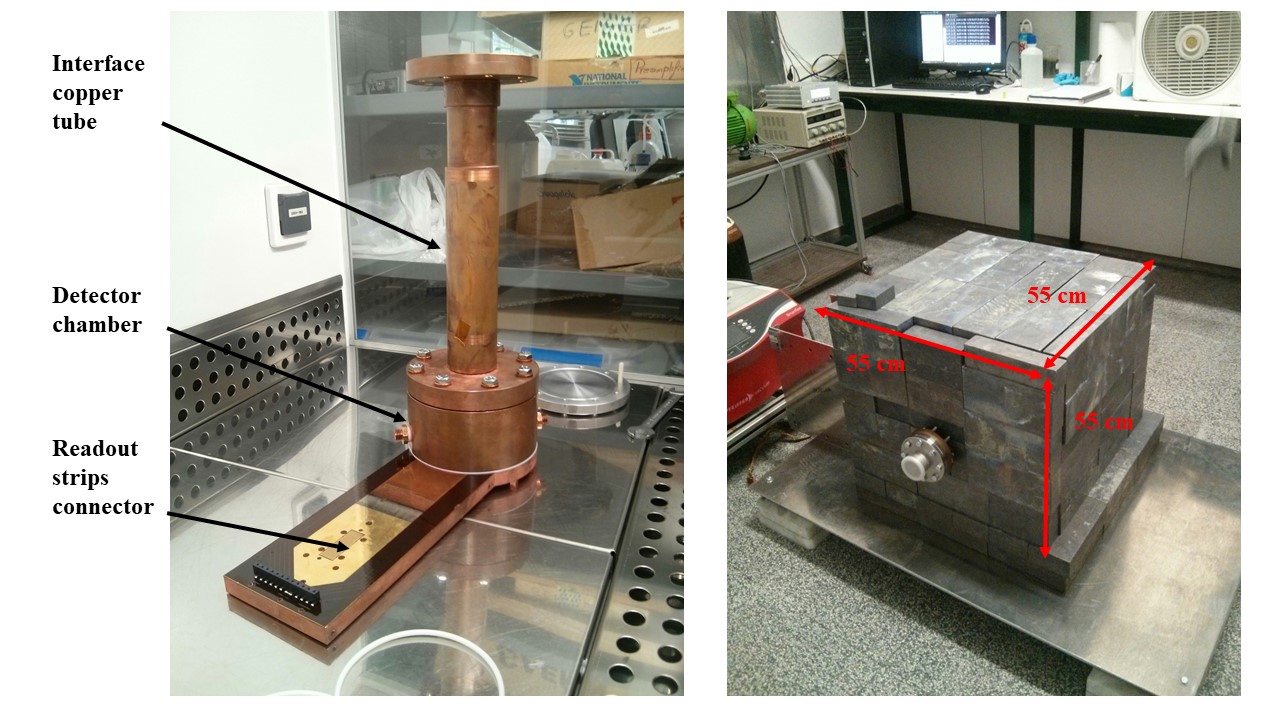}

\vspace{0.1cm}
\includegraphics[width=10.2cm,trim=10cm 0 30cm 0,clip]{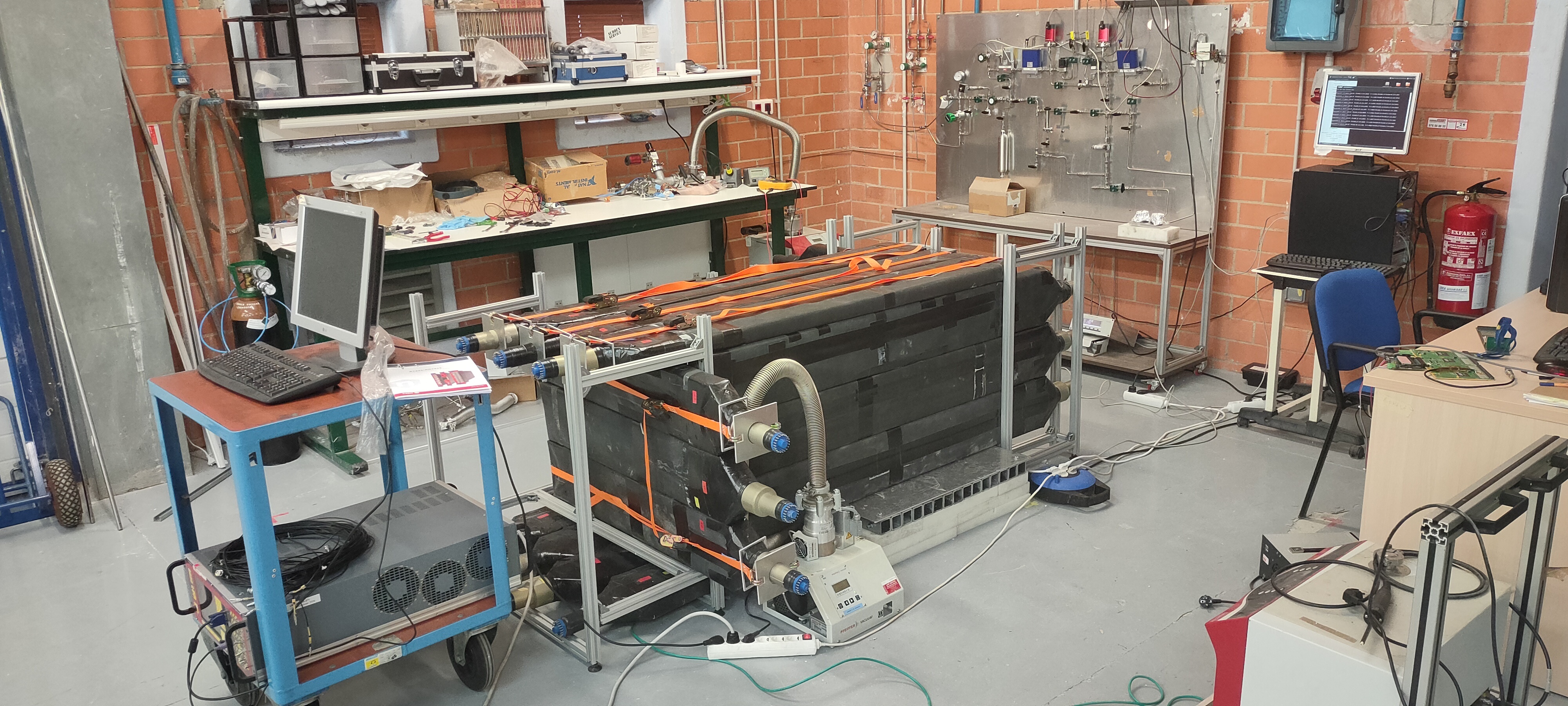}
\caption{\label{fig:D0Zaragoza} Top-left: Photograph of the IAXO-D0 detector chamber assembled on the readout support, both made of copper. The detector is coupled to a copper interface tube for coupling to the optics system. The high density rectangular connector for the conection to the front electronics can also be seen in the middle of the golded plated zone in the readout support. Top-right: Photograph of the detector surrounded by the lead shielding. Bottom: IAXO-D0 setup surrounded by the $\sim4\pi$ plastic scintillator system to veto cosmic muons.}
\end{figure}

One such detector has already been installed in a dedicated test setup, dubbed IAXO-D0, at the University of Zaragoza. The detector has been shielded with a \SI{20}{cm} thick lead castle. The system has been equipped with new AGET-based DAQ electronics~\cite{Baron:2017kld} (see below), and a new gas system with closed recirculation (in view of the use of Xe-based gas mixture). Still without a veto system, the background achieved is compatible with the CAST result mentioned above~\cite{ElisaphD}. Fig.~\ref{fig:D0Zaragoza} shows a photo of the detector chamber and  the detector surrounded by  the  lead shielding.

The IAXO-D0 setup will serve as the starting point to implement the roadmap of improvements considered above, and to consolidate a detector configuration for BabyIAXO. In the following paragraphs we describe in some detail the main areas of work contemplated. 

\subsubsection{Active and passive shielding}

The implementation of a highly-efficient, almost 4$\pi$ coverage, muon veto system is the most important background improvement with respect the status described above. Such a system is being implemented in IAXO-D0 and its effect on the background will soon be tested. A \SI{99}{\%} muon tagging efficiency is targeted.  Different technological options are under consideration. A first version of a customized-geometry plastic-scintillator muon veto system for IAXO-D0 is being commissioned at the University of Zaragoza (see picture at the bottom of Figure~\ref{fig:D0Zaragoza}). A more optimized design will be developed for the final installation in BabyIAXO after the results at IAXO-D0. An alternative approach based on a combination of highly efficient and space optimized plastic scintillators using silicon PMs and large Micromegas detectors for an active muon tracking is being developed at Johannes Gutenberg University (Mainz). This system can be used to characterize detectors, not only Micromegas but also alternative detector technologies.  It will be very useful to validate simulation models to background data. 

The current passive shielding design including a \SI{20}{cm}-thickness lead wall  should in principle be sufficient to reduce external radiation down to levels well below the BabyIAXO target background. However, some improvement could be needed for subsequent more ambitious steps. They include modifications in the way the pipe towards the magnet is shielded/vetoed to prevent secondaries (from e.g. external gamma radiation) entering the setup. 

An additional active shielding component designed to tag high-energy neutrons is currently under study, and will depend on the outcome of the simulations work and tests with IAXO-D0 to assess the effect of cosmic neutrons on the background of the detector.

\subsubsection{Gas system}

The gas system built for IAXO-D0 includes a closed-loop gas recirculation system with adequate pumping and filtering components, suitable to work with Xe-based gas mixtures. Operation with Xe could potentially lead to background improvements by preventing the presence of the argon escape peak at \SI{3}{keV} in the middle of the energy RoI, as well as the contribution of the beta decay of the $^{39}$Ar isotope (dominant part of the intrinsic radioactive background according to current simulation work). Moreover, Xe offers higher stopping power to x-rays than Ar, and therefore for similar detection efficiency lower gas pressure is needed. In particular we target operation with \SI{500}{mbar} of Xe (to be compared with the current \SI{1.4}{bar} of Ar). This lower pressure reduces the mechanical stress of the x-ray window, potentially allowing for thinner windows and better transmission at low energies. 

\subsubsection{Ultra thin windows}
\label{sec:windows}
The CAST Micromegas and the first GridPix detectors were equipped with mylar foils of 2--4\,$\mu$m thickness as x-ray windows.  Their purpose is to maintain the pressure difference between the detector gas and the vacuum while allowing the low energy x-rays (0--10\,keV) to enter the detector.  For this, the foil is supported by a metallic strongback with a spider-web design which leaves a central 8\,mm hole that conveniently leaves room for the focused spot. Although this approach has been successfully applied at CAST, the detector threshold is mostly determined by the transmission of the window, which quickly drops for energies below 1.5\,keV.   Moreover, the leak tightness of the windows is not very good contaminating the vacuum with the detector gas.

\begin{figure}
	\centering
		\includegraphics[width=\textwidth]{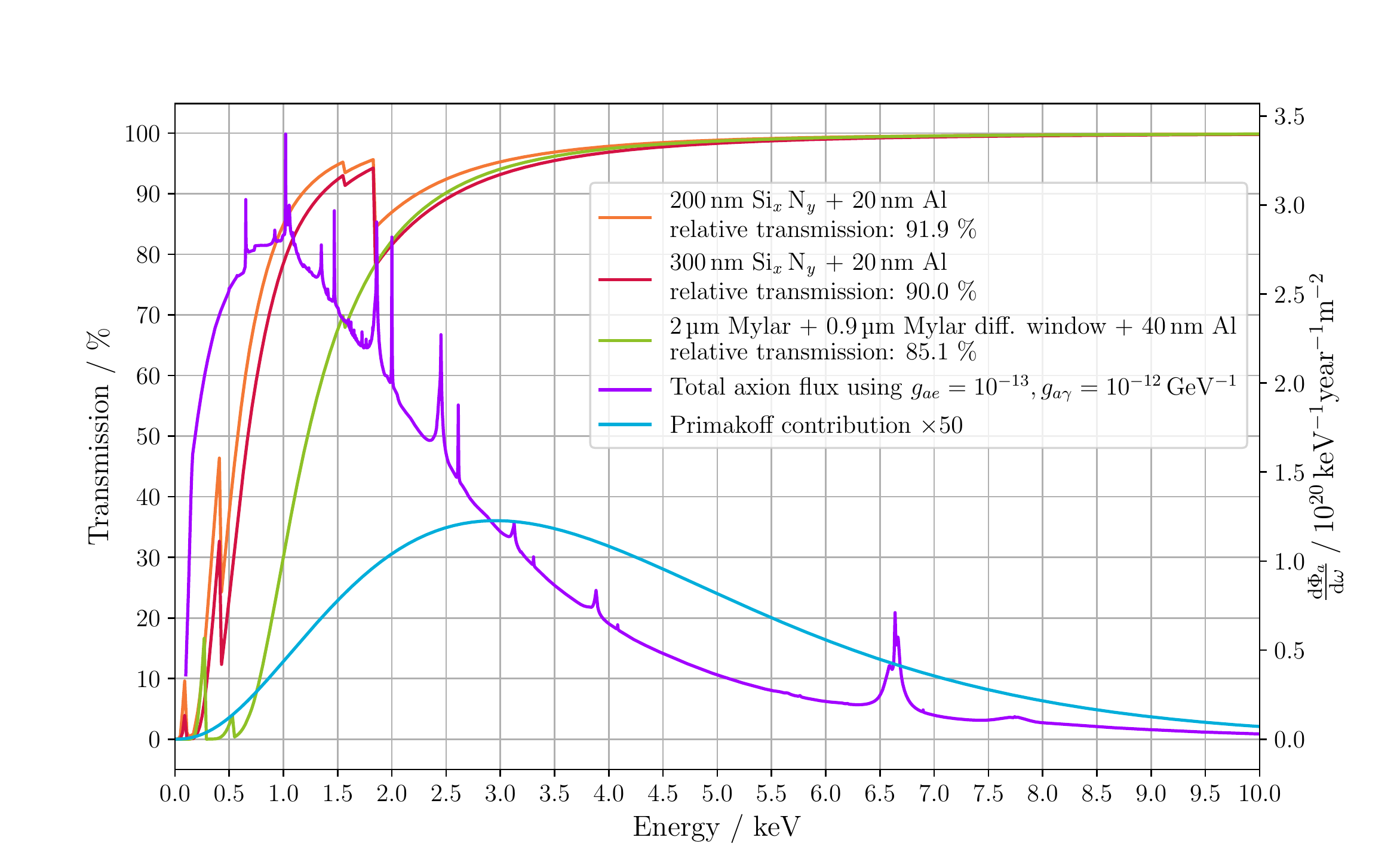}
		\includegraphics[width=0.6 \textwidth]{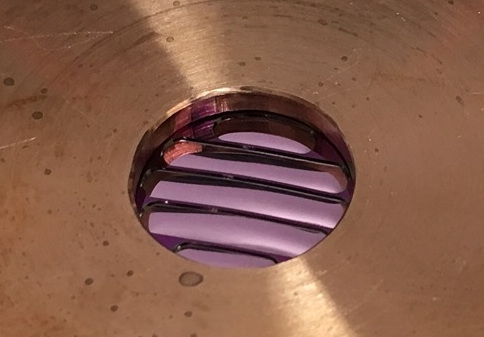}
	\caption{Top: Transmission of silicon nitride windows in comparison to mylar. The expected axion spectrum is also shown. Bottom: A 300\,nm silicon nitride window under 1500\,mbar pressure.}
	\label{fig:transmission}
\end{figure}

To circumvent these problems, the University of Bonn is developing ultra thin windows made of silicon nitride. These windows consist of a silicon strongback covered with a silicon nitride membrane only a few hundred nm thick, leading to a much improved transmission in the range below 1.5\,keV (Fig. \ref{fig:transmission} left). This development led so far to windows with an open diameter of 14\,mm and a membrane thickness of 300\,nm (Fig. \ref{fig:transmission} right), which have been used for the second GridPix detector in CAST, proving the operability. These windows survived a pressure difference of 1500\,mbar and have a leak tightness better than $8\times 10^{-8}$\,mbar\,l$/$s. Further improvements will be made towards thinner membranes, better strongback designs and potentially larger areas.

\subsubsection{Mechanical design of the detector system and assembly proposal of the detector system}

Mechanical design of the coupling of the detector to the x-ray optics is 
studied taking into account alignment considerations. The 
installation procedure of the detector with the shielding and the 
services (gas and electronics) is also considered.

\begin{figure}[b!]
\begin{center}
\includegraphics[width=0.8\textwidth]{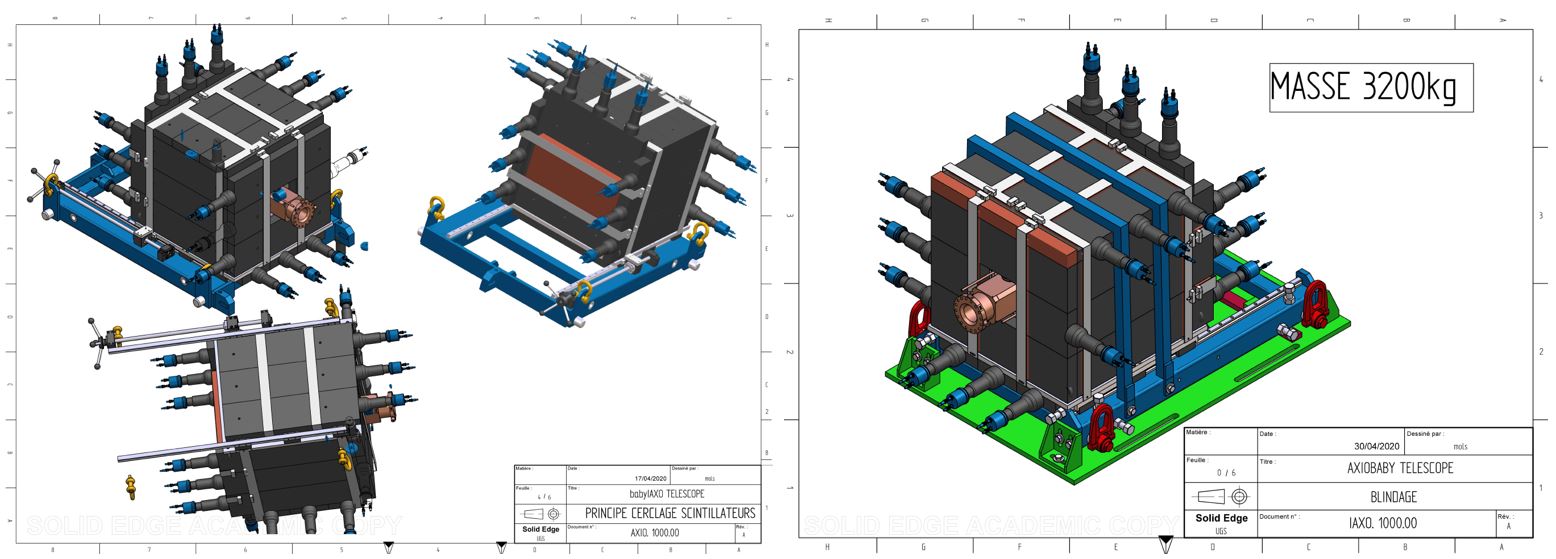}
\caption{Left: Different views of the detector with the passive and active 
shield showing the 4$\pi$\,sr coverage. Right: Perspective of the 
whole assembly with the indicated weight.}
\label{fig:mechanical-design1}
\end{center}
\end{figure}

The passive shielding will be a cast lead block of \SI{20}{cm}-thick walls 
specially manufactured for this purpose, and a relatively hermetic 
plastic envelope around this passive shield with a continuous N$_2$(g) flow 
injected will be included. The active shield will be composed of plastic 
scintillators covering almost completely the detector following a 4$\pi$ coverage strategy (with the only exception of the x-ray 
beamline from where the axion signal arrives) as shown in figure~\ref{fig:mechanical-design1}.

\begin{figure}[b!]
\begin{center}
\includegraphics[width=\textwidth]{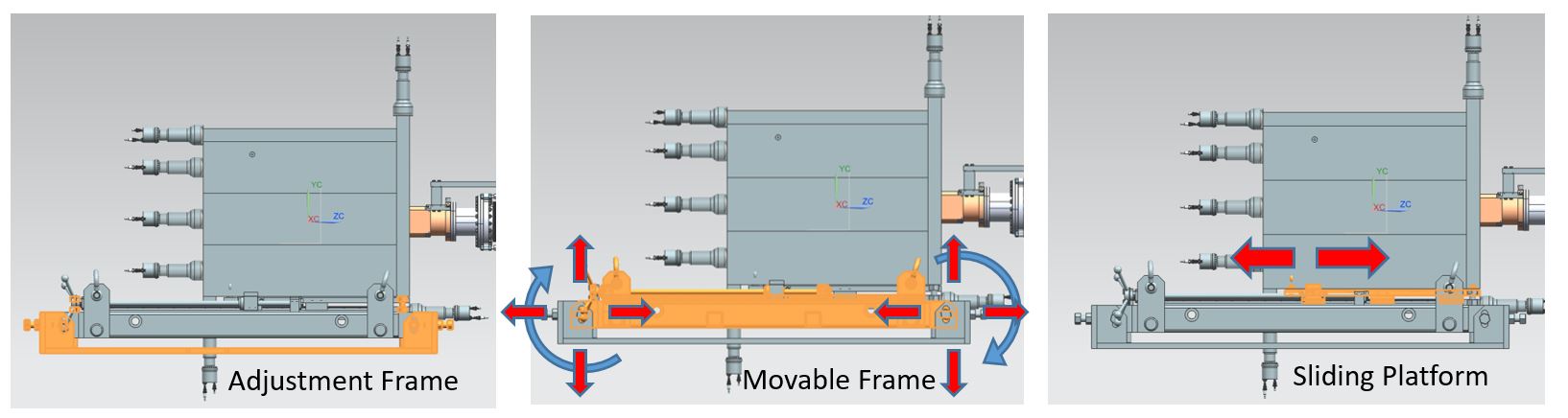}
\caption{The components of the frame assembly with their movement degrees of freedom.}
\label{fig:platform-movements}
\end{center}
\end{figure}

\begin{figure}[b!]
\begin{center}
\includegraphics[width=15cm]{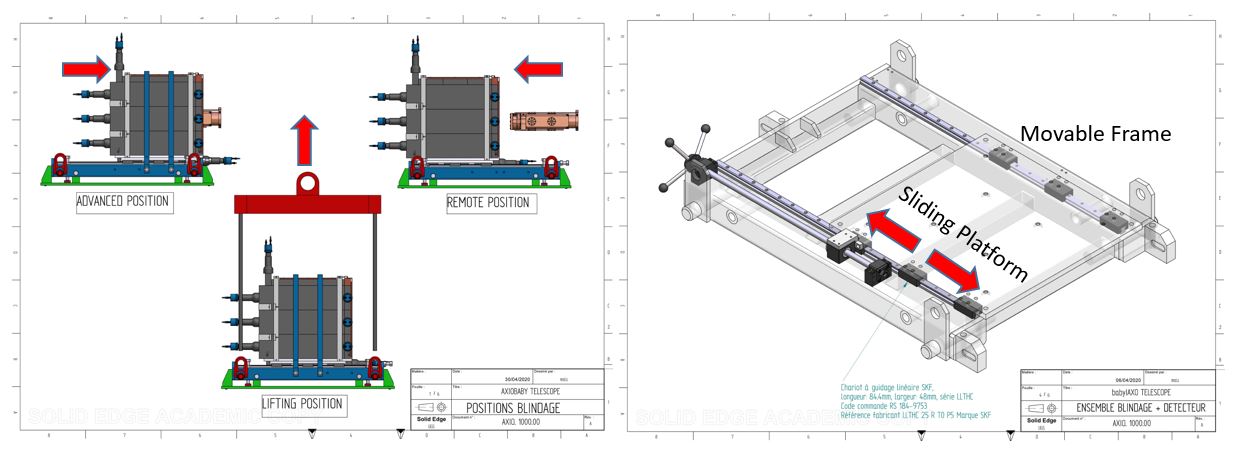}
\caption{Left: Different movements of the Micromegas Detector shield assembly. Right: Perspective with a detailed view of the sliding platform and the movable frame components}
\label{fig:Lifting-platform}
\end{center}
\end{figure}

The BabyIAXO Micromegas detector assembly uses a combination of frames, as shown in 
Fig.~\ref{fig:platform-movements}, in order to achieve the following objectives. First, for flexible positioning to meet the alignment requirements: ($\pm$\SI{25}{mm} in $x$, $y$, $z$) + \SI{200}{mm} in $x$-axis of the detector/shielding once it is 
placed on the BabyIAXO platform to cope with deviations from the nominal position. The shielding structure must absorb these tolerances generated in different steps of the whole BabyIAXO experiment assembly process: possible examples are the tolerances at the BabyIAXO platform caused by 
the welding process of the beam profiles or deviations on the expected telescope focal point location that could appear. Secondly, to get 
direct access to the detector once it is installed onto the BabyIAXO platform (for commissioning/maintenance operations) by means of a  sliding horizontal movement ---shown in figure~\ref{fig:platform-movements}--- (from 
data-taking position to parking/remote position). Finally, to enable direct transportation of the whole assembly onto/from the BabyIAXO platform during the installation/dismantle by means of lifting lugs and/or a spreader beam (Fig.~\ref{fig:Lifting-platform} left).

The detector chamber positioning can be assured by means of a fixed  rigid assembly to the movable frame of the Micromegas detector shield assembly  or with an actuator. Once the detector is correctly located, the bolted unions lock the position of the whole assembly in order to prevent any free movement, and, by means of a locking break, also the sliding movement of the detector shield is eliminated and consequently, it remains at the 
data-taking position. The location and position of all the components (sliding platform, movable frame and detector) can be determined with the help of geometers, using three cross-wire markers protruding by means of metallic extensions out in a position where it can be accordingly checked. Consequently, the detector position can be calculated with only the uncertainty of the manufacturing tolerances of the mentioned extensions of the cross-wire markers. This configuration can be used routinely for 
the position cross-checking of the chamber.

\begin{figure}[b!]
\begin{center}
\includegraphics[width=15cm]{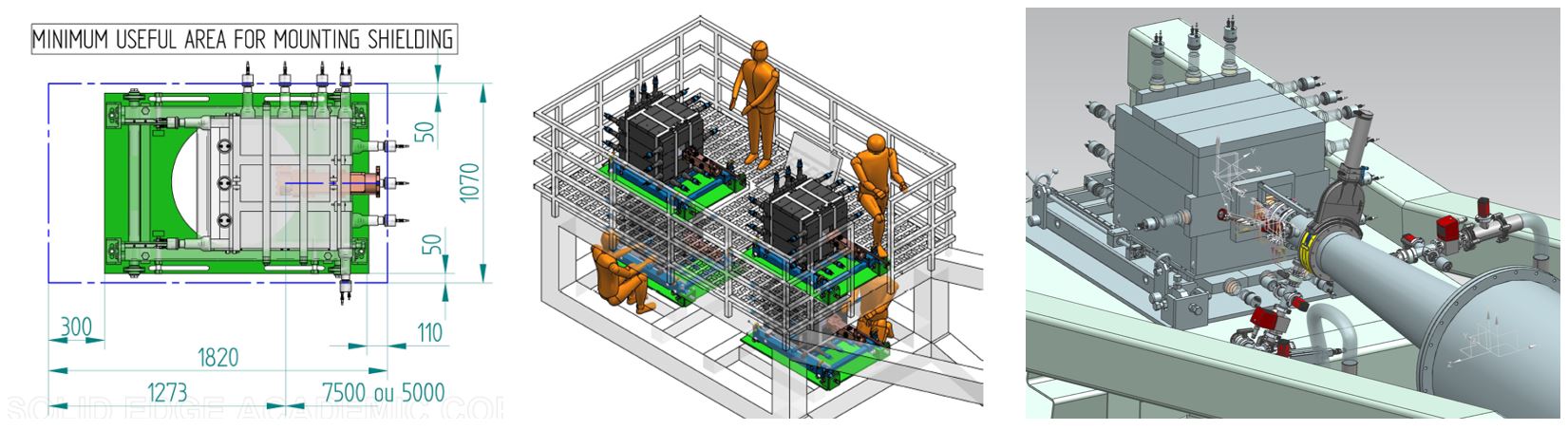}
\caption{Left: Dimensions of the Micromegas detector assembly on the 
BabyIAXO platform. Middle: Preliminary detector layout onto the BabyIXO platform with two different focal lengths: \SI{7.5}{m} and \SI{5}{m}. Right:
Tentative representation with the detector and the beamline components.}
\label{fig:Detector-on-platform}
\end{center}
\end{figure}

The weight of the system, including the detector, the active and passive shieldings and the adjustable set of frames, is estimated at 3.2\,t (figure~\ref{fig:mechanical-design1} right) with a volume of about \SI{1.8}{m}$\times$ \SI{1.3}{m} $\times$ \SI{1.3}{m} (including the DAQ electronics volume and gas routing). figure~\ref{fig:Detector-on-platform} shows a preliminary distribution onto the platform for different focal lengths.

Between the optics and the detector, a calibration system will be 
installed. A bellow will be added in order to ease the alignment process and a pneumatic gate valve for safety reasons, as shown in figure~\ref{fig:Detector-on-platform} right. At the BabyIAXO platform at DESY, the different services (gas, high voltage, electronics signals) will be routed along the platform to reach electronics racks, pumping systems attached to the platform for vacuum generation in Beam Line in Vacuum Phase and gas lines to reach the Gas cabinet of the experimental hall.

\begin{figure}[b!]
\begin{center}
\includegraphics[width=0.8\textwidth]{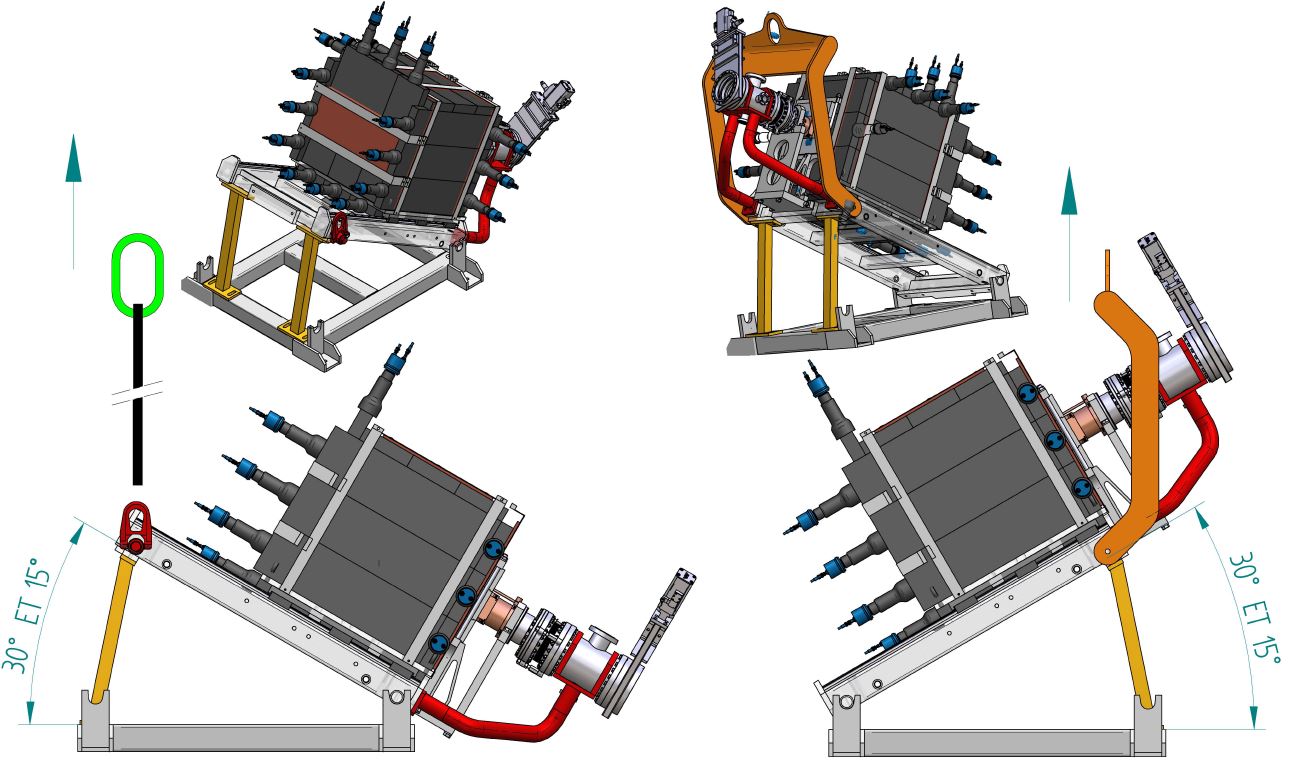}
\caption{The setup to check the detector performance at
different slope angles ($\pm\ang{15}$ degrees and $\pm\ang{30}$ degrees).}
\label{fig:Inclination-platform}
\end{center}
\end{figure}

However, prior to the detector installation at the BabyIAXO platform at DESY, two different setups are projected to check the assembly performance considering the different slope angles that the detector will be positioned following the Sun trajectory ($\pm\ang{25}$).
To validate the mechanical assembly as well as to study the background dependence as a function of the inclination angle, a new inclination platform has been conceived. The different slope angles are achieved by means of hydraulic actuators or with a crane as shown in figure~\ref{fig:Inclination-platform}.

%
\subsubsection{Data acquisition electronics}

\begin{figure}[htb]
\begin{center}
\includegraphics[width=0.9\textwidth]{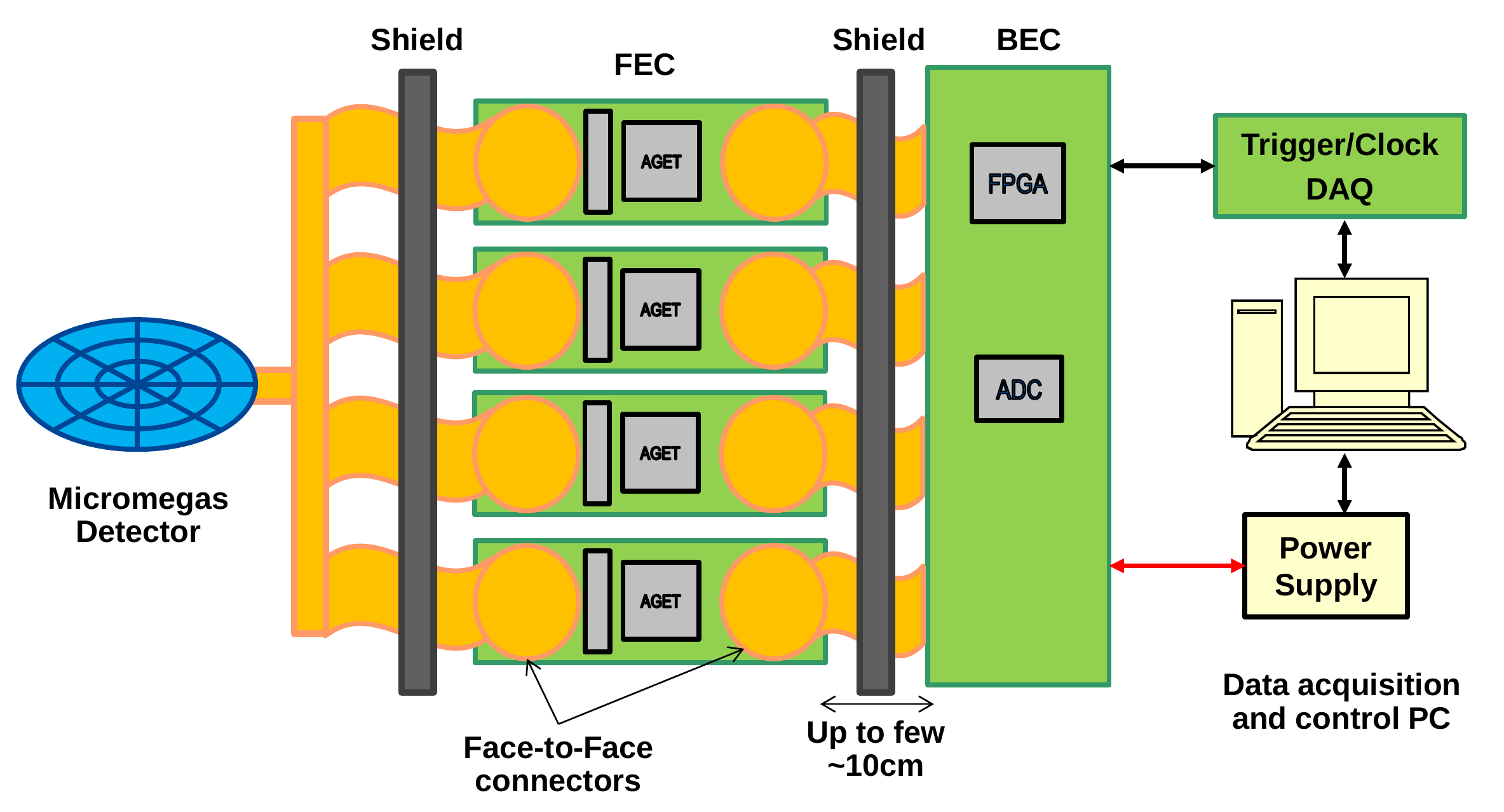} 
\end{center}
\caption{Electronics readout block diagram.}
\label{fig:electronicsReadoutMicromegas}
\end{figure}

The readout electronics for the BabyIAXO Micromegas detectors must allow to extract the maximum information available in the detector, as the ability to discriminate background events offline will rely on it. Modern electronics chips conceived for TPCs allow to digitize full temporal waveforms of a large number of channels. The modest size and readout area of the detector and the low acquisition rate expected does not pose big challenges to the standard electronics parameters (sampling rate, number of channels, shaping, reading time, etc.). However, we ideally require readout noise as low as possible, as this would directly contribute to lower energy threshold of the detector. Reducing the (usually high) radioactivity of the front-end components would allow to place the electronics closer to the detector (in current designs, like that of CAST or in IAXO-D0, it is outside the shielding) further reducing electronics noise. 
The AGET chip~\cite{Baron:2017kld}, currently used in the DAQ electronics for the IAXO-D0 prototype, 
features an equivalent noise charge of about 900 $e^-$ for an typical input capacitance of 30 pF and a peaking time of 230 ns~\cite{Anvar2011}, values representative of this application. This chip provides autotrigger capabilities for each electronic channel. This feature should help lower the energy thresholds (\SI{<1}{keV}) with respect to previous configurations where the detector was triggered with the Micromegas mesh signal with a much higher capacitance than each individual strip. 

The proposed AGET-based electronics for the BabyIAXO detectors are based on a modular and flexible general purpose readout system for small to medium size gaseous detectors designed at IRFU, CEA Saclay. The boards layout is depicted in Fig.~\ref{fig:electronicsReadoutMicromegas}. 
The 240 channels of the Micromegas readout detector active area will be connected to 4 Front-End Cards (FEC) populated with one AGET chip each. The output of the four FEC will be connected to the Back-End Board (BEC) which will include the ADC and an FPGA to control the system. Lastly, the BEC will communicate with a remote DAQ PC via a standard Gigabit Ethernet link and optionally receives global clocking and triggering information from a central Trigger and Clock Module (TCM). 

In order to reduce the effects on noise, the FEC is expected to be placed as close to the detector as possible , at least, at mid distance inside the shielding, between the detector and the BEC. For that, we plan to minimize the radiactive impurities by selecting its components and a Kapton based board. Some of the components needed include capacitors, resistors and diodes for a spark protection circuit and the AGET coupling capacitors the radiopurity of which is being studied. In the case of the AGET chip, the possibility of using naked chips and wire bonding them directly to the board is being investigated. Lastly, the connectors are based on customized solutions in which two Kapton printed circuit boards (PCBs) are simply put in "face-to-face" contact, without solder or intermediate connector pieces, developed and tested in the University of Zaragoza in the context of previous low-background Micromegas applications~\cite{Chen:2016qcd}.


Simulations are being carried out to study the effects of the electronics on the detector. In these, we replicate the design containing the boards and electronic components and study emission events generated via Monte Carlo. From such studies we can estimate the energy deposited on the detector due to FE impurities. These simulations are performed by means of the REST software~\cite{REST_TPCconf},
extensively used for both background models and data analysis in the
CAST experiment by the group of the University of Zaragoza. Several positions of the board and distances from the detector are being checked in order to define the FEC optimal placement. The elements included in the FEC simulation and their activity correspond to known measurements of the components and materials obtained from a screening program of the TREX project~\cite{Castel:2018gcp}. In case their measurements are not available yet, worst cases of known similar components are used.

\subsection{Control of radiopurity and backgrounds}

To go beyond the preliminary understanding of backgrounds exposed before, and advance towards better quantifying background sources, further low-background studies will be needed. A working package on low-background techniques and radiopurity has been set to coordinate these studies in the collaboration. The goals are to measure the radioactive
contamination of virtually any material used in the detector setups as well as to quantify the environmental backgrounds in the experimental site, to guide detector groups in the selection of the radiopure detector components and to complete a detailed background model using this information as input.

Although enough information already exists on the current background sources of the Micromegas detectors to define relevant detectors at the \SI{e-7}counts~keV$^{-1}$~cm$^{-2}$~s$^{-1}$ level for BabyIAXO, this line of work will be relevant to further improve the detectors in view of IAXO. Also it will serve to assess, and if possible improve, the low-background of the additional detection technologies under consideration for IAXO that are described below. 

More specifically the actions here contemplated are:

\begin{itemize}
\item Radioassays of any material near the detector. For the case of the Micromegas detectors, a great deal of information is already available for the radiopurity of the readout planes themselves and related componentes, developed mainly in the context of the TREX project \cite{Cebrian:2010ta,1748-0221-8-11-C11012,Iguaz:2015myh,Irastorza:2015dcb,Castel:2018gcp}. A specific study to reduce the $^{40}$K content of microbulk micromegas is underway, screening samples after different treaments carried out at CERN.
Different complementary techniques could be considered: germanium gamma-ray
spectrometry carried out deep underground in the LSC, offering several ultra-low background HPGe detectors for radiopurity assessment, together with other techniques like glow discharge mass spectrometry (GDMS), inductively coupled plasma mass spectrometry (ICPMS) and 
measurements using the BiPo-3 detector operating in Canfranc~\cite{BiPo_detector}, offering a high sensitivity for thin samples to quantify the activity of the lower part of the $^{238}$U and $^{232}$Th chains. All the obtained results could be stored in a database for internal use within the collaboration.
\item Improvements in radiopurity of components of the detector. An extensive screening program will be essential to identify detector components with the highest activity level, and they could be replaced in detector design updates if more radiopure alternatives are available.
\item Control of procedures for cleaning and storage of materials. Once the most adequate components in terms of radiopurity are selected, the storage and cleaning of the materials must be controlled in order to avoid possible contaminations for example from radon plate-out.
\item Measurements of environmental backgrounds. The characterization of the external background components at the experimental site, including radon levels and gamma and neutrons fluxes will allow to adjust the design of the different elements of the active and passive shieldings to optimize its efficiency in the background reduction.
\item Development of a detailed background model for the detector. In collaboration with the software group, background models will be elaborated taking as input all the activity and background level results obtained from the dedicated measurements.
\item Improvement of the offline discrimination algorithm. A further reduction of background is possible thanks to the implementation of more effective discrimination algorithms.
\end{itemize}
Though all these results come from the analysis of data and simulations of Micromegas detectors, some
of them can be extrapolated to alternative detector technologies: the control of the radiopurity, the understanding
of signal and background event topologies for discrimination and the use of active and passive
shielding will be compulsory and need to be studied.

\subsection{R\&D of additional detector technologies}
  Micromegas detectors are the baseline technology for the x-ray detectors of BabyIAXO. However, technologies like GridPix,  Metallic Magnetic Calorimeters (MMC), Neutron Transmutation Doped sensors (NTD), Transition Edge Sensors (TES) and Silicon Drift Detectors (SDD) are interesting due to their excellent energy resolution, energy threshold, efficiency and the possibility to use ultra-pure materials with respect to radioactive impurities. Their background rejection in the region of interest has never been studied in detail. We foresee to perform R\&D tests of the developed detector systems at the BabyIAXO telescope  to demonstrate that competitive performances will be obtained. The advantages of proposing five suited technologies for IAXO are the following:
 \begin{itemize}
 \item
 They can improve the energy threshold, of great interest for additional ALPs searches (hidden photons, chameleons…), and permit the investigation of fine structures in the axion spectrum originating from processes involving electrons, bremsstrahlung, Compton and axio-recombination and therefore extending the physics case of IAXO.
\item
In addition, in case of a positive signal in the vacuum phase, as axions are not massless, the measured axion spectrum will be different from the expected photon spectrum, especially at low energies. Having detectors with low threshold and good energy resolution, would allow to determine $m_a$ with some sensitivity, which is not the case with the baseline Micromegas detectors.
These complementary technologies would allow to carry out axion precision measurements.
\item
If equivalent performances can be achieved with these alternative technologies, the ideal configuration for IAXO will be a combination of these different technologies; in case a signal is detected, such a configuration will minimize systematic effects and reinforce the claim significance.
\end{itemize}

\begin{figure}[t!]
\begin{center}
\includegraphics[width=14cm]{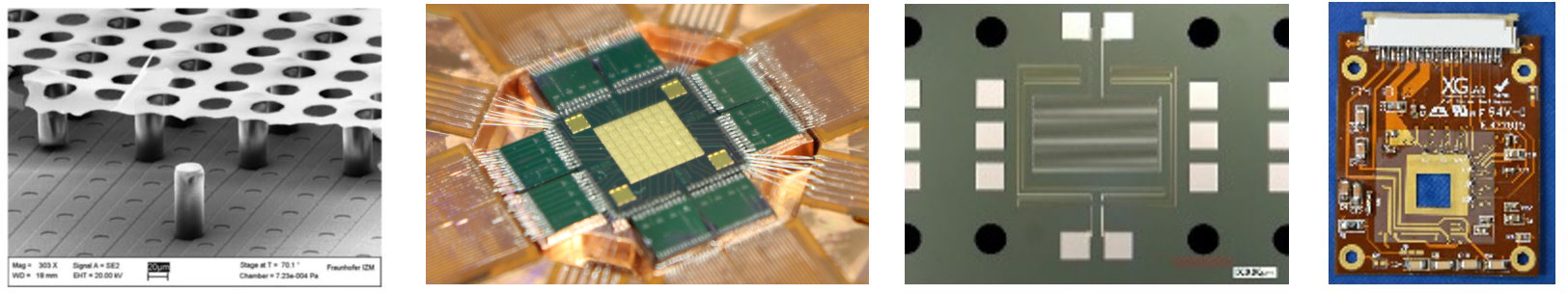}
\caption{Active zone of a Gripix detector (left), MMC array (center-left),  NbSi TES sensor (center-right) and SDD (right). }
 \label{fig:alternativedetectors}
 \end{center}
\end{figure}

\subsubsection{GridPix}
GridPix detectors are an evolution of the Micromegas technology where the Micromegas mesh is produced by photolithographic postprocessing techniques on top of a pixelized readout chip allowing small feature sizes and precise alignment~\cite{Krieger:2018nit}. Each grid hole of the mesh is aligned to one pixel allowing single electron detection. The first GridPix detector, developed and operated by the IAXO Bonn group, has been installed and operated in the CAST experiment in 2014--2015 with an energy threshold of \SI{300}{eV} and achieving background levels of $\sim $\SI{e-5}counts~keV$^{-1}$~cm$^{-2}$~s$^{-1}$~\cite{Krieger:2018nit,Anastassopoulos:2018kcs}.
For the  2017--2018 CAST data taking different improvements were implemented. First, active cooling was introduced in order to avoid performance dependence with temperature. Second, to reduce the muon background contribution, three main improvements were introduced:

\begin{itemize}
\item Two scintillators were installed in order to tag muons entering the detector from the beamaxis and crossing the detector; 
\item The central chip was surrounded by six veto GridPixes to allow rejection of muon tracks hitting only the corner of the central chip mimicking an x-ray event;
\item The induced signal from the central grid was recorded for triggering purposes and for signal shape background discrimination. 
\end{itemize}

Third, a neural network approach is being investigated giving promising results on the background rejection.  All these improvements will result in a lower background level for the last CAST data taking campaign.
In order to further improve the background redution, the radiopurity of the GridxPix detector will be optimised.  In this context, the University of Bonn and the University of Siegen are developing new polyimide PCBs. Care will also be taken in finding radiopure materials for the mounting of the detector on the beamline and for sealing materials for the gas volume. In addition the successor of the TimePix, TimePix3 will be introduced into the next detector generation. With TimePix3 a fully three-dimensional reconstruction of the charge cloud associated with the x-ray conversion can be exploited
for improved background rejection. Furthermore, dead-time free readout can be achieved.
With the combination of all these efforts, background levels similar to the ones obtained with the Micromegas detector should be at reach.

\vspace{5mm}

\subsubsection{MMCs}
MMCs are energy dispersive detectors operated at temperatures below 100\,mK~\cite{Fleischmann}. Their resolving power E/$\Delta$E  above 5000, the intrinsic response time well below 1\,$\mu$s and the excellent linearity make magnetic micro-calorimeters very attractive for numerous experiments~\cite{Pies2012,Gastaldo:2017edk}. For soft x-rays detectors having a quantum efficiency of 99\% at 6\,keV, an energy resolution of 1.6\,eV (FWHM) has already been demonstrated~\cite{Kempf2018}. 
The IAXO Heidelberg group, leader in the development and operation of MMCs, is developing a flexible detector platform allowing for the installation of different systems based on MMCs for BabyIAXO. A first measurement gives a preliminary estimation on the background in the order of 10$^{-5}\,$keV$^{-1}\,$cm$^{-2}$\,s$^{-1}$ between 6 and 10\,keV. This measurement was done above ground and with neither muon veto nor shielding. Further improvements besides the use of active and passive shields are expected from selection of radiopure setup materials (e.\,g. for circuit boards, copper holder parts and superconducting shield). 
If signals which could be interpreted as derived from axion conversion would be identified, a second detector system characterized by very high energy resolution could be installed using the same platform allowing for spectroscopy study. 
Aspects like performance stability of the detector technology operated on a moving platform or implementation of the coupling of the detector system to the optics are under investigation.

\vspace{5mm}

\subsubsection{NTDs}
NTDs~\cite{1994InPhT..35..127H} are among the most robust and widely used thermistors technologies of cryogenic rare event search detectors (operated typically at $\sim$10-20 mK)~\cite{Pirro:2017ecr,Poda:2017jnl}. In particular, performance and radiopurity of NTD Ge instrumented thin Ge bolometers, e.g. used in CUPID-0 \cite{Azzolini:2019tta} and CUPID-Mo \cite{Armengaud:2020a} double-beta decay experiments for the detection of optical photons (needed for the active background rejection), could match the IAXO demands on detector technology. Indeed, a very low noise ($\sim$30-60\,eV RMS) and high energy resolution ($\sim$0.2--0.4\,keV FWHM at 5.9\,keV x-rays of~$^{55}$Fe) have been already reproducibly achieved in aboveground and underground operations of these devices~\cite{Poda:2017jnl,Armengaud:2017,Armengaud:2020a}. Moreover, a very preliminary estimation of the background level show the counting rate of $\sim $10$^{-5}\,$keV$^{-1}\,$cm$^{-2}$\,s$^{-1}$ in the energy range of 2--8\,keV, measured above ground at IJCLab (Orsay, France) using Ge bolometers in a pulse-tube-based cryogenic set-up partially shielded from environmental radioactivity (e.g. \cite{Bekker:2016}). The same detectors operated underground in the Gran Sasso and Modane underground laboratories~\cite{Armengaud:2017,Berge:2018} –--in a low radioactivity environment--– exhibited a background of  $\sim$ 10$^{-6}\,$keV$^{-1}\,$cm$^{-2}$\,s$^{-1}$~\cite{Poda:2017jnl} (no anticoincidence cut with coupled massive bolometers has been applied yet). Considering that no special precaution was taken in the light detector assembly and material selection (as the conducted tests were mainly technical), we deem these results quite promising. Operation at sub-K temperatures is no more an issue as several commercially available solutions exist and have proven to be reliable. 

\vspace{5mm}

\subsubsection{TESs}
TESs are widely used as low temperature detectors (operated typically at ~20-100\,mK) in major x-ray imaging applications~\cite{Ravera,2008JLTP..151...82C}. They offer the possibility of precise single photon counting with zero dark count rate, energy threshold below 10 eV and high resolution (E/$\Delta$E $\sim 1000$). The photon absorber material and dimensions can be adjusted to the requirements of the considered experiment in order to get high intrinsic photon absorption efficiency, close to 100\%. The challenge for IAXO is to reduce the background of the detection to an extremely low level in the region of interest, below 8\,keV. In fact, these devices, even if they adopt a different temperature-sensor technology, are very similar to the NTD-instrumented thin Ge bolometers that we envisage for axion detection in terms of materials, total surface and sensitivity. Detectors based on TES allow a very flexible design and can be fabricated using radioactive-pure materials. They are already operating in several DM research projects requiring very low background like CRESST or CDMS and are being considered by the IJCLab group for the EDELWEISS (dark matter) and LUMINEU and CROSS (neutrino physics) projects.

\vspace{5mm}

\subsubsection{SDDs}
SDDs collect signal in small readout nodes by internal static fields and are used in x-ray spectroscopy. They are currently used in the TRISTAN Project~\cite{1475-7516-2015-02-020}. The interest of this type of detector is that due to their very low capacitance, nearly independent of their active surface, the detector can reach a very good energy resolution (130\,eV FWHM at 6\,keV)~\cite{Mertens_2019} with a low threshold ($< 500$ eV), no extra entrance window and little cooling requirements. Moreover, their design is very flexible allowing active areas ranging from mm$^2$ to cm$^2$ with a flexible number of pixels. An existing detector was placed underground and a first background evaluation of $\sim 6 \times 10^{-4}\,$keV$^{-1}\,$cm$^{-2}$\,s$^{-1}$ was obtained in the region of interest. Monte Carlo simulations using radiopurity measurement of the detector electronics predict an intrinsic background on the order of $\sim 10^{-6}\,$keV$^{-1}\,$cm$^{-2}$\,s$^{-1}$ between 0 and 40 keV. This first measurement will be  improved by using a dedicated test-bench under construction in a shallow underground laboratory.

%% file: sections/BabyIAXO_DM_detectors.tex
\subsubsection{Axion DM detectors in BabyIAXO and IAXO}

Although the focus of BabyIAXO is on the ``axion helioscope'' configuration, it is clear that the BabyIAXO magnet (and eventually the IAXO magnet) constitutes a remarkable infrastructure to implement additional setups to search for axions/ALPs in alternative ways. A most appealing option is to implement DM axion detectors that could exploit the particular features of the (Baby)IAXO magnet. The basic idea of the axion-haloscope-technique proposed in~\cite{Sikivie:1983ip} is to combine a high-$Q$ microwave cavity inside a magnetic field to trigger the conversion of axions of the DM
halo into photons. A straightforward implementation of this concept in one bore of the (Baby)IAXO magnet gives competitive sensitivity~\cite{redondo_patras_2014} even using relatively conservative values for detection parameters, thanks to the large $B^2 V$ available, in the approximate mass range of 0.6--2~\si{\micro\eV}. Another recent concept proposes the use of a carefully placed pick-up coil to sense the small oscillating magnetic field that could be produced by the DM axion interacting in a large magnetic volume~\cite{Sikivie:2013laa}. This technique is best suited for even lower $m_a$ and is better implemented in toroidal magnets~\cite{Kahn:2016aff,Silva-Feaver:2016qhh}. The large size and toroidal geometry of IAXO suggest that it could host a very competitive version of this detection technique.

However, as argued e.g. in \cite{Irastorza:2018dyq}, there is a strong motivation to explore the mass range for QCD axion DM well above \SI{e-5}{eV}.
The difficulty in searching in the ``high-mass'' range can be understood from
the fact that the figure of merit of scanning with haloscopes for axion DM goes
with the square of the cavity volume times the $Q$ factor of the cavity.
Most existing setups use solenoidal magnets and cylindrical
cavities. The diameter of the cylinder sets the frequency scale of the resonance
and thus the axion mass scale to which the experiment is sensitive. Thus going to
high mass means going to small diameters. In addition, the cavity quality factor $Q$ typically decreases for smaller cavities.

\begin{figure}[ht!]
\centering
\includegraphics[width=0.55 \textwidth]{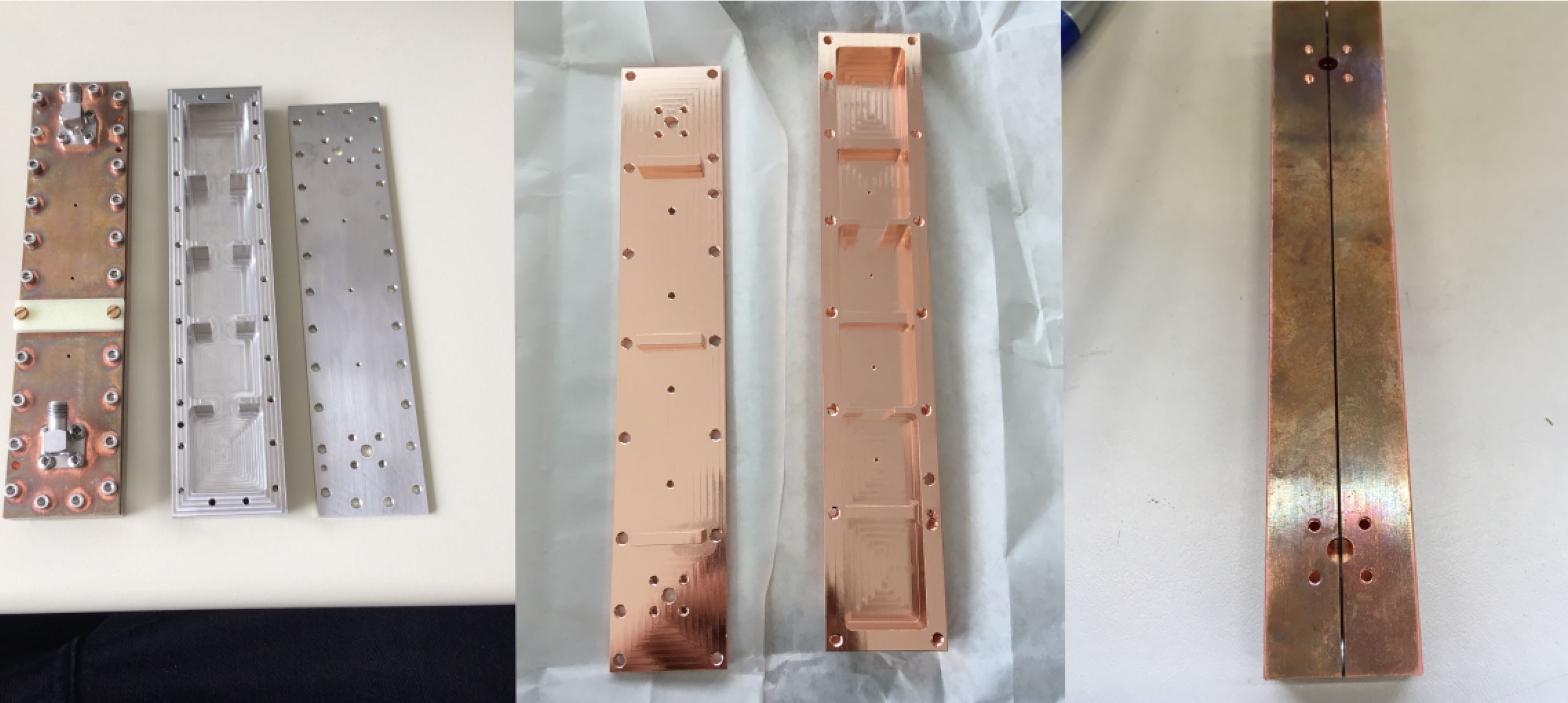}
\includegraphics[width=0.048 \textwidth]{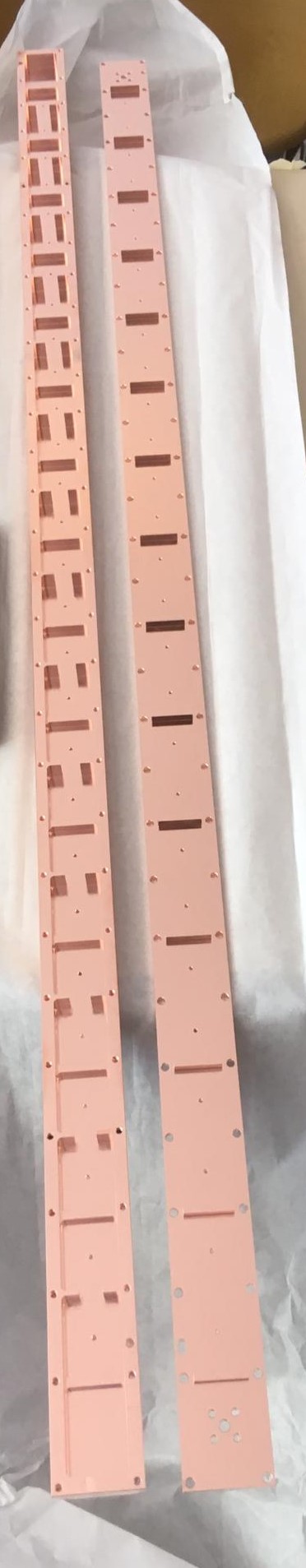}
	\caption{\label{fig:RADESstructures} 
	RADES cavities (stainless steel with copper coating) resonating around 8--9 GHz.
From left to right: first 5 sub-cavities structure with inductive couplings (non-tunable, used in data-taking 2018 in CAST), alternating inductive- and capacitive-irises sub-cavities structure (non-tunable, used for proof-of-principle of behavior of alternating cavity types), 5 sub-cavities vertical cut structure (tunable, used for cryogenic measurements for the proof-of-principle of tuning), alternating 30-cavities structure (non-tunable, used in data-taking 2020 in CAST). The first three structures are each about 15 cm long, the 4th one is of about  1 m length,
see ~\cite{Melcon:2018dba,Melcon:2020xvj,Cuendis:2019qij} for details.
}
\end{figure}

To tackle higher masses, different strategies are being developed in the community~\citep{Irastorza:2018dyq}. For instance, to compensate the loss in $V$ one can go to very strong magnetic fields and/or use superconducting cavities to keep $Q$ large. Another avenue is to decouple $V$ from the resonant frequency and go for large $V$ structures resonating at high frequency. Several strategies in this direction are being explored. One of them, currently being tested
in exploratory setups at the CAST experiment at CERN~\cite{Melcon:2018dba,Fischer:2289074}, employs long rectangular cavities~\cite{Baker:2011na}.  The particular advantage is that the volume can be kept very large (long cavity), while the resonance frequency can be rather high, through the usage of relatively thin cavities. 
The R\&D strategy of the ``RADES group'' is
to segment the long rectangular cavities in smaller sub-cavities trough irises ( see figure~\ref{fig:RADESstructures} for the various prototypes in past and present use). The design challenge here in going to long structures is to well separate the mode coupling to the axions from other modes in such microwave filters. Recent progress on design advances and the construction of a 1-m-long filter structure, as well as first steps towards a tuning system through splitting the cavities in two halves are detailed in ~\cite{Melcon:2020xvj} and \cite{Cuendis:2019qij}, respectively.

Some of these developments are performed by groups belonging to the IAXO collaboration, with the long-term goal to assess the possibility of (Baby)IAXO hosting a multitude of such rectangular cavities. The status of these developments are at too early a stage to define any credible time plan for implementation in BabyIAXO. However, depending on the progress of this R\&D, the collaboration will consider the possibility of haloscope setups as part of a post-baseline data taking plan in the experiment. 

%% file: sections/BabyIAXO_SupportDriveSystem.tex
\subsection{Requirements}

\begin{figure}[b!]
\centering
 \includegraphics[trim={1cm 5mm 7mm 5mm},clip,width=0.49\linewidth]{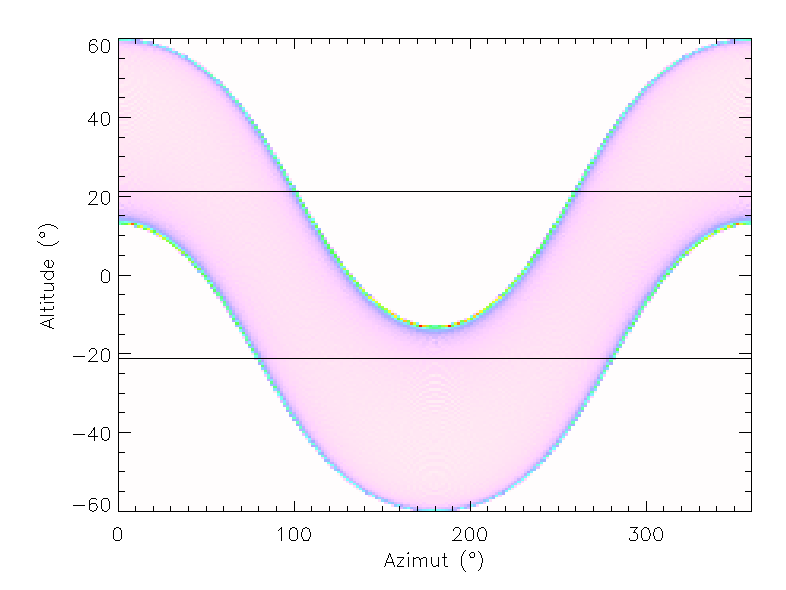}
 \includegraphics[trim={1cm 5mm 7mm 5mm},clip,width=0.49\linewidth]{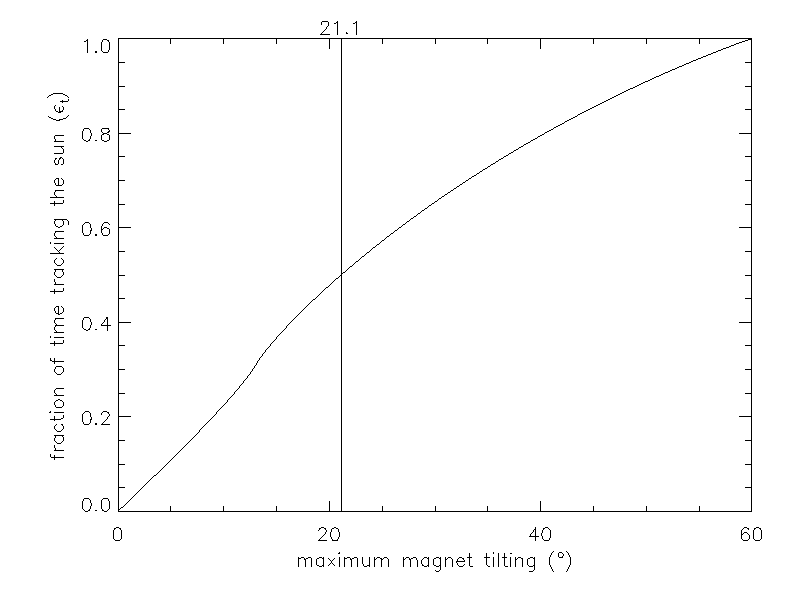}
\caption{Left: region of the azimuth-altitude coordinates of the Sun position along the year for DESY geographical coordinates. The horizontal lines correspond to an elevation of \SI{+-21.1}{\degree}. Right: fraction of the time the Sun is reachable versus maximum elevation. The horizontal line indicates that an elevation of \SI{+-21.1}{\degree} would correspond to a \SI{50}{\%} of average exposure over the year.  }
    \label{fig:sun_alt_azth}
\end{figure}

In order to search for solar axions, the full BabyIAXO assembly (magnet, telescopes and detectors) 
has to track the sun for long periods, for which a support and drive system is needed. Given that non-tracking detector operation time is also needed, for calibration and background measurements, we target a minimum of 50\% sun tracking capability for the drive system. The plot on the left of figure~\ref{fig:sun_alt_azth} shows the path of the sun in azimuth-altitude coordinates in one full year for the geographical position of DESY. On the right of the same figure, the integrated sun tracking time for a maximum elevation shows that a minimum elevation of \SI{+-21.1}{\degree} is required to assure a \SI{50}{\%} of exposure time over the year. 
The magnet needs to be rotated by \SI{360}{\degree} during normal sun tracking, before it revolves
back to its starting position at a faster speed. The movement can be performed using a rotating platform with an elevation drive, or a tower with both azimuth and elevation drive systems. The main dimensions and requirements on the drive system are summarized in table~\ref{tab:drive1}.

\begin{table}[]
\centering
\begin{tabular}{m{3cm} l c l c} \hline
         &  \textbf{BabyIAXO}       &         &    &  \textbf{CTA MST}     \\ \hline
\textbf{Technical Data}     &     &    &     &      \\ 
         & Magnet length      &  \SI{11}{m}  &  Diameter   & \SI{12}{m}        \\ 
         & Total length          & \SI{21}{m}   &  Focal length  &  \SI{16}{m}     \\ \
         & Weight of magnet  &  \SI{35}{\tonne}   & optical system  &   \SI{53.6}{\tonne}    \\ 
         & Load on drive system   &  \SI{71.6}{\tonne}    &                    &  \SI{53.6}{\tonne}     \\ \hline
\textbf{Requirements on drive system }  &          &                &    &      \\ 
           & Movement in altitude   & \SI{+-25}{\degree}    &  &  \SIrange{-2}{95}{\degree}      \\
           & Movement in azimuth   & \SI{360}{\degree}    &   &  \SI{360}{\degree} (\SI{540}{\degree})   \\
           &  Speed of movement  &  &   & \\ 
           &      - normal tracking     & speed of Sun &  & speed of stars     \\
           &        - fast movement  & & &  \SI{<90}{\second} \\
           & Pointing precision & & & \\
           & - during tracking &  \SI{<0.01}{\degree} & & \SI{<0.1}{\degree} \\
           & - RMS post-calibration & & & \SI{<7}{\arcsecond} (\SI{<0.002}{\degree})  \\ \hline
\end{tabular}
\caption{\label{tab:drive1}Comparison of the overall dimensions and requirements on the 
	drive system for BabyIAXO and the CTA MST.}
\end{table}

\begin{figure}[t!]
\begin{center}
 \includegraphics[width=\linewidth]{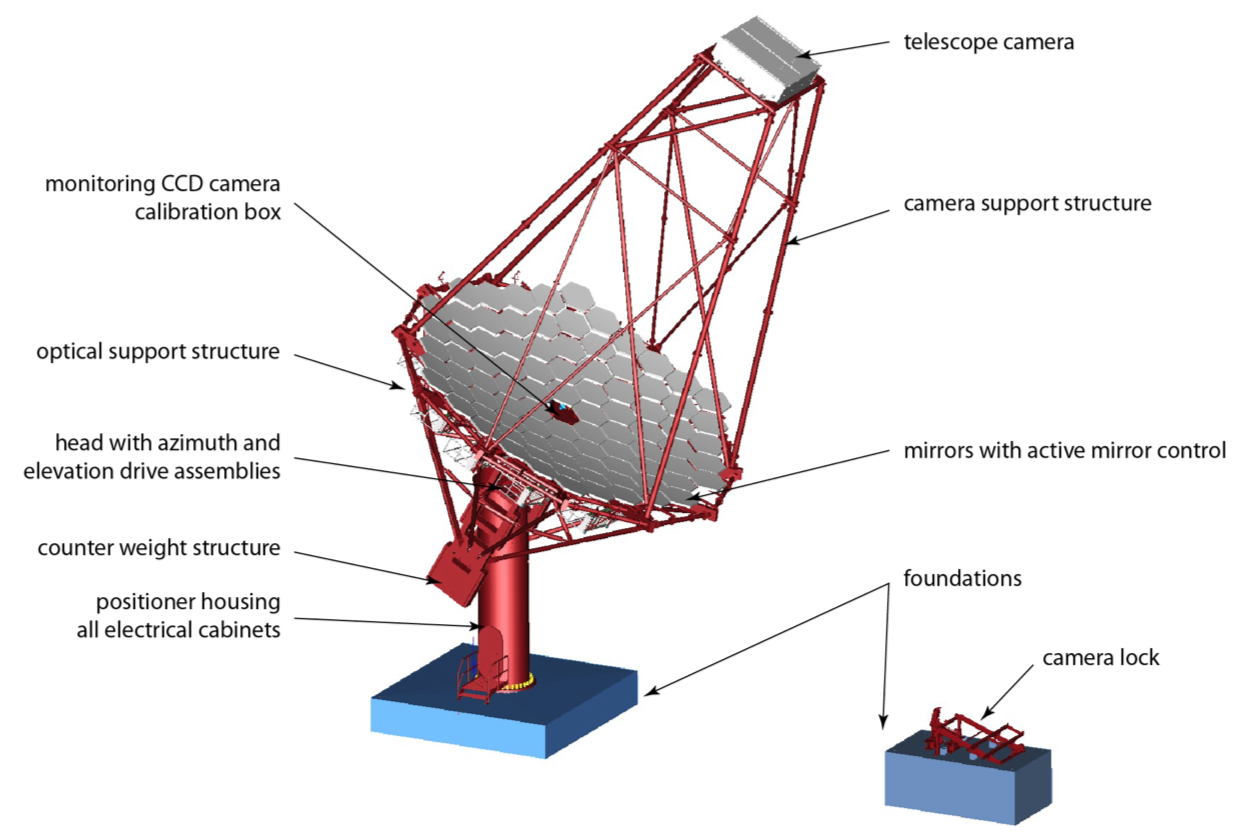}
\caption{Overview of the complete MST assembly~\cite{Garczarczyk:2015zya}. }
    \label{fig:drive1b}
\end{center}
\end{figure}

The DESY CTA group in Zeuthen has been leading the effort of designing  
the tower and positioning system of the Medium-Sized Telescope (MST)~\cite{Garczarczyk:2015zya}, which is part of the Cherenkov 
Telescope Array (CTA). 
Figure~\ref{fig:drive1b}  shows a schematic overview of the MST with the main component. 
CTA will comprise of 40 MST telescopes, 70 Small-Sized Telescopes plus some Large-Sized Telescopes.  
As part of the design, the Zeuthen group has played a leading role in building, commissioning
and testing a prototype MST  telescope located in Berlin-Adlershof. 
The goals of the prototype were to test and optimise the behaviour of the mechanical structure under loads and environmental
influences, the behaviour of the drive and safety concepts and in addition a complete checkout 
of the functionality of the mirror control and alignment system.
A lot of experience was gained during the operation of the prototype. The design goals were achieved.

The weights and loads of  BabyIAXO and CTA MST are compared in table~\ref{tab:drive1}. They are
similar, including the total weight of both structures. 
A detailed mechanical study has been in progress integrating the MST and BabyIAXO designs. 
Details will  be presented  in the following sections. 
Figure~\ref{fig:drive2a} shows the CAD model of the  BabyIAXO magnet, telescopes and detectors supported by an outer support
frame  mounted onto the MST positioner~\cite{Garczarczyk:2015zya}. 
All components in red are part of the original MST design. 
The MST positioner is very well suited to be used and adapted for BabyIAXO, and will actually be used.  
The positioner was disassembled in February 2020. All items to be used for BabyIAXO were shipped
to DESY, Hamburg in April and moved to the HERA South Hall.

  \begin{figure}[!t]
\begin{center}
\includegraphics[width=0.8\textwidth]{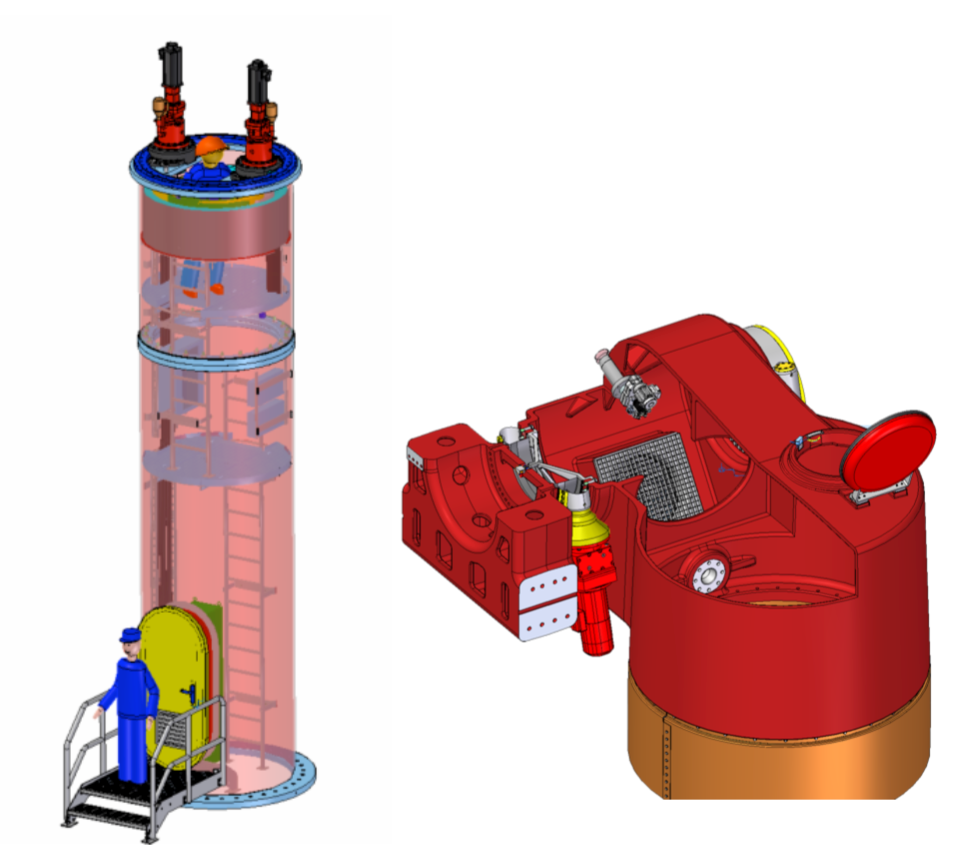}
\caption{CAD view of the MST tower and head~\cite{Garczarczyk:2015zya}. The scale of the two views is not the same.}
    \label{fig:drive2}
\end{center}
\end{figure}
The complete BabyIAXO structure and drive system will consist of several components: 
the foundation, the tower, rotating head with yoke including the elevation
drives, two counter weights and a large support frame supporting the magnet, telescopes, 
detectors and infrastructure items. 
These components will be described in the following sections. 
The information concerning MST is taken from the CTA MST Technical Design Report~\cite{Garczarczyk:2015zya}.

\subsection{Foundations}

\begin{table}[t!]
\centering
\begin{tabular}{llccc} \hline
                                      &  component & number  &   weight (t)     &    weight (t)    \\ \hline 
\textbf{Load on support frame}  &        &                &                      &                 \\ 
                                      &  magnet      &  1            & 35   &  35     \\ 
                                      &  cryo coolers &   7        & 0.3   &     2.1   \\ 
                                      &  telescopes   &   2        & 0.5   &      1      \\ 
                                      & detector + shielding  &  2 & 4   &   8    \\ 
                                      &  vacuum system &      2    & 1.5   &   3     \\
                                      &  muon veto   &   2        & 0.5   &     1     \\ 
                                      &  electronics   &    1       &  1    &  1     \\ 
                                      &  platforms   &  1           &  1    &   1     \\ 
                                      &  sum           &               &        &   52.1    \\ 
\textbf{Additional load on yoke} &     &           &    &       \\ 
                                      &  support frame      &   1    &  25  &  25     \\ 
                                      &  counter weights   &  2     & 10   &  20     \\ 
                                     &  sum             &   &    &   45    \\                                                                            
                                     &   &       &    &       \\ 
\textbf{Additional load on floor}  &   &       &    &       \\ 
                                     &  yoke* &   2    & 2   &  4     \\ 
                                     &  head* &   1    &  5  &  5     \\ 
                                     &  tower* &    1   &  7  &  7     \\ 
                                     &  auxiliary assemblies* &  1     & 3.5   & 3.5      \\ 
                                     &  sum  &       &    &   19.5    \\ 
                                     &    *CTA MST &       &    &       \\                                                                                                                                                                                                                                                                                                  
 \textbf{Total sum}         &   &       &    &    117    \\                                                                                                                                                                                                                                                                                                                                       
 \hline			
\end{tabular}
\caption{\label{tab:drive2} Summary of weight and load of the BabyIAXO components.}
\end{table}


Table~\ref{tab:drive2} summarizes the total weight and load of  BabyIAXO 
including the complete structure and drive system. 
The loads onto the positioner are \SI{960}{kN} for BabyIAXO and \SI{540}{kN} for the MST,   
and on the foundation of \SI{1170}{kN} and \SI{855}{kN} for BabyIAXO and MST, respectively.
In case of  MST, the role of the tower foundation is to transfer the load of the  telescope to the soil,
in order to ensure the stability of the telescope against overturning during operation and extreme environmental
conditions (wind, ice load and seismic activities). 
The foundation consists of a \SI{1.3}{m} thick reinforced concrete slab of \SI{7x7}{m} size. 
The tower flange is bolted to the foundation using 36 M48 (10.9) bolts. 
The safety parameters of the tower foundation are safety factors against overturning of
3.15 for self-weight of the full structure, and 2.15 in case of extreme wind conditions. 

In case of BabyIAXO, the tower will be directly mounted onto the \SI{1}{m} thick concrete slab of the
HERA hall floor.  
The pressure on the hall floor is \SI{730}{kN/m^2},  which is only \SI{9}{\%} of the allowed surface pressure on
the concrete slab without taking the reinforcement of the concrete into account~\cite{Platzer}. 
For comparison, the maximal load of  the former ZEUS detector in the HERA South Hall was up to  \SI{2.0}{MN/m^2}. 
In principle, 36 M20 bolts with dowels would be sufficient to fix the tower to the floor. 
We are planning to use 36 M30 bolts, since they are easily available. 
Although the total mass of BabyIAXO is somewhat larger than the MST values, the overall 
requirements on the BabyIAXO support are relaxed w.r.t. the MST, since
 forces due to extreme environmental conditions (wind and ice loads) do not apply
inside a hall.  
In addition, seismic forces are not an issue, since there are no seismic activities in Hamburg
according to DIN EN 1998-1/NA.

\subsection{Positioner: tower, head and yokes}

The main functions of the tower are:
\begin{itemize}
\item Transfer of the load of the head to the foundation;
\item Provide resistance against bending and dynamic swinging (high stiffness);
\item Provide storage for the electrical equipment and allow for maintenance activities.
\end{itemize}

The tower has a cylindrical shape with an outer diameter of \SI{1.80}{m} and a wall thickness of \SI{20}{mm}. 
It carries the azimuth bearing, which is also the interface to the head. 
The interface of the tower to the foundation is based on a flange with pre-stressed bolts, 
allowing alignment of the tower tilt during the assembly process. 
Access to the tower is possible through a door at ground level or via two hatches on top of the head.
There are floors in the tower, used as maintenance platforms providing access to electrical distributions
and electronics.
The top level below the head allows for maintenance of the azimuth drive system. 
The cables guide through the head are arranged in a way that a rotation  of the azimuth axis by more
than \SI{360}{\degree} is possible, actually up \SI{560}{\degree} is possible in case of the MST. 

\begin{figure}[t]
\begin{center}
 \includegraphics[width=0.49\textwidth]{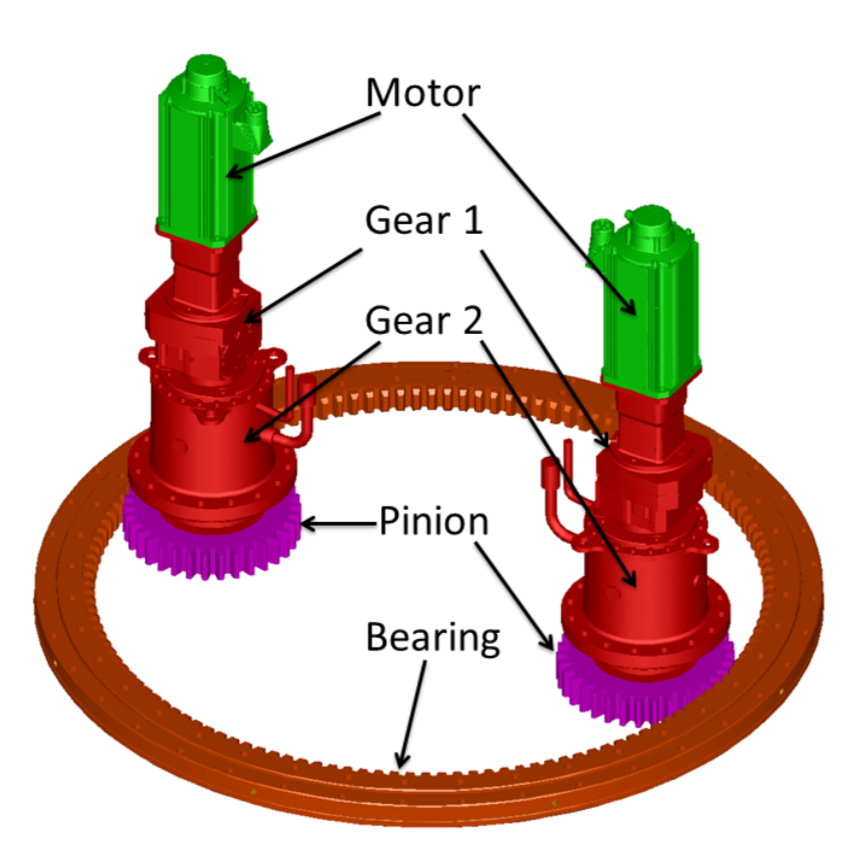}
 \includegraphics[width=0.49\textwidth]{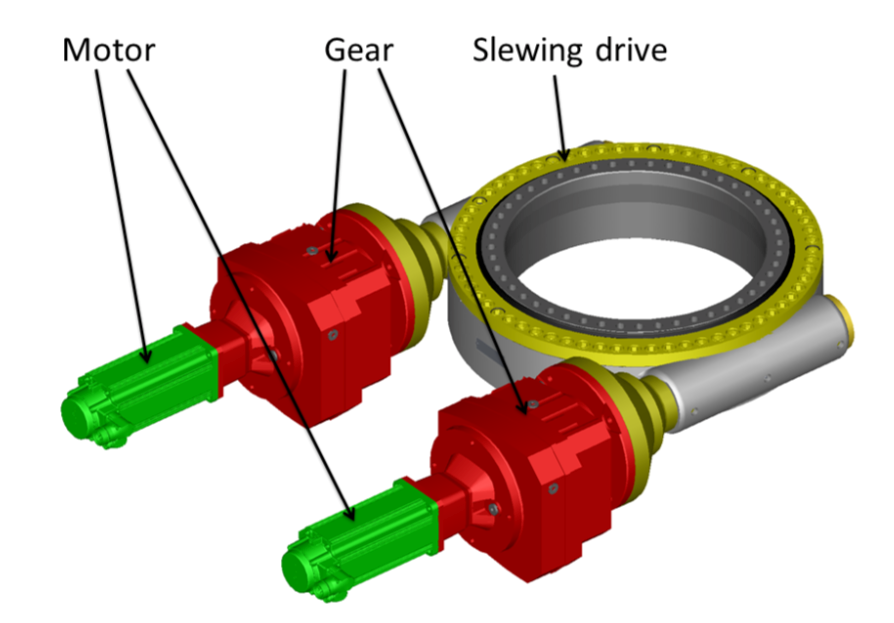}
\caption{Left: azimuth drive sub-assembly~\cite{Garczarczyk:2015zya}. Right: elevation drive sub-assembly~\cite{Garczarczyk:2015zya}. }
    \label{fig:drive4}
\end{center}
\end{figure}

The head is the interface connecting the tower through two yokes with the BabyIAXO support frame. 
The azimuth drive assembly (motors, gears and encoders) is located inside the head.
The specifications for the machined precision and strength of the head material are very high 
and have been demonstrated with the prototype in Adlershof. 
It has to connect and hold the large diameter bearing and should show limited deformations due
to forces caused by gravity and movement momenta. Additional forces taken into account
for MST, such as strong winds and ice on the structure,  are of course not an issue for BabyIAXO.
The function of the yokes is to provide a connection between the head and the support frame,
as well as for the counter weight structure. 
Figure~\ref{fig:drive4} shows a 3D model of these two components.

\subsection{Mechanical elements of the drive assemblies}

The components of the drive assembly were carefully selected by the CTA design team in order to 
fulfill the requirements of the original the MST. A detailed Drive-Failure Mode and Effects Analysis was performed in order to study all failure modes. 
The main components of the azimuth system are shown in figure~\ref{fig:drive4}, 
containing a ball bearing slewing ring, two pinions and two motors with integrated planetary gears.
Two motors were chosen in order to obtain high reliability, accuracy and flexibility. 
The function of the ball bearing is to transfer the axial and radial force, as well as the tilting 
momenta. The ball race diameter is \SI{1800}{mm}, with an axial and radial run-out of less than \SI{0.1}{mm} without
any backlash. 
A gear ratio of 1:4 between the bearing and pinion was chosen. Including the gear ratios of the
two gears of 986 the total gear of the azimuth axis is 3944. 
The standard synchronous motor from Bosch-Rexroth can supply torque at standstill and show
a good running smoothness even at a speed of 1 rpm.

The elevation drive sub-assembly connects the head with the yoke and provides the rotation of
the support frame. 
It is split into two separate sub-assemblies, left and right.
Each side contains a slewing drive, two gears and two motors as shown in figure~\ref{fig:drive4}. 
Similar to the azimuth drive system, the elevation system uses two motors per slewing drive resulting in a higher reliability, accuracy and
flexibility. The bearing has an axial and radial run-out of less than \SI{0.1}{mm}.
A gear ratio of 216.1 was selected for motor speeds ranging between 1 and 3600 rpm.

Redundant encoders are used for each axis in order to ensure that the orientation is known in case 
one of the encoder fails. 
The accuracy of the encoders is \SI{+-20}{arcsec}.




\subsection{Support frame}

\begin{figure}[!b]
	\centering
		\includegraphics[width=7cm,valign=c]{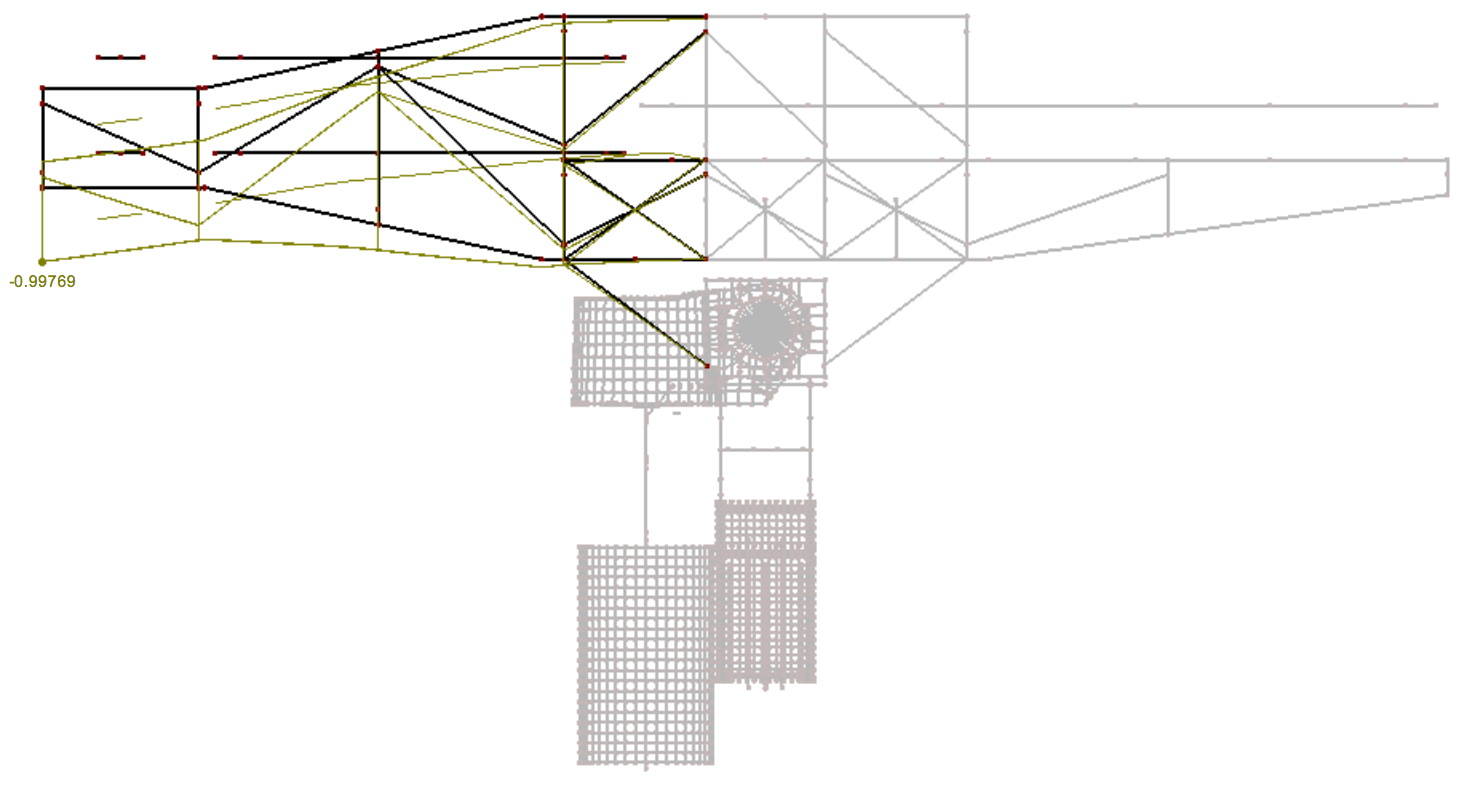}
		\includegraphics[width=7cm,valign=c]{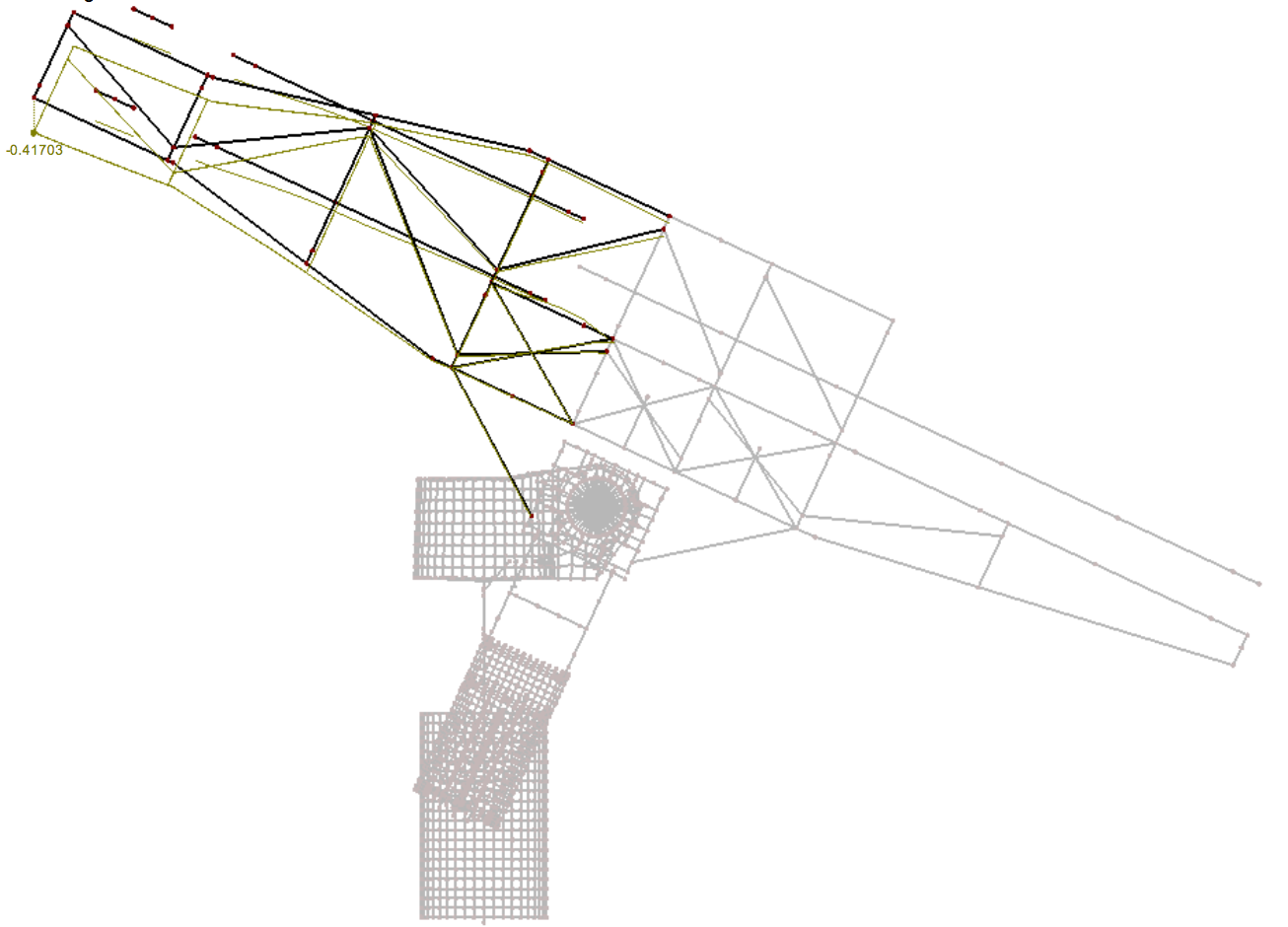}
	\caption{Deformation of the backside of the support frame due to the gravitational load at \SI{0}{\degree} and \SI{25}{\degree}.  For illustration, 
	the deformations are scaled up by a factor of 1000 (left) and 1200 (right figure). }
	\label{fig:drive6}
\end{figure}

The support frame is the interface between the tower/head/yoke and the BabyIAXO assembly consisting 
of magnet, telescopes, detectors and other items. 
The design requirements are:
\begin{itemize}
\item Support magnet, optics, vacuum system and detectors with equipment;
\item Small differential deformation during movement;
\item Provide access for personnel to magnet service box, telescopes, vacuum system, detectors, 
muon veto, electronics and other equipment via platforms and catwalk.
\end{itemize}
  
The requirements of deformation resulting in displacements of the telescopes and detector are:
\begin{itemize}
\item The detector position has to be determined with a precision of \SI{1}{mm} for  the elevation range  of \SI{+-25}{\degree}. This is mainly a requirement on the survey, although a small displacement of a few mm would be desirable.
\item The requirements on the mounting of the optics are more stringent  in order to maintain high efficiency. 
The movement of the spot position in the detector due a rotation of  the telescope
should to be less than \SI{1}{mm} during all movements.
\end{itemize}

A preliminary design of the support frame is shown in figure~\ref{fig:drive2a}. It consists of rectangular profiles, \SI{300 x 300 x 8}{mm}, made out of stainless steel  1.4301.
The frame is bolted to the yoke and supports the magnet, telescopes, detectors and other equipment
if necessary. The total  weight of the frame itself is \SI{17}{t}.
The horizontal deformation of the frame at the very end behind of the detectors is \SI{-1.0}{mm} at \SI{0}{\degree} and \SI{-0.4}{mm} for the elevation of \SI{25}{\degree} as shown in figure~\ref{fig:drive6}, i.e. a change of the detector position of \SI{0.6}{mm} due to the elevation change. For the full movement  from \SI{+25}{\degree} and \SI{-25}{\degree} this amounts to \SI{1.2}{mm}.
At the location of the optics, the bending of the frame corresponds to a
rotation of \SI{0.0046}{\degree}  at \SI{0}{\degree} elevation. The position of the optics can  be adjusted
to take this in account. At \SI{25}{\degree} elevation the rotation is \SI{0.0014}{\degree}. 
The difference between \SI{0}{\degree} and \SI{+25}{\degree}  results in a rotation of the optics of \SI{0.0032}{\degree}. This corresponds to spot movement at the detector of \SI{+0.4}{mm} (at \SI{8}{m} distance).
The maximum stress due to the load is \SI{40}{N/mm^2}, much less than the yield strength of \SI{190}{N/mm^2}.

Counter weights are attached to the yokes for balancing the torque on the elevation system introduced by the support frame. 
This is necessary for safe operation of the motors, gears and bearings and to reduce the power consumption of the motors
during operation.

%% file: sections/BabyIAXO_infra.tex

In addition to the specific infrastructure associated with the BabyIAXO magnet, the BabyIAXO setup will require 
a large experimental hall to host the experiment and including its structure and drive system, 
a counting room to host the main controls and data acquisition systems, as well as ancillary working space to 
assist the operations in the main hall. It will also require the availability of a minimal technical force in situ, to support 
the construction and maintenance of the infrastructure and coordinate the activities coming from the collaboration institutes. 

\subsection{Civil engineering}

\begin{itemize}
\item A main experimental hall to host BabyIAXO, with a footprint encompassing a circle of 21 m diameter and a height of 13 m. This hall must also:
\begin{itemize}
\item have easy access for big transports or cranes (the 10 m magnet must be entered to the hall) with a sufficiently large entrance;
\item be equipped with a crane for a max. weight of 50 tons;
\item have a standard ventilation system to smooth temperature variations (but without particular stability specification);
\item have a concrete slab that can take the weight of the whole experiment (about 120 tons).
\end{itemize}
\item A counting room adjacent to the main hall (and ideally near to the experiment) of about 30~m$^2$, to host the computers to 
control the experiment's systems and acquire data, and space for the experiment operators and shiftcrew.  
The room must have standard high speed internet connection for data transfer; 
\item A small gas cabinet, located outside, able to host several gas bottles with the appropriate gas system to bring gas supply to the inside halls;
\item A second experimental room (or a series of rooms amounting to the equivalent area) to provide some working area for support to the main experiment, and including some services like: 
\begin{itemize}
\item A laminar flow cabinet of at least 4~$\times$~4~m$^2$ area, for clean interventions (detectors, optics,...);
\item A small mechanical workshop to support quick in-situ interventions (\SI{15}{m^2});
\item A gas detector working area with gas supply line from the gas cabinet (\SI{15}{m^2});
\item A general purpose working area (e.g. for electronics tests, detector calibrations, etc.), with several working benches, general storage, etc. (\SI{40}{m^2});
\item A cryogenics supply area might be needed e.g. containing liquid N$_2$ storage tank, in order to reduce time for cool down of the magnet. 
\end{itemize} 
\item Some office space for the local team and visiting scientists from the collaboration. 
\end{itemize}

\subsection{HERA South Hall}

\begin{figure}[t]
	\centering
	\includegraphics[width=0.49\textwidth,valign=c]{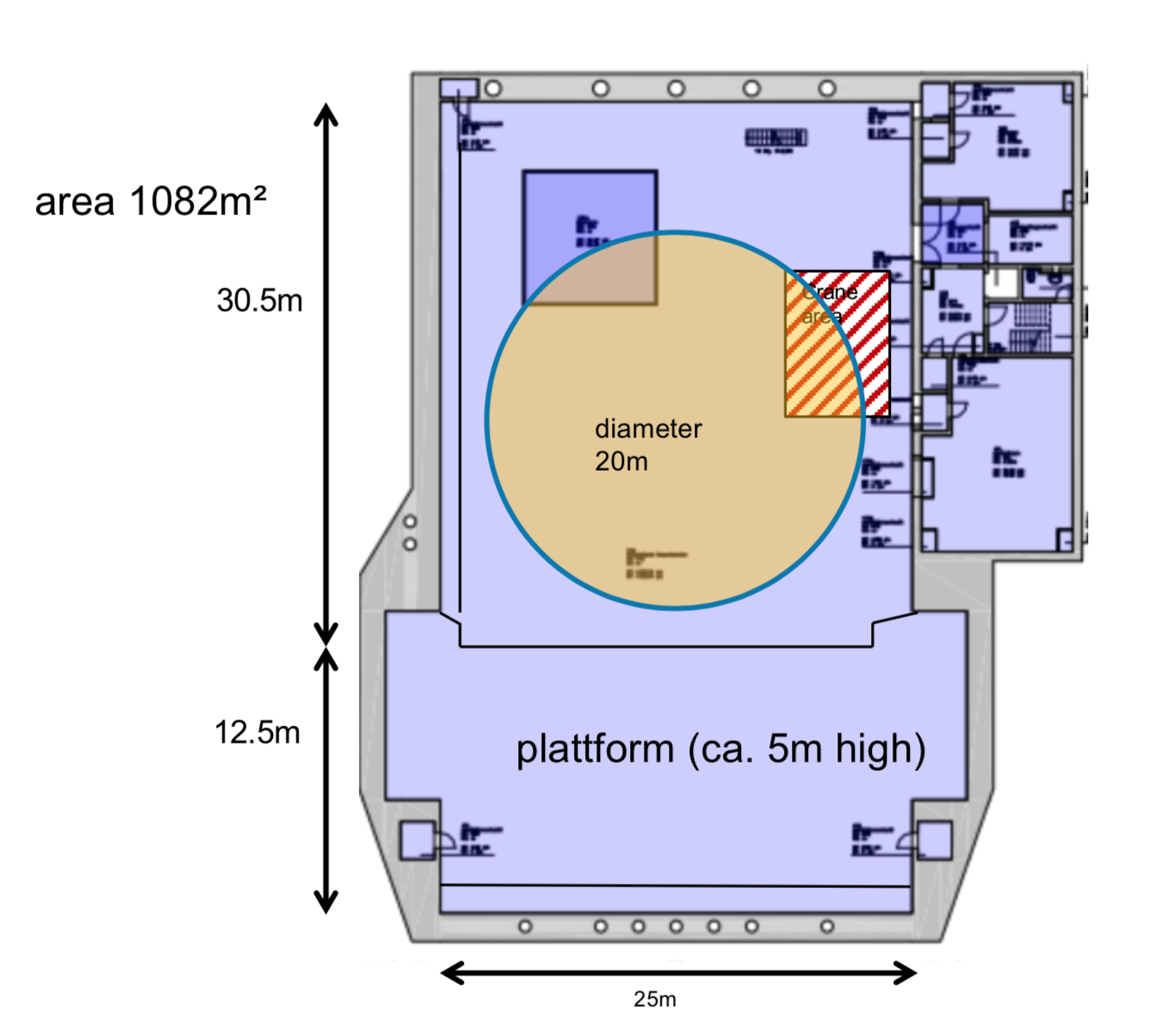}
	\includegraphics[width=0.49\textwidth,valign=c]{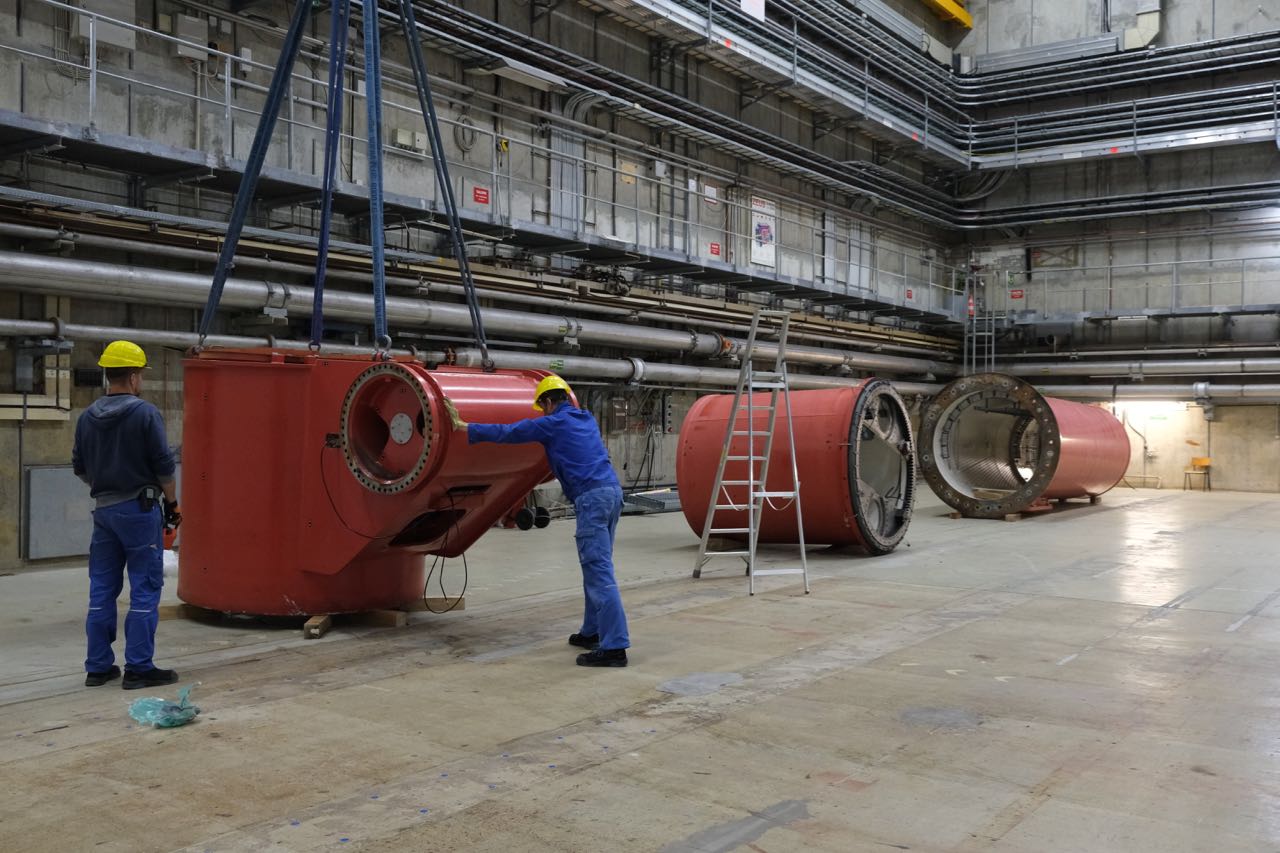}
	\caption{Left: Floor plan of the HERA South Hall including a circle of \SI{20}{m} diameter indicating the movement of BabyIAXO. Right: Photograph of the South Hall craning in the head of the structure and drive system. }
	\label{fig:southhall_footprint}
\end{figure}

The availability and suitability of the HERA halls has been studied, since the surface halls at DESY are all being used. While the underground HERA halls West and North are used for the detector lab, PR exhibition (HERA West) and the ALPS II and the proposed MADMAX experiments (HERA North Hall), the HERA South and East halls are in principle available for future activities. 

Both underground experimental halls have a size of \SI{43x25}{m}, a height of \SI{15}{m}.
The South Hall (the location of the former ZEUS detector) was chosen for two reasons.  It is 
equipped with two \SI{40}{\tonne} cranes, 
whereas the East Hall only has one \SI{40}{\tonne} crane.
In addition, using the existing plumb line point and erecting two new
points would be straight forward in the South Hall, and would not be
possible in the East Hall. 

Items can be lowered into the hall by two \SI{20}{\tonne} shaft cranes. The ground floor was designed to support the \SI{3000}{\tonne} ZEUS detector. In addition to the experimental hall, the building has several other rooms which were used as control room of the ZEUS experiment, meeting room, mechanical workshop, space for power supplies and air-conditioning as well as rooms for computing, storage and gas supplies. Since 2010, the hall was used for activites of the European X-Ray Free-Electron Laser Facility (XFEL), which were finished in 2019.
The hall is still in very good shape, and ideally suited for housing BabyIAXO, except that liquid He and cooling water is no longer available since the shut-down of the HERA accelerator in 2007. The liquid He supply line was cut to be used for the XFEL. A refurbishment would require significant investments. The heat loss of the He line from the West to the South Hall would be considerably higher compared to the requirements of the BabyIAXO magnet. 
As described in the magnet section \ref{sec:BabyIAXO_magnet}, we have therefore decided setting up a small, independent cryocooling system for the magnet.

A new, dedicated water cooling system could be installed if needed for the operation of BabyIAXO. This could be part of the refurbishment of the HVAC of the HERA halls, which is in preparation. 

As  mentioned in section \ref{sec:BabyIAXO_drive}, the MST components have already been moved to the HERA South Hall. After maintenance and a few modifications they will be ready to be assembled in the HERA hall.   


\subsection{Data acquisition and control}

Data acquisition and control systems raise a significant challenge to modern experiments in particle physics. On one hand, readout and control systems are no longer in domain of scientists and are mostly standardized and implemented by manufacturing companies. On the other hand, we seldom use unique devices that require unique control and readout procedures and integrations. The BabyIAXO case is complicated by the fact that it will use several different detectors developed by different teams and those detectors should be used interchangeably.

In order to both implement an industrial-level control system and be able to incorporate individual unique devices, BabyIAXO will feature a DAQ and control integration mechanics developed in close cooperation by MIPT, DESY and other industrial partners. The integration scheme is based on reactive event streams. Each subsystem used to read the data or control devices has two "connector" streams of events: one for input and one for output. Those streams could be connected to a central event dispatching service (or multiple distributed services). In order to integrate a new device or system, one needs only a tiny service that "translates" events from the language, specific to the device to universal language understood by the framework. The same works for commands sent by other system users (for example system operators, or other devices). Those commands are translated in-flight into internal commands understood by the device.

The parts of the control system dedicated to standard equipment like cryogenics, mechanics, etc will be implemented using the DOOCS framework (\cite{DOOCS}). The unique device drivers will be managed by the DataForge-control API implemented with kotlin-multiplatform technology (the prototype is in the active development and is available at \cite{dataforge-control}). The interconnection between different systems and the interchange event formats is currently being developed by a Waltz-controls collaboration (\cite{waltz-rfc}).

%% file: sections/BabyIAXO_conclusions.tex
In this paper we have presented the conceptual design of BabyIAXO, an intermediate-scale axion helioscope to be hosted at DESY. The project is based on the 15-years long experience acquired in CAST, as well as on state-of-the-art technologies on superconducting magnets, x-ray optics and low background detectors. Despite the considerable scaling-up step with respect to CAST, the project design does not rely on untested solutions or pending R\&D. BabyIAXO is conceived as an intermediate stage in size, and will allow to test all subsystems (magnet, optics and detectors) at a relevant scale, further mitigating risks of the final IAXO.

Apart from being a technological prototype of IAXO, BabyIAXO itself will already produce relevant physics outcome. The latest result from CAST has improved, for the first time for an axion helioscope, the best limits on $\gagamma$ from astrophysics, and has marginally entered into unexplored ALP and axion parameter space. BabyIAXO, and later on IAXO, will go well beyond this point and venture deep into unexplored area, in particular probing a large fraction of QCD axion models in the meV to eV mass band. This region encompasses the values invoked to solve the transparency anomaly, and the ones solving the stellar cooling anomalies, as well as a number of other possible hints. It also includes some of the axion models that can account to the totality of DM.

Most of this region is only realistically at the reach of the helioscope technique, and therefore the BabyIAXO/IAXO program is unique in the wider landscape of experimental axion searches. Given that the potential of the experiment does not rely of assuming DM is made of axions, in case of a non-detection, robust limits on the axion parameters will be set. In the case of a positive detection, (Baby)IAXO will permit high-precision measurements campaigns to determine axion parameter and getting further insight on the underlying axion model. In any case, IAXO will likely play an important role in the future of axion and ALP research, with potential for a discovery, even already at the BabyIAXO stage.